\documentclass[twoside,11pt]{article}
\usepackage{amsmath}
\usepackage{jmlr2e}
\usepackage{subfigure}

\newtheorem{assumption}{Assumption}
\newtheorem{condition}{Condition}

\newcommand{\ba}{\mbox{\bf a}}
\newcommand{\bb}{\mbox{\bf b}}
\newcommand{\bc}{\mbox{\bf c}}
\newcommand{\be}{\mbox{\bf e}}
\newcommand{\bff}{\mbox{\bf f}}

\newcommand{\bk}{\mbox{\bf k}}
\newcommand{\bI}{\mbox{\bf l}}

\newcommand{\bu}{\mbox{\bf u}}
\newcommand{\bv}{\mbox{\bf v}}

\newcommand{\bx}{\mbox{\bf x}}

\newcommand{\bz}{\mbox{\bf z}}
\newcommand{\bm}{\mbox{\bf m}}
\newcommand{\bA}{\mbox{\bf A}}
\newcommand{\bB}{\mbox{\bf B}}
\newcommand{\bC}{\mbox{\bf C}}

\newcommand{\bD}{\mbox{\bf D}}
\newcommand{\bK}{\mbox{\bf K}}

\newcommand{\bE}{\mbox{\bf E}}
\newcommand{\bF}{\mbox{\bf F}}

\newcommand{\bM}{\mbox{\bf M}}

\newcommand{\bP}{\mbox{\bf P}}
\newcommand{\bQ}{\mbox{\bf Q}}
\newcommand{\bR}{\mbox{\bf R}}

\newcommand{\bU}{\mbox{\bf U}}
\newcommand{\bV}{\mbox{\bf V}}
\newcommand{\bW}{\mbox{\bf W}}
\newcommand{\bX}{\mbox{\bf X}}

\newcommand{\bY}{\mbox{\bf Y}}
\newcommand{\bZ}{\mbox{\bf Z}}

\newcommand{\bzero}{\mbox{\bf 0}}
\newcommand{\bveps}{\mbox{\boldmath $\varepsilon$}}

\newcommand{\bbeta}{\mbox{\boldmath $\beta$}}
\newcommand{\btheta}{\mbox{\boldmath $\theta$}}
\newcommand{\bgamma}{\mbox{\boldmath $\gamma$}}

\newcommand{\bxi}{\mbox{\boldmath $\xi$}}

\newcommand{\bLambda}{\mbox{\boldmath $\Lambda$}}

\newcommand{\bOmg}{\mbox{\boldmath $\Omega$}}

\newcommand{\bSigma}{\mbox{\boldmath $\Sigma$}}

\newcommand{\RR}{\mathbb{R}}

\newcommand{\var}{\mathrm{var}}
\newcommand{\cov}{\mathrm{cov}}

\newcommand{\Sig}{\mathbf{\Sigma}}

\newcommand{\diag}{\mathrm{diag}}

\newcommand{\supp}{\mathrm{supp}}

\DeclareMathOperator*{\argmin}{arg\,min}
\DeclareMathOperator*{\argmax}{arg\,max}

\def\spacingset#1{\renewcommand{\baselinestretch}%
{#1}\small\normalsize} \spacingset{1}

\jmlrheading{1}{2020}{}{0/00}{0/00}{hft19}{Xu Han, Ethan X. Fang and Cheng Yong Tang}
\ShortHeadings{PROD for High Dimensional Linear Regression}{Han, Fang and Tang}
\firstpageno{1}

\begin{document}


\title{Pre-processing  with Orthogonal Decompositions for High-dimensional Explanatory Variables }

\author{\name Xu Han \email hanxu3@temple.edu \\
       \addr Department of Statistical Science\\
       Fox Business School, Temple University\\
       Philadelphia, PA 19122, USA
       \AND
       \name Ethan X. Fang \email xxf13@psu.edu \\
       \addr Department of Statistics\\
       The Pennsylvania State University\\
       University Park, PA 16802, USA
       \AND
       \name Cheng Yong Tang \email yongtang@temple.edu\\
       \addr Department of Statistical Science\\
       Fox Business School, Temple University\\
       Philadelphia, PA 19122, USA}

\editor{}

\maketitle

\begin{abstract}

Strong correlations between explanatory variables are problematic for  high-dimensional regularized regression methods.  
Due to the violation of the Irrepresentable Condition,  the popular LASSO method may suffer from false inclusions of inactive variables. 
In this paper, we propose pre-processing with orthogonal decompositions (PROD) for the explanatory variables in high-dimensional regressions. 
The PROD procedure is constructed  based upon a generic orthogonal decomposition of the design matrix.    
We demonstrate by two concrete cases  that the PROD approach can be effectively constructed for improving the performance of high-dimensional penalized regression.  
Our theoretical analysis reveals their  properties and benefits for high-dimensional penalized linear regression with LASSO. Extensive numerical studies with simulations and data analysis show the promising performance of the PROD.

\end{abstract}








\section{Introduction}
 
Identifying important variables  associated with the response variable of interest from a large number of explanatory variables has been a fundamental problem in high dimensional inference, with wide applications to many scientific areas, such as biology, genetics, medicine, neuroscience and finance. We refer to the overviews by \cite{FanLv_2009_Sinica}, \cite{Fanetal_2011_ARE},  the book length monographs of \cite{Buhlmannvan_2011}, \cite{Hastieetal_2015}, and references therein for a big picture of the recent  literature. 
Additionally,    modern large and complex data sets are often  accompanied by complicated and informative structures so that dependence between variables are common due to various reasons; see, for example, the discussion of challenging big data problems in \cite{Fanetal_2014_NSR}. 

We consider the linear regression model
\begin{equation}\label{lm}
\bY=\bX\bbeta+\bveps,
\end{equation}
where $\bY$ is an $n$-dimensional vector of the observed response variables, $\bX$ is an $n\times p$ matrix containing the observations of the $p$ explanatory variables -- referred to as the model matrix,  $\bbeta$ is a $p$-dimensional unknown sparse vector, and $\bveps$ is the random error.  
We are interested in the  penalized regression methods that minimize 
\begin{align}\label{eq:preg}
n^{-1}\|\bY-\bX\bbeta\|_2^2+\sum_{j=1}^p P_\lambda (|\beta_j|)
\end{align}
with respect to $\bbeta$, where $P_\lambda(\cdot)$ is some penalty function with tuning parameter $\lambda$.  When $P_\lambda(|\beta|)=\lambda|\beta|$,  (\ref{eq:preg}) becomes the LASSO \citep{Tibshirani_1996_JRSSB}. Other popular choices of the penalty functions include the SCAD \citep{FanLi_2001_JASA},  the elastic net \citep{Zou2005}, the adaptive LASSO \citep{Zou_2006_JASA},  the bridge estimator \citep{Huangetal_2008_AOS}, the adaptive elastic-net \citep{ZouZhang}, the folded concave penalty \citep{Lv1}, the MCP \citep{Zhang_2010}, the hard thresholding \citep{Zengetal_2014}, the CARDS \citep{Keetal_2015}, the LAMP \citep{Fengetal_2018}, among others. To avoid digression, we focus our discussions on the LASSO methods in this study.   

It is documented in the literature that strong correlations between explanatory variables are problematic for regularized  regression methods, and can  lead to substantial amount of false positives (inactive variables selected as active variables). For example, for the LASSO method \citep{Tibshirani_1996_JRSSB}, theoretical analysis shows that  the so-called  Irrepresentable Condition  \citep{ZhaoYu_2006_JMLR,Wainwright_2009} is required so that the sparse estimator can correctly identify the inactive variables by its zero components. The Irrepresentable Condition is a constraint on the correlation level between the two unknown sets of explanatory variables: the active ones and the inactive ones; see also \cite{Buhlmannvan_2011}, \cite{Hastieetal_2015} for more discussions on this condition.  Analogous conditions are  assumed for other regularized regression methods;   
see, for example,  \cite{Zou_2006_JASA},  \cite{Lv1}, and \cite{lv2}.  

Recognizing the impact from the aforementioned strong correlations,  
various remedies have been investigated in the literature.  
\cite{Jia_2015} proposed  to simultaneously  left multiplying  a matrix on both the response vector and the model matrix containing explanatory variables.  The  matrix multiplied is calculated from the singular value decomposition of the model matrix. Targeting at partitioning the model matrix for parallel computing,  \cite{Wang2016} studied  decorrelation methods for regularized regression. 
\cite{Paul_2008} proposed to incorporate a latent variable model and then construct ``pre-conditioned'' response variable for  regularized regression analysis.  Also facilitated by some unknown latent structure, \cite{Wang_2012} considered incorporating a  factor model  for the purpose of variable screening.  Other structures have also been considered, e.g. clustering  in \cite{Buhlmannetal_2013}, graphs in \cite{Keetal_2014} and hierarchical structure in \cite{Haoetal_2018}. 
Recently,  \cite{Fanetal_2019} considered decorrelating the explanatory variables by incorporating a latent factor model 
in constructing the regularized likelihood. 



In this paper, we propose \underline{pr}e-processing  with  \underline{o}rthogonal \underline{d}ecompositions (PROD), a generic procedure for the high-dimensional explanatory variables.  
We show that the orthogonal decompositions of $\bX$ in PROD always exist by projecting the original model matrix  to the linear space spanned by some pre-specified variables and its complementary space respectively.  Thanks to the orthogonality in PROD, regressions with the processed variables after the PROD inherit the interpretations associated with the original  model without altering the variance-covariance structure of the model errors. As a resulting benefit, all  the methods  developed under the original model can be  equally applied on the variables constructed from PROD.    

For regularized linear regressions, we analyze with detail two specific cases of the PROD. The first one is by applying the principal component analysis to reduce the level of correlations  in the model matrix by  accounting for the contributions  from some  common latent factors.  
When the model matrix does not possess a clear model structure, we recommend our second proposal of PROD based on random generations, by constructing the orthogonal decomposition based on some randomly generated variables.  Our theory confirms the effect of a reduced level of correlations between the explanatory variables, which is promising for regularized regression methods to potentially reduce the false positives. Our comprehensive comparative numerical studies illustrate the merits of PROD procedure by showing better variable selection performances in various scenarios.

In a related study to ours, \cite{cevid2020spectral} recently considered a pre-processing  framework with a rich class of spectral transformations by ``trimming'' the singular values of $\bX$ while keeping its left and right eigenvectors intact. Though both our PROD and the framework of \cite{cevid2020spectral} include the principal component analysis method as a special case, they have different focuses methodologically. In particular,  the setting of \cite{cevid2020spectral} concerns the impact from some confounding variables that are correlated with $\bX$ such that issues with regularized  regression methods may arise; see also \cite{chernozhukov2017lava}.   In contrast, our framework attempts to construct effective approaches for reducing the level of correlations in $\bX$, so as to alleviate the restrictions due to the Irrepresentable Condition. As a merit, because of the generality and flexibility of the PROD, one can  try different orthogonal decompositions for  incorporating multiple considerations on $\bX$ in practice, while maintaining the same goal for recovering the sparse support of $\bbeta$. Theoretically, \cite{cevid2020spectral} established results concerning the estimation errors of penalized regression estimators.   While in our theory,  the primary  concerns are on sparse support recovery -- weakening the strength of the Irrepresentable Condition with PROD.  



The rest of this paper is organized as follows. The setting, our methods, and main results are elaborated in Section \ref{s2}.
Extensive numerical evidence by simulations and data analysis are reported in Section \ref{s4}.  We conclude with some discussions in Section \ref{s5}. All the technical proofs are relegated to the Appendix.

Throughout the paper, we use the notations as follows: for a vector $\bx=(x_1,\cdots,x_p)'$, $\|\bx\|_2=\sqrt{\sum_{i=1}^px_i^2}$, $\|\bx\|_1=\sum_{i=1}^p|x_i|$; for a matrix $\bA=(a_{i,j})\in\mathbb{R}^{m\times n}$, the Frobenius norm $\|\bA\|_F=\sqrt{trace(\bA'\bA)}$, the $l_1$ norm $\|\bA\|_1=\max_{1\leq j\leq n}\sum_{i=1}^m|a_{i,j}|$, and the $l_{\infty}$ norm $\|\bA\|_{\infty}=\max_{1\leq i\leq m}\sum_{j=1}^n|a_{i,j}|$.






\section{Methodology and Main Results} \label{s2}

\subsection{Irrepresentable Condition}
For model (\ref{lm}), we denote by $\bbeta_0=(\beta_{01},\cdots,\beta_{0p})'$  the truth of the unknown model parameter $\bbeta=(\beta_1,\dots, \beta_p)'$, where $\bA'$ is the transpose of $\bA$. We consider that the model error $\bveps\sim N(\bzero, \sigma^2 \bI)$ for some $\sigma^2>0$. We assume that all predictors are centered with mean zero. 

Let $S=\text{supp}(\bbeta_0)=\{j: \beta_{0j}\neq 0, 1\leq j\leq p\}$ be the set collecting the indices of the nonzero components in $\bbeta_0$, and $S^c$ be its complement. Let $k=|S|$ be the cardinality of $S$. The Lasso estimator is defined as 
\begin{equation}\label{lasso}
\widehat{\bbeta}_{Lasso}\equiv \argmin_{\bbeta\in\mathbb{R}^p}\{n^{-1}\|\bY-\bX\bbeta\|_2^2+\lambda_n\|\bbeta\|_1\}
\end{equation}
where $\widehat{\bbeta}_{Lasso}=(\widehat{\beta}_{Lasso,1},\cdots,\widehat{\beta}_{Lasso,p})'$, and  $\widehat{S}_{Lasso}=\{j: \widehat{\beta}_{Lasso,j}\neq 0, 1\leq j\leq p\}$. Theoretically,  the variable selection consistency is important: 
\begin{equation*}
P(\widehat{S}_{Lasso}=S)\rightarrow1, \quad\text{ as } n\rightarrow\infty. 
\end{equation*}
The famous Irrepresentable Condition below is known to be sufficient and essentially necessary for the Lasso method to be variable selection consistent, see Chapter 2 in \cite{Buhlmannvan_2011}, \cite{ZhaoYu_2006_JMLR} and \cite{Wainwright_2009}.   
\begin{condition}[Irrepresentable Condition]\label{ir}
There exists $\gamma$ $(0<\gamma\leq 1)$ such that 
\begin{equation}\label{eq1}
\max_{j\in S^c}\|(\bX_S'\bX_S)^{-1}\bX_S'\bx_j\|_1\leq1-\gamma
\end{equation}
where $\bx_j$ denotes the $j$-th column of $\bX$. 
\end{condition}
Clearly, the Irrepresentable Condition imposes the constraint on the level of correlations between the two unknown sets of variables in $S$ and $S^c$. 
When there are strong dependence between them, the Irrepresentable Condition is often violated. As a consequence,    Lasso performs poorly,  often with a substantial amount of false inclusions of variables in $S^c$; see the demonstration with our numerical studies in Section \ref{s4}.   

\subsection{Pre-processing Explanatory Variables }
The foundation of our PROD procedure is the following orthogonal decomposition:
\begin{align}
\bX=\bP_Z\bX+(\bI-\bP_Z)\bX, 
\label{eq:decom}
\end{align}
where $\bP_Z=\bZ(\bZ'\bZ)^{-}\bZ'$ is the projection matrix, $\bA^{-}$ denotes the Moore-Penrose pseudoinverse of a generic matrix~$\bA$, and $\bZ$ is a  $n\times q$ matrix containing  a generic set of variables that can  be, for example,  component(s)  of $\bX$, linear combination(s) of $\bX$,  more general functions of $\bX$, and even a new  set  of constructed variables.   

We define 
\begin{equation}\label{eq5}
\widehat{\btheta}\equiv\argmin_{\btheta\in\mathbb{R}^p}\{n^{-1}\|\bY-(\bI-\bP_Z)\bX\btheta\|_2^2+\lambda_n\|\btheta\|_1\}. 
\end{equation} 
to be the PROD estimator. 
We first demonstrate that it is appropriate to use $\widehat{\btheta}$ for estimating  $\bbeta_0$ in model (\ref{lm}). 
Let a vector $\bz\in\mathbb{R}^p$ be a sub gradient for $\bbeta\in\mathbb{R}^p$, written as $\bz\in \partial\|\bbeta\|_1$. Then the Karush-Kuhn-Tucker (KKT) condition for  (\ref{eq5}) shows that $\widehat{\btheta}$ satisfies
\begin{equation}\label{t1}
2n^{-1}\bX'(\bI-\bP_Z)(\bY-(\bI-\bP_Z)\bX\btheta)+\lambda_n\widetilde{\btheta}=0, 
\end{equation}
where $\widetilde{\btheta}\in\partial\|\widehat{\btheta}\|_1$. Note that $\bY=\bP_Z\bX\bbeta_0+(\bI-\bP_Z)\bX\bbeta_0+\bveps$, by the orthogonality of $\bP_Z\bX$ and $(\bI-\bP_Z)\bX$, then (\ref{t1}) is equivalent to 
\begin{equation*}
2n^{-1}\bX'(\bI-\bP_Z)((\bI-\bP_Z)\bX\bbeta_0+\bveps-(\bI-\bP_Z)\bX\btheta)+\lambda_n\widetilde{\btheta}=0, 
\end{equation*}
which is the KKT condition of 
\begin{equation}\label{of}
\argmin_{\btheta\in\mathbb{R}^p}\{n^{-1}\|\bY_M-(\bI-\bP_Z)\bX\btheta\|_2^2+\lambda_n\|\btheta\|_1\}
\end{equation}
where $\bY_M=(\bI-\bP_Z)\bX\bbeta_0+\bveps$. Therefore, $\widehat{\btheta}$ obtained via (\ref{eq5}) can be viewed as a Lasso-type estimator of the true parameter $\bbeta_0$ via the objective function (\ref{of}).  

The Irrepresentable Condition  (\ref{eq1}) in the context of the PROD estimator  \eqref{eq5} using $(\bI-\bP_Z)\bX$ then becomes
\begin{condition} \label{m}
There  exist some $0<\gamma\leq 1$ such that 
\begin{equation}\label{eq10}
\max_{j\in S^c}\|({\bX}_S'(\bI-\bP_Z){\bX}_S)^{-1}{\bX}_S'(\bI-\bP_Z){\bx}_j\|_1\leq1-\gamma. 
\end{equation}
\end{condition}

Our PROD framework provides a flexible framework by different choices of $\bZ$ for the construction of the orthogonal decomposition in (\ref{eq:decom}). We  illustrate in the following Sections 2.3 and 2.4 that the corresponding Condition \ref{m} based on $(\bI-\bP_Z)\bX$ is weaker than the Irrepresentable Condition Condition \ref{ir}  based on the original $\bX$ under reasonable conditions.    This serves as a theoretical justification that our PROD method is a useful approach in attempting to alleviate the impact due to strong dependence among $\bX$.   

\subsection{PROD with Least Important Components} \label{pca}

The principal component analysis (PCA) is a popular method to decompose the correlation structure of $\bX$.  
A common observation in practice  is that the first a few principal components,  say $q$ of them, account for a substantial portion of the covariances between the components of $\bX$. Hence, after removing the impact from the  first $q$ principal components, the correlation level of the remaining components is expected to be substantially reduced. 
Thus PCA  provides a natural device for PROD, that is, one may identify $\bZ$ as the first $q$ principal components.  

Specifically,  let $\bX=\bU\bD\bV'$  be  the singular value decomposition (SVD) where $\bU$ and $\bV$ are respectively $n\times r$ and $p\times r$  matrices with orthonormal columns,  $r\leq \min(n, p)$ is  the rank of $\bX$, and $\bD$ is the diagonal matrix containing the singular values of $\bX$. By letting $\bU=(\bu_1,\dots,\bu_r)$, $\bV=(\bv_1,\dots,\bv_r)$, and $\bD=\text{diag}(d_1,\dots, d_r)$,  we have $\bX=\sum_{i=1}^r d_{i}\bu_i\bv_i'$. In the decomposition  (\ref{eq:decom}), let $\bZ=(d_1\bu_1,\cdots,d_q\bu_q)$, correspondingly $\bP_Z\bX=\sum_{i=1}^q d_{i}\bu_i\bv_i'$ and $(\bI-\bP_Z)\bX=\sum_{i=q+1}^{r} d_{i}\bu_i\bv_i'$. Note that $(\bI-\bP_Z)\bX$ is based on the last $(r-q)$ components, which distinguish from the first $q$ principal components. Thus, we call our method in this scenario as PROD with least important components. 

This approach works particularly well when $\bX$ follows some factor model structure. Driven by some underlying common factors, $\bX$ usually has several large eigenvalues with other eigenvalues much smaller. Such factor models have been widely applied in financial econometrics and biomedical studies. We will show that under the factor model structure, Condition \ref{m} based on $(\bI-\bP_Z)\bX$ can be easily satisfied. We consider in this case that $\{X_j\}_{j=1}^p$ are realizations from the following factor model 
\begin{equation}\label{eq19}
X_j=\bb_j'\bff+k_j, \quad\quad 1\leq j\leq p
\end{equation}
where $\bff$ contains the latent $s$-dimensional factors with $s\ll p$,   $\bB=(\bb_1,\cdots,\bb_p)'$ is the $p\times s$ factor loading matrix, and $\bK=(k_1,\cdots,k_p)'$ contains the so-called idiosyncratic contributions from each of the $p$ components in the predictors.  Let $\bSigma_K=(\sigma_{K,ij})_{p\times p}$ be the population covariance matrix of $\bK$. In classical multivariate analysis, $\bSigma_K$ is assumed to be diagonal implying uncorrelated components of $\bK$.  We show that as long as the correlation level of $\bK$ is low at the population level,  then PROD with least important components is expected to work well. In particular, under the factor model (\ref{eq19}), we have the following theorem. 
\begin{theorem}\label{th2}
Suppose $\bbeta_0$ in (\ref{lm}) has $k$ nonzero entries. If $\{\sigma_{K,ii}\}$ are all bounded away from zero, and the population correlations associated with $\bSigma_K$ are bounded by $\frac{c}{2k-1}$ for some $0\leq c<1$. Let $q=s$, under  the Assumptions \ref{a1}-\ref{a5} in the appendix, Condition \ref{m} based on $(\bI-\bP_Z)\bX$ is satisfied with high probability. 
\end{theorem}

In practice,  the number of factors $s$ is unknown; we can apply the eigenvalue ratio (ER) method \citep{Ahn2013} to estimate the number of factors in model (\ref{eq19}). More specifically, the ER estimator is defined as $\widehat{k}_{ER}=\argmax_{\{l: 1\leq l\leq l_{\max}\}}(d_l/d_{l+1})$, where $d_i$ is the $i$th largest singular value of $\bX$ and $l_{\max}$ is the maximum possible number of factors. \cite{Ahn2013} has shown that under some mild regularity conditions, the number of factors can be consistently estimated. %

Under the conventional setting with diagonal $\bSigma_K$, we can show that Condition \ref{m} based on $(\bI-\bP_Z)\bX$ is asymptotically weaker than Condition \ref{ir} based on $\bX$, making it easier to satisfy, as in the following theorem.  

\begin{theorem}\label{th5}
In model (\ref{eq19}), assume $\cov(\bff,\bff)=\bSigma_f$, and $\bSigma_K$ is diagonal. Under  the Assumptions \ref{a1}-\ref{a5} in the appendix, if $p, k, s$ are fixed, as $n\rightarrow\infty$, 
\begin{eqnarray*}
&&\max_{j\in S^c}\|(\bX_S'\bX_S)^{-1}\bX_S'\bx_j\|_1\stackrel{{\mathcal{P}}}\rightarrow A_1,\\
&&\max_{j\in S^c}\|(\bX_S'(\bI-\bP_Z)\bX_S)^{-1}\bX_S'(\bI-\bP_Z)\bx_j\|_1\stackrel{{\mathcal{P}}}\rightarrow A_2;
\end{eqnarray*}
Let $\bB_S$ be the sub matrix of $\bB$ with rows corresponding to the set $S$. If $\bB_S\bSigma_f\bB_{S^c}'\neq \bzero$, then $A_1>A_2$. 
\end{theorem}

The following Theorem \ref{thm5} gives an extensive result with diverging dimensions. 
\begin{theorem}\label{thm5}
Suppose $\bSigma_K$ is diagonal, under the assumptions \ref{a1}-\ref{a5} in the appendix. If $k=o(p^{1/3})$ and $k^3\log p=o(n)$, then 
\begin{equation*}
\max_{j\in S^c}\|(\bX_S'(\bI-\bP_Z)\bX_S)^{-1}\bX_S'(\bI-\bP_Z)\bx_j\|_1=o_p(1). 
\end{equation*}
If $\bX_S'\bX_{S^c}\neq \bzero$, then $\max_{j\in S^c}\|(\bX_S'\bX_S)^{-1}\bX_S'\bx_j\|_1>0$. 
\end{theorem}

The convergence in Theorem \ref{th5} is based on the sample mean convergence to the population mean. Thus, the convergence rate is at the order of $n^{-1/2}$. The proof of Theorem \ref{thm5} shows that the convergence rate is $O_p(\frac{k^{3/2}}{\sqrt{p}}+k^{3/2}\sqrt{\frac{\log p}{n}})$, which is slower since we consider the diverging dimensions here. Moreover, the conditions and results in Theorems \ref{th5} and \ref{thm5} are also different. For example, in Theorem \ref{th5} we have an asymptotic limit result for the irrepresentable condition on $\bX$, which is useful for  discussing the minimum signal strength, while in Theorem \ref{thm5} we do not have such a limit result. 

\subsection{ PROD with Random Generations}


Practically,  strong empirical correlations  between components of $\bX$ may be resulting from multiple reasons such that 
a low-rank common components may not always fit; see, for example, the study of \cite{FanZhou_2016_JMLR}. 
Such a practical reality motivates us to explore additional possibilities for PROD. If we treat the columns of $\bX$ as realizations of random variables, the variables that are expected to have minimal associations with $\bX_S$ are independent ones of those from $\bX_S$. A straightforward choice is just to independently generate a new set of random variables in $\bZ$, or  construct $\bZ$ as some random linear combinations of $\bX$. We refer to this approach as random generation. Since random generations require no prior information, they can be implemented multiple times at one's own discretion, making them very suitable for initial exploration before deciding a final choice. 

In our implementations, we consider two instances with random generations. The first one is to construct $\bZ$ as a random linear combination of $\bX$, that is, $\bZ=\bX\bB$ with columns of $\bB$ randomly generated. 
We denote this method as Random Generation with B (RGB). 
The second one is to randomly generate all components in $\bZ$ from a pre-specified distribution. We denote  this method as Random Generation with Z (RGZ). 
There are multiple devices for randomly generating $\bB$ and $\bZ$; we refer to \cite{Vempala_2005} and reference therein.

 
Our methods differ from the so-called random projection methods in the literature -- a class of computationally efficient dimension reduction methods that are shown to have good geometric properties; see, for example, \cite{Vempala_2005}.  
Conventional random projection method typically regress the response variable $\bY$ on $\bQ=\bX_{n\times p}\bR_{p\times q}$ instead of $\bX$, where each element in $\bR$ are randomly generated. However, the success of our PROD method is from the orthogonal decomposition of $\bX$, a property that  
the random projection method does not satisfy.  

PROD with random generations have good properties. That is, under some regularity conditions, Condition \ref{m} after the PROD with random generations is easier to be satisfied than Condition \ref{ir}. For our technical analysis, let $\bX=(\widetilde\bX_1',\dots, \widetilde\bX_n')'$, and we assume that the $p$-dimensional vectors $\widetilde \bX_i'$ $(i=1.\dots, n)$ are independent realizations from some zero mean distribution with  positive definite  covariance matrix $\bSigma\in {\mathbb R}^{p\times p}$.  Without loss of generality, for the orthogonal decomposition in PROD, we let $\bZ=(\bgamma_1,\cdots,\bgamma_q)$, where $(\bgamma_1,\cdots,\bgamma_q)$ are orthogonal of each other and independent of $\bX$. Then we have the following theorem. 


\begin{theorem}\label{thm3}
Suppose $(\bX_S'(\bI-\bP_Z)\bX_S)$ is invertible. If $q+k\leq n$, $EX_j^4<\infty$ for $1\leq j\leq p$ and $\sum_{i=1}^nE[\gamma_{li}^2\gamma_{mi}^2]=O(n^{-\delta_1})$, $\sum_{i=1}^n\cov(\gamma_{li}^2,\gamma_{mi}^2)=O(n^{-\delta_2})$ for some $\delta_1>0, \delta_2>0$ and for $1\leq l\neq m\leq q$, then if $p, k$ are fixed, as $n\rightarrow\infty$ and  $q/n\to \xi$, 
\[
\max_{j\in S^c}\|({\bX}_S'(\bI-\bP_Z){\bX}_S)^{-1}{\bX}_S'(\bI-\bP_Z)\bx_j\|_1-(1-\xi)\max_{j\in S^c}\|(\bX_S'\bX_S)^{-1}\bX_S'\bx_j\|_1\stackrel{{\mathcal{P}}}\rightarrow  0. 
\]
\end{theorem}

For random generation method RGZ in Section 2.4, $(\bgamma_1,\cdots,\bgamma_q)$ are chosen independently of $\bX$, satisfying the conditions in Theorem \ref{thm3}. When $\xi>0$, Theorem \ref{thm3} shows that asymptotically Condition \ref{m} is easier to be satisfied than Condition \ref{ir}, which provides theoretical justification why our PROD approach is advantageous over conventional regularized regression with $\bX$ as the model matrix. 
%
The conditions $\sum_{i=1}^nE[\gamma_{li}^2\gamma_{mi}^2]=O(n^{-\delta_1}), \sum_{i=1}^n\cov(\gamma_{li}^2,\gamma_{mi}^2)=O(n^{-\delta_2})$ are very mild. Note that $\sum_{i=1}^n\gamma_{li}^2=1$ for any $1\leq l\leq q$. Therefore, $\gamma_{li}^2$ is small being close to $O(n^{-1})$. Correspondingly, $\sum_{i=1}^nE[\gamma_{li}^2\gamma_{mi}^2]$ is also small being close to $O(n^{-1})$. Similarly, $\sum_{i=1}^ncov(\gamma_{li}^2, \gamma_{mi}^2)$ is also small.  
As a special case, $(\bgamma_1,\cdots,\bgamma_q)$ can be chosen as unit basis vectors, then $\sum_{i=1}^nE[\gamma_{li}^2\gamma_{mi}^2]=0$ and $\sum_{i=1}^n\cov(\gamma_{li}^2,\gamma_{mi}^2)=0$ automatically satisfy the conditions in Theorem \ref{thm3}. 

 
 Next, we provide the convergence rates results with diverging dimensions

\begin{theorem}\label{thm6}
If the row vectors of $\bX$ are independently and identically distributed as $N(\bzero,\bSigma)$, $\lambda_{\min}(\bSigma_{SS})\geq c_{\min}>0$, and $k^{3/2}\sqrt{\frac{\log p}{n}}=o(1)$, $q/n=\xi$, then 
\begin{equation*}
\big|\|(\bX_S'\bX_S)^{-1}\bX_S'\bX_{S^{c}}\|_1-\|\bSigma_{SS}^{-1}\bSigma_{SS^{c}}\|_1\big|=o_p(\|\bSigma_{SS^{c}}\|_1);  
\end{equation*}
furthermore, if $k^{5/2}\sqrt{\frac{\log p}{n}}=o(1)$, then 
\begin{eqnarray*}
&&\big|\|(\bX_S'(\bI-\bP_Z)\bX_S)^{-1}\bX_S'(\bI-\bP_Z)\bX_{S^{c}}\|_1-(1-\xi)\|\bSigma_{SS}^{-1}\bSigma_{SS^{c}}\|_1\big|\\
&=&o_p(\|\bSigma_{SS^c}\|_1+\|\bSigma_{SS}\|_1(\|\bSigma_{SS}\|_1+1)(\|\bSigma_{SS^c}\|_1+1)). 
\end{eqnarray*}
\end{theorem}
The multivariate normality and the assumption on $\bSigma_{SS}$ have  been considered in \cite{Wainwright_2009}. As an example, assume $\bSigma_{i,j}=\rho^{|i-j|}$, then $\|\bSigma_{SS}\|_1=O(1)$ and $\|\bSigma_{SS^c}\|_1=O(1)$, which results in $o_p(1)$ on the right handed sided of the above expressions in Theorem \ref{thm6}. The detailed convergence rates depend on $k$, $p$ and $n$, which can be found in the proof. For other $\bSigma$, the requirements on $k$, $p$ and $n$ can be adjusted to achieve the consistency. 

Comparing with Theorem \ref{thm6}, Theorem \ref{thm3} is established under a much weaker condition for the distribution of $\bX$, which requires that $EX_j^4<\infty$, while Theorem \ref{thm6} assumes multivariate normal distribution for $\bX$. Furthermore, the convergence rates in Theorem \ref{thm3} are at the order of $n^{-1/2}$ due to the sample mean convergence, while the convergence rates in Theorem \ref{thm6} are much slower. For example, assume $\bSigma_{i,j}=\rho^{|i-j|}$, then the second convergence rate in Theorem \ref{thm6} is $O_p(k^{5/2}\sqrt{\frac{\log p}{n}})$. 

 
%
%

\subsection{Minimum Signal Strength}

The foremost benefit of using $(\bI-\bP_Z)\bX$ instead of $\bX$ for linear regression is  from reduced level of correlations between the explanatory variables.  On the other hand, however, removing the first $q$ principal components may also weaken the capability for detecting the signals from $\bbeta$, especially those weak signals from the components of $\bbeta$ with small values.   The reason is intuitively clear:  using $(\bI-\bP_Z)\bX$ may introduce larger variance. To see that,  
the variance of  the un-penalized least squares  estimator for an estimable function $\bc' \bbeta$ is $\sigma^2 \bc' (\bX'(\bI-\bP_Z)\bX)^-\bc$, which  may become larger than $\sigma^2 \bc' (\bX'\bX)^-\bc$.   
This can be viewed as a result from information loss -- if $(\bI-\bP_Z)\bX$ becomes less informative, the variance becomes larger. 
A larger  variance implies increased difficulty for the estimation. 

When applying regularized methods for high-dimensional regression,  the aforementioned  effect  is also the case, which may increase the difficulty of detecting weak signals in the model parameter.  
From (\ref{eq:preg}), and denote by $\widehat\bbeta$ as its minimizer,  the zero sub-gradient condition implies that
\begin{align}\label{eq:subg}
n^{-1}\bX' (\bY-\bX\widehat\bbeta)+\lambda \widehat \bz=\bzero
\end{align}
where $\widehat\bz=(\hat z_1,\dots, \hat z_p)'$ with $\hat z_j=\text{sign}(\hat\beta_j)$ $(j=1,\dots,p)$, and $\text{sign}(0)\in [-1,1]$.  Without loss of generality, we decompose the model matrix as $\bX=(\bX_S,\bX_{S^c})$ where $S=\text{supp}(\bbeta_0)$. Then (\ref{eq:subg}) becomes 
\begin{align}\label{eq:block}
-n^{-1}\begin{pmatrix}
\bX_S'\bX_S& \bX_S' \bX_{S^c}\\
\bX_{S^c}'\bX_S& \bX_{S^c}\bX_{S^c}
\end{pmatrix} 
\begin{pmatrix}
\widehat \bbeta_S-\bbeta_S\\
\widehat\bbeta_{S^c}
\end{pmatrix}
+n^{-1}
\begin{pmatrix}
\bX_S' \bveps\\
\bX_{S^c}' \bveps
\end{pmatrix}+\lambda \begin{pmatrix}
\widehat \bz_S\\
\widehat \bz_{S^c}
\end{pmatrix}=\bzero. 
\end{align}
If $\widehat\bbeta_{S^c}=\bzero$, then we have
\begin{align}\label{eq:1}
\widehat\bbeta_S-\bbeta_S&=(n^{-1}\bX_S'\bX_S)^{-1} (n^{-1}\bX_S'\bveps) +\lambda (n^{-1}\bX_S'\bX_S)^{-1}\text{sign}(\widehat\bbeta_S), \\\label{eq:2}
\widehat\bz_{S^c}&=\bX_{S^c}'\bX_S(\bX_S'\bX_S)^{-1}\text{sign}(\widehat\bbeta_S) +\lambda^{-1}n^{-1}\bX_{S^c}'(\bI-\bP_{X_{S}})\bveps.
\end{align}
Here (\ref{eq:1}) leads to the error bound of the LASSO estimator for the nonzero components. 
When the irrepresentable condition holds with $\max_{j\in S^c}\|(\bX_S'\bX_S)^{-1}\bX_S'\bx_j\|_1\leq1-\gamma$ for some uniform $\gamma>0$, the second term in (\ref{eq:2}) is bounded by $\gamma$ with high probability by appropriately choosing $\lambda$, ensuring that zero components of $\bbeta_0$ are correctly estimated as zeros.
Furthermore, to ensure that all components of $\bbeta_S$ are not mistakenly estimated as zero, (\ref{eq:1}) implies that some minimal signal strength condition is needed: $\lambda \|(n^{-1} \bX_S'\bX_S)^{-1}\|_\infty$ cannot dominate the smallest component of $\bbeta_S$.  

As one referee pointed out, there is a delicate trade-off among the required minimum signal strength, the tuning parameter for the penalty, and the irrepresentable condition. Indeed,  due to such a trade-off, cautions are needed when deciding which PROD may best help. As such, one needs to consider the impact  on the minimal signal strength when applying PROD,  which is an important aspect besides the impact from the Irrepresentable Condition. 

According to \cite{Wainwright_2009} Theorem 1, if the irrepresentable condition of the original $\bX$ is bounded by $1-\gamma_1$, and there exists some $C_1>0$ such that $\Lambda_{\min}(n^{-1}\bX_S'\bX_S)\geq C_1$ where $\Lambda_{\min}$ denotes the smallest eigenvalue, then if the regularization parameter $\lambda_1>\frac{2}{\gamma_1}\sqrt{\frac{2\sigma^2\log p}{n}}$, the minimum signal strength required for support recovery  is 
\begin{equation*}
\lambda_1\Big[\|(\bX_S'\bX_S/n)^{-1}\|_{\infty}+\frac{4\sigma}{\sqrt{C_1}}\Big]. 
\end{equation*}
For the PROD method, we replace $\bX$ by $(\bI-\bP_Z)\bX$. Suppose the irrepresentable condition of $(\bI-\bP_Z)\bX$ is bounded by $1-\gamma_2$, and there exists some $C_2>0$ such that $\Lambda_{\min}(n^{-1}\bX_S'(\bI-\bP_Z)\bX_S)\geq C_2$, then the required minimum signal strength with $\lambda_2>\frac{2}{\gamma_2} \sqrt{\frac{2\sigma^2\log p}{n}}$ is
\begin{equation*}
\lambda_2\Big[\|(n^{-1}\bX_S'(\bI-\bP_Z)\bX_S)^{-1}\|_{\infty}+\frac{4\sigma}{\sqrt{C_2}}\Big]. 
\end{equation*} 

We consider the setting in Theorem 2 and Theorem 4 for the illustration. Suppose $n^{-1}\bX_S'\bX_S\stackrel{{\mathcal{P}}}\rightarrow\bSigma_1$, $n^{-1}\bX_S'(\bI-\bP_Z)\bX_S\stackrel{{\mathcal{P}}}\rightarrow\bSigma_2$, $\Lambda_{\min}(\bSigma_1)\geq C_{\min}$ and $\Lambda_{\min}(\bSigma_2)\geq C_{\min}^*$. Furthermore, for the irrepresentable condition, suppose $(\bX_S'\bX_S)^{-1}\bX_S'\bX_{S^c}\stackrel{{\mathcal{P}}}\rightarrow\widetilde{\bSigma}_1$, $({\bX}_S'(\bI-\bP_Z){\bX}_S)^{-1}{\bX}_S'(\bI-\bP_Z)\bX_{S^c}\stackrel{{\mathcal{P}}}\rightarrow\widetilde{\bSigma}_2$, $\|\widetilde{\bSigma}_1\|_1\leq1-\gamma_X$ and $\|\widetilde{\bSigma}\|_1\leq1-\gamma_{PROD}$. For the ease of presentation, we will evaluate 
\begin{equation*}
q_1=\frac{1}{\gamma_X}[\|\bSigma_1^{-1}\|_{\infty}+\frac{4\sigma}{\sqrt{C_{\min}}}]
\end{equation*} 
and 
\begin{equation*}
q_2=\frac{1}{\gamma_{PROD}}[\|\bSigma_2^{-1}\|_{\infty}+\frac{4\sigma}{\sqrt{C_{\min}^*}}]
\end{equation*}
for the comparison. We have the following two concrete examples. 

{\bf Example 1}  For PROD with least important components, under the conditions of Theorem 2 with $\bSigma_K$ and $\bSigma_f$ as identity matrices. By singular value decomposition, $\bB_S=\bU\bSigma_S\bV'$, where $\bU$ is a $k\times k$ matrix containing the left eigenvectors, $\bV$ is a $q\times q$ matrix containing the right eigenvectors, and $\bSigma_S$ can be further expressed as $\bSigma_S=\left(\begin{array}{c}\bLambda \\ \bzero\end{array}\right)$, in which $\bLambda$ is a $q\times q$ diagonal matrix. Analogously, by singular value decomposition, $\bB_{S^c}=\widetilde{\bU}\widetilde{\bSigma}_{S^c}\widetilde{\bV}'$, where $\widetilde{\bU}$ is a $(p-k)\times(p-k)$ matrix containing the left eigenvectors, $\widetilde{\bV}$ is a $q\times q$ matrix containing the right eigenvectors, $\widetilde{\bSigma}_{S^c}=\left(\begin{array}{c}\widetilde{\bLambda} \\\bzero\end{array}\right)$, in which $\widetilde{\bLambda}$ is a $q\times q$ diagonal matrix. To simplify the discussion, we further assume that both $\bU$ and $\widetilde{\bU}$ are identity matrices, and $\bV=\widetilde{\bV}$.  Then $q_2<q_1$. 

{\bf Example 2}  For PROD with random generations, under the conditions of Theorem 4, if $\gamma_X\leq \min(\frac{1-\xi}{2-\xi}, \frac{\xi\sqrt{1-\xi}}{1-(1-\xi)^{3/2}})$, then $q_2<q_1$. 

The proofs of the two examples are given in the Appendix Section B. Examples 1 and 2 show that PROD does not necessarily increase the minimal signal strength required for support recovery. 

In summary, when PROD is applied, the effects are two-fold. First, the Irrepresentable Condition is more likely to be valid with reduced level of correlations between variables in $(\bI-\bP_Z)\bX$ because $\max_{j\in S^c}\|(\bX_S'(\bI-\bP_Z)\bX_S)^{-1}\bX_S'(\bI-\bP_Z)\bx_j\|_1$ becomes smaller. 
Second, $\|(n^{-1} \bX_S'(\bI-\bP_Z)\bX_S)^{-1}\|_\infty$ changes, which  may require different minimum signal strength condition to correctly identify the nonzero components in the model parameter.  Hence, there could be two consequences after applying PROD. First, the number of false inclusions of variables should become smaller, due to the reduction in the level of correlations between variables. 
Second, when the signal strength is weakened, the number of false exclusions of variables may become larger compared with the conventional Lasso. In Section \ref{s4},  our numerical investigations consider two types of settings for nonzero $\beta$'s: ``Dirac" and ``Unif". For Dirac, nonzero $\beta$'s are set as 0.5. For Unif, nonzero $\beta$s are realizations from Uniform[0,1]. In both settings, the signal strength is relatively weak. As a promising observation, the true positives of our PROD are still comparable to those of Lasso, suggesting the good performance of PROD in practice.

\section{Numerical studies}\label{s4}

In this section, we conduct extensive numerical investigations on the proposed methods using both synthetic and real datasets. As we shall see, in comparison with the popular $\ell_1$-penalized method, our methods provide substantially better results in support recovery. To better reveal the big picture,  we also compare different choices of the information criteria for choosing the tuning parameters in the penalized least squares methods. We present the results corresponding to least important component method (LICM) discussed in Section 2.3, and the random generation methods, RGZ and RGB, discussed in Section 2.4. 

\subsection{Comparison with Lasso}
We first conduct numerical studies using synthetic data, and we consider three generating schemes. The scenarios of the data generating processes are given as follows. 

{\bf Model 1} (Factor Model 1): We consider the classical three-factor model setting. Specifically, we first generate a loading matrix $\bA\in\RR^{p\times 3}$, where each entry of $\bA$ is independently sampled from the  standard normal distribution. Then, we permute the rows of $\bA$ to make that the first $k$ rows, denoted as $\ba_1$,..., $\ba_k$, to have the smallest $\ell_2$-norms. Next, for each $i = 1,..., n$, we independently generate the sample by 
$$
\bx_i = \bA\bff_i + \be_i,
$$
where $\bff_i \sim N(\bzero,\bI_3)\in\RR^3$ and $\be_i\sim N (\bzero,\bI_p) \in\RR^p$. Finally, we generate the response variable $y_i$ by
$$
y_i = \bx_i'\bbeta + \epsilon_i, \text{ where }\epsilon_i \sim N(0,0.5^2),
$$
for each $i = 1,...,n$.   This is a setting that the conventional factor model holds, so that the PROD-LICM is expected to work reasonably well. 

{\bf Model 2} (Factor Model 2): We consider another three-factor model. Following the previous setup, we first generate a loading matrix $\bA \in \RR^{p\times 3}$, where each entry is independently sampled from the  standard normal distribution. Then, we permute the rows of $\bA$ to ensure that the first $k$ rows to have the smallest $\ell_2$-norms. In addition, we  set $a_{11}, a_{21},\ldots,a_{51} = 0$, i.e., the first 5 variables of the explanatory variables have no loadings on the first factor.  
We further standardize the variance of all the explanatory variables that we set $\var(X_1) = \var(X_2) = \cdots = \var(X_p) = \sigma^2$.  The rationale is that the covariances between the first five explanatory variables and others have no contributions from the first factor. 

Specifically, we generate the factors by letting each $\bff_{i} \sim N(\bzero,\bI_3)$ for $i=1,...,n$, and let the error $\be_i \sim N\big(\bzero,\diag(\sigma_1^2,...\sigma_p^2)\big)$. It is not difficult to see that $\var(X_j) = \|\ba_j\|^2_2 + \sigma_j^2$. We then choose $\sigma^2 = 1.5 \cdot \max_j\{ \|\ba_j\|^2_2\}$ and let $\sigma_j^2 = \sigma^2 -  \|\ba_j\|^2_2$ for all $j =1,\ldots d$. Then, the design matrix and response variables are generated following the previous scheme. 

{\bf Model 3} (Toeplitz Design): We generate the covariates from a multivariate Gaussian distribution, where the mean is 0 and the covariance matrix is a Toeplitz matrix. In particular, we sample each $\bx_i \sim N(\bzero,\Sig)$, and $\Sig_{jl} = \rho^{|j-l|}$ for some $\rho \in (0,1)$, and we let each response be $y_i = \bx_i'\bbeta + \epsilon_i$, where $\epsilon_i \sim N(0,1)$ for all $i=1,\ldots,n$.  This is a setting when there are no low-dimensional factors contributing to the covariances between the explanatory variables. 

In all three schemes, we consider two settings to generate  $\bbeta$. In the first setting, the first $k$ components of $\bbeta$ are all set as 0.5, and we refer this setting as ``Dirac". In the second setting, we sample the first $k$ components of $\bbeta$ independently from $\text{Unif}[0,1]$, and we refer this setting as ``Unif". The rest components of $\bbeta$ are set as 0's.   Throughout this section, we fix sample size $n = 250$, and we consider different combinations of dimension $p$ and sparsity level $k$. 

For each setting, we repeat the simulation schemes 100 times and report the support recovery results using different methods and different information criteria. Define the true positive (TP) discoveries as $\supp(\bbeta)\cap \supp(\hat\bbeta^\lambda)$, and false positive (FP) discoveries as $\supp(\bbeta)^C \cap \supp(\hat\bbeta^\lambda)$. We report the averaged true positive rates (TPR) and false positive rates (FPR), where $TPR=|TP|/k$ and $FPR=|FP|/(p-k)$. We consider three different information criteria to select the tuning parameter $\lambda$ including five-fold cross validation (5-CV), Bayesian information criterion (BIC) by \cite{schwarz1978estimating}, and the generalized information criterion (GIC) by \cite{fan2013tuning}. We report the results in Tables \ref{tab:factor1}, \ref{tab:factor2} and \ref{tab:top} corresponding to Factor Model 1, Factor Model 2 and Toeplitz Design, respectively.  In all settings, it is seen that the Lasso method achieves slightly better true positive support recovery than the PROD with least important components method (LICM). However, our proposed methods PROD by LICM achieves substantially better false positive control in comparison with Lasso method.  

\begin{table}[ht!]
\centering
\caption{\footnotesize Quantitative comparisons of the LICM, Lasso, RGZ and RGB. We report the averaged true positive rates and false positive rates after repeating each simulation setup 100 times with $n=250$ under Factor Model 1.  }
\footnotesize
\begin{tabular}{r*{11}{r}}
\hline\hline
& &  \  & \multicolumn{2}{c}{LICM}  & \multicolumn{2}{c}{Lasso} & \multicolumn{2}{c}{RGZ} & \multicolumn{2}{c}{RGB}\\
 \cline{4-11}
 $(k,p)$ & \text{Coef} &  \text{TM}  & TPR &  FPR & TPR &  FPR & TPR &  FPR & TPR &  FPR \\
\hline
(15,200) & \text{Dirac} & 5-CV  &100.0\%   &5.0\%  &100.0\%  &13.2\%  &100.0\% &8.9\% &100.0\% &5.7\% \\
& & \text{BIC} &100.0\% &2.6\% &100.0\% &4.1\% &100.0\% &2.0\% & 100.0\% & 2.1\% \\
& & \text{GIC}  &100.0\% &2.2\% &100.0\% &3.5\% & 100.0\%&1.6\% &100.0\% &1.8\%  \\ 
& \text{Unif} & 5-CV &83.2\%  & 3.2\% & 92.7\% &10.1\%&95.5\% &8.0\% &93.2\% &5.8\%  \\ 
& & BIC  & 83.3\% &1.3\%  &86.7\%  &3.1\% &92.0\%&7.3\% &90.3\% &2.4\% \\
& & GIC  &77.8\%   &1.0\%   &85.9\%   &2.6\%   &90.6\%  &1.5\%  &89.1\% &2.0\% \\ 
(15,500) & \text{Dirac} & 5-CV & 100.0\%  &1.1\%  &100.0\% &6.2\%   &100.0\%  &3.4\%  & 100.0\%  &2.1\% \\
& & \text{BIC}  &100.0\%   &0.5\%  & 100.0\%& 1.0\% &100.0\%  &0.7\%  &100.0\%   &0.8\% \\
& & \text{GIC}  &100.0\%   &0.3\%   &100.0\%   &0.6\%   &100.0\% & 0.3\% &  100.0\%  & 0.5\% \\ 
& \text{Unif} & 5-CV  &92.5\% &1.0\%  & 87.5\%  &5.5\% & 95.7\%  &2.9\%   &93.9\%  & 2.2\%\\
& & BIC  & 85.1\%   & 0.3\%  & 90.6\% &0.8\%  & 92.4\%  & 0.7\%   & 90.4\%   & 0.9\% \\
& & GIC  &82.9\%   &0.1\%   &87.9\%   &0.6\%   &90.7\%   & 0.6\%   &89.1\%   & 0.8\%   \\
(20,500) &  \text{Dirac} & 5-CV  &100.0\%  &1.1\%  &100.0\%  &7.5\%   &100.0\%  &5.0\%  &100.0\%  &3.0\% \\
& & \text{BIC} &100.0\% & 0.5\% &100.0\% &1.4\% & 100.0\%& 1.1\%& 100.0\%& 1.1\%\\
& & \text{GIC}  & 100.0\% &0.4\% &100.0\% &1.1\% &100.0\% &0.9\% &100.0\% & 1.0\% \\ 
& \text{Unif} & 5-CV  &82.7\%  &0.9\%  &95.2\% &6.9\% & 94.7\%&4.0\% &91.1\% &2.8\% \\ 
& & BIC  & 84.8\%   &0.4\%  &89.5\% &1.3\% &91.2\% &1.3\% & 87.5\%&1.4\% \\
& & GIC  &82.8\%   &0.3\%  &86.6\% &1.0\% &88.6\% &1.0\% &85.0\% & 1.0\%\\ 
(10,1000) & \text{Dirac} & 5-CV  &100.0\%  &0.7\%  &100.0\%   &2.9\% &100.0\% &2.0\% &100.0\% &2.0\%  \\
& & \text{BIC}  &100.0\%   &0.1\%  &100.0\%  &0.3\%&100.0\%&0.2\%&100.0\%& 0.3\%\\
& & \text{GIC}  & 100.0\%  &0.1\%  &100.0\%  &0.2\%  &100.0\%  &0.1\%  &100.0\%  &0.1\% \\
& \text{Unif} & 5-CV  &92.6\%  &0.9\%  &96.5\%  &2.6\%  &94.3\%  &1.7\%  &82.4\%  &1.5\%\\ 
& & BIC  &84.4\%   &0.1\%  &89.1\%  &0.3\%   &87.9\%   &0.2\%   &87.3\%  &0.3\%\\
& & GIC  &81.2\%   &0.0\%   &84.8\%   &0.1\%   &84.3\%   &0.1\% &85.3\%  &0.1\%\\
(15,1000) & \text{Dirac} & 5-CV  &100.0\%   &1.0\% &100.0\% &4.0\%&100.0\%&2.4\%&100.0\%&2.0\%\\
& & \text{BIC}  & 100.0\%  &0.3\%  &100.0\%   &0.6\%   &100.0\%  &0.5\%   &100.0\%  &0.6\% \\
& & \text{GIC}  & 100.0\% &0.2\%  &100.0\%   &0.4\%   &100.0\%  &0.4\%   &100.0\%   & 0.4\%\\
& \text{Unif} & 5-CV  & 87.5\%  &0.8\%  &94.3\%  &3.7\%   &94.9\%   & 1.4\% & 87.6\% & 1.8\%\\
& & BIC  & 83.8\% &0.2\%  &88.5\%   &0.4\%   &90.9\%   &0.3\%   & 88.3\% &0.4\%\\
& & GIC  &81.8\%  &0.1\%  &85.7\%   &0.3\%   &87.7\%   &0.3\%   &86.5\%   &0.3\%\\
\hline\hline
 \end{tabular}
 \label{tab:factor1}
 \end{table}

 \begin{table}[ht!]
\centering
\caption{\footnotesize Quantitative comparisons of the LICM, Lasso, RGZ and RGB. We report the averaged true positive rates and false positive rates after repeating each simulation setup 100 times with $n=250$ under Factor Model 2. }
\footnotesize
\begin{tabular}{r*{11}{r}}
\hline\hline
& &  \ & \multicolumn{2}{c}{LICM}  & \multicolumn{2}{c}{Lasso} & \multicolumn{2}{c}{RGZ} & \multicolumn{2}{c}{RGB}\\
 \cline{4-11}
 $(k,p)$ & \text{Coef} &  \text{TM} & TPR &  FPR & TPR &  FPR & TPR &  FPR & TPR &  FPR \\
\hline
(15,200) & \text{Dirac} & 5-CV  &100.0\%   &2.5\%  &100.0\%   & 13.6\%   &100.0\%   &8.7\%  &100.0\%   &6.2\% \\
& & \text{BIC} &100.0\%   &1.2\%   &100.0\%  &2.4\%  &100.0\%   &1.5\%   &100.0\%   &2.2\%  \\
& & \text{GIC} & 100.0\% & 0.7\%  & 100.0\%  &1.5\%   &100.0\%   &1.7\%   &100.0\%   &2.2\%  \\
& \text{Unif} & 5-CV  &89.9\%  &2.5\%  &88.1\%   &12.4\%   &95.7\%  &7.3\%   &94.2\%  &5.6\% \\
& & BIC  & 84.5\% & 0.5\% & 90.1\% & 1.7\%  &92.3\%  &1.8\%   &91.1\%   &2.2\% \\
& & GIC  &84.2\%  &0.3\%  &88.1\%  &1.3\%   &90.7\%  &1.2\%   &89.6\%   &1.8\%  \\
(15,500) & \text{Dirac} & 5-CV  &100.0\%  &1.1\%   &100.0\%  & 5.9\% & 100.0\% &4.1\%   &100.0\%  &3.0\% \\
& & \text{BIC}  &100.0\%   &1.4\%  &100.0\%   &2.4\%   &100.0\%   &2.5\%   &100.0\%   &3.3\%  \\
& & \text{GIC}  &100.0\%   &0.3\%   &100.0\%  &0.5\%   &100.0\%   &0.5\%  & 100.0\%  & 0.7\% \\
& \text{Unif} & 5-CV  &92.3\%  & 0.3\% & 91.4\%  &9.1\%   &93.9\%   &3.6\%   &92.0\%  &2.9\%  \\ 
& & BIC  &88.2\%  &0.1\%  &87.3\% &0.8\% &89.3\%  &0.7\%  & 87.9\% &0.8\%\\
& & GIC &87.9\%  &0.1\%  &87.2\%   &0.6\%   &87.8\%   &0.5\%  &86.0\%   &0.5\%  \\
(20,500) &  \text{Dirac} & 5-CV  &100.0\%   &1.2\%  &100.0\%  &8.0\%   &100.0\%  &4.7\%  &100.0\%  &3.1\% \\
& & \text{BIC}  & 100.0\%  &0.6\%  &100.0\%  &1.7\%   &100.0\%  &1.2\%  &100.0\%  &1.5\%  \\
& & \text{GIC}  &100.0\%  &0.3\%  &100.0\%  &0.9\%   &100.0\%   &0.9\%  &100.0\%  &1.4\%  \\
& \text{Unif} & 5-CV  &90.2\%   &3.1\%   &95.9\%   &8.1\%   &95.1\%   &4.0\%  &91.0\%   &2.8\% \\ 
& & BIC  &86.1\%  &0.6\%  &89.6\%  &1.5\%   &91.5\%   &1.0\%   &87.4\%   &0.6\%  \\
& & GIC  &81.8\% & 0.2\% &87.1\%   &0.4\%   &88.7\%   &0.6\%   &85.5\%   &0.6\%  \\ 
(10,1000) & \text{Dirac} & 5-CV & 100.0\% & 1.1\%  &100.0\%   &3.1\%   &100.0\% & 1.9\%&100.0\% &1.7\% \\
& & \text{BIC} &100.0\%  &0.2\%  &100.0\%   &0.3\%  &100.0\%  &0.2\%  &100.0\%  &0.2\%  \\
& & \text{GIC} &100.0\%  &0.1\%  &100.0\%  &0.2\%   &100.0\%  &0.1\%   &100.0\%   &0.1\%  \\
& \text{Unif} & 5-CV  & 92.6\% &1.2\%  &96.5\%   &2.9\%   &94.4\%   &1.6\%   &92.5\%   &1.5\%   \\ 
& & BIC& 84.4\%  &0.1\%  &89.1\%  &0.4\%   &87.9\%   &0.2\%   &87.4\%   &0.2\%   \\ 
& & GIC &81.2\%  &0.1\%  &84.8\%   &0.2\%   &85.5\%   &0.1\%   &85.2\%  &0.1\%  \\ 
(15,1000) & \text{Dirac} & 5-CV  &100.0\%  &1.3\%  &100.0\%  &4.1\%  &100.0\%  & 2.3\% & 100.0\% & 2.5\% \\
& & \text{BIC}  &100.0\%  &0.4\%  &100.0\%  &0.7\%   &100.0\%   &0.3\%   &100.0\%  &0.4\%   \\
& & \text{GIC} & 100.0\% &0.3\% &100.0\%  &0.5\%  &100.0\%  &0.2\%  &100.0\%  &0.2\% \\
& \text{Unif} & 5-CV & 90.1\% &1.2\%  &96.7\%  &4.1\% &94.6\%& 2.1\% &92.9\%  &1.9\% \\
& & BIC  &82.1\%  &0.2\%  &88.9\%  &0.4\%  &89.7\%  &0.3\%  &100.0\%  &0.3\% \\
& & GIC  &83.0\% &0.1\%  &86.2\%  &0.3\%   &87.1\% &0.2\% &85.9\%   &0.2\% \\
 \hline\hline
 \end{tabular}
 \label{tab:factor2}
 \end{table}

\begin{table}[ht!]
\centering
\caption{\footnotesize Quantitative comparisons of the LICM, Lasso, RGZ and RGB. We report the averaged true positive rates and false positive rates after repeating each simulation setup 100 times with $n=250$ under Toeplitz Design. }
\footnotesize
\begin{tabular}{r*{11}{r}}
\hline\hline
& &  \ & \multicolumn{2}{c}{LICM}  & \multicolumn{2}{c}{Lasso} & \multicolumn{2}{c}{RGZ} & \multicolumn{2}{c}{RGB}\\
 \cline{4-11}
 $(k,p,\rho)$ & \text{Coef} &  \text{TM}  & TPR &  FPR & TPR &  FPR  & TPR &  FPR & TPR &  FPR \\
\hline
(15,200,0.5) & \text{Dirac} & 5-CV  &100.0\%  &1.6\% & 100.0\%&9.4\%  & 100.0\% &8.0\%   &100.0\%  &5.1\%  \\
& & \text{BIC}  & 100.0\% &0.8\%  &100.0\%   &2.3\%   &100.0\%   &1.3\%   &100.0\%  & 1.7\%  \\
& & \text{GIC} &100.0\%  &0.6\%  &100.0\%   &2.1\%   &100.0\%   &0.8\%   &100.0\%   &1.3\%   \\
& \text{Unif} & 5-CV  & 84.1\% & 2.0\%  &88.1\%   &12.7\%  &93.8\%   &6.5\%   &93.2\%  &4.8\%   \\ 
& & BIC  & 82.1\%  &1.0\%  & 88.0\% & 3.0\%  &92.2\%   &1.1\% &91.6\%  &1.4\%  \\
& & GIC  &81.4\%  &0.7\%  &85.6\%   &2.6\%   &91.5\%   &0.7\%   &91.5\%   &1.0\%  \\ 
(15,500,0.85) & \text{Dirac} & 5-CV  &100.0\%   &0.6\%  &100.0\%   &7.2\%   &100.0\%   &1.1\%   &100.0\%   &0.9\%  \\
& & \text{BIC} &100.0\%   &0.3\%   &100.0\%   &1.1\%   &100.0\%   &0.2\%   & 100.0\%  &0.4\%  \\ 
& & \text{GIC}  &100.0\%   &0.3\%   &100.0\%  & 0.7\%  & 100.0\%  &0.1\%   &100.0\%   &0.3\%  \\ 
& \text{Unif} & 5-CV  &90.7\%   &0.9\%  &90.8\%   &6.2\%   &92.9\%   &1.0\%   &92.9\%  & 0.7\% \\
& & BIC  & 88.4\% & 0.3\% & 92.6\%  &1.0\%   &91.9\%   &0.2\%   &91.8\%  &0.4\%   \\
& & GIC  & 86.4\% &0.2\%  &87.9\%  &0.7\%  &91.5\%  &0.1\%   &91.5\%   &0.3\%  \\
(20,500,0.7) &  \text{Dirac} & 5-CV  &100.0\%& 1.1\%  &100.0\%  &6.7\%   &100.0\%   &2.0\%   &100.0\%   &1.4\%  \\
& & \text{BIC} &100.0\%  &0.4\% &100.0\%   &1.4\%   &100.0\%   &0.4\%   &100.0\%   &0.5\%  \\
& & \text{GIC} & 100.0\% & 0.3\% &100.0\%  &1.2\%   &100.0\%   &0.2\%   &100.0\%   &0.3\%  \\ 
& \text{Unif} & 5-CV  & 93.2\% &0.6\%   & 96.3\%  &7.5\% & 94.8\% & 1.8\% & 94.9\%  &1.1\% \\
& & BIC  &85.8\%  &0.3\%  &90.1\%   &1.4\%  &91.8\%  &0.3\%   &91.8\%   &0.5\%  \\
& & GIC  &84.4\%  &0.3\%  &89.0\%   &1.1\%  &91.5\%   &0.2\%   &91.3\%    &0.4\%   \\
(10,1000,0.6) & \text{Dirac} & 5-CV  &100.0\%  &0.7\%  &100.0\%   &3.3\%   &100.0\%   &1.2\%  &100.0\%   &1.1\% \\
& & \text{BIC}  &100.0\%   &0.1\%  &100.0\%  &0.2\%   &100.0\%   &0.1\%  &100.0\%  &0.1\% \\
& & \text{GIC} & 100.0\% & 0.1\% & 100.0\% & 0.2\%  &100.0\%   &0.0\%  &100.0\%  &0.1\%  \\
& \text{Unif} & 5-CV  &91.8\% &0.8\%  & 95.9\%  &2.8\%   &89.9\%   &1.0\%  &89.7\%  & 0.8\% \\ 
& & BIC  &86.2\%  &0.1\%   &88.5\%  &0.3\%  &87.8\%  &0.1\%  &87.4\%  &0.1\%  \\
& & GIC  &83.6\%  &0.1\%  &86.2\%   &0.1\%   &87.0\%   &0.0\%   &86.5\%  & 0.0\% \\
(15,1000,0.9) & \text{Dirac} & 5-CV  &100.0\%  &1.0\%  &100.0\%   &4.1\%   &100.0\%   &0.5\%   &100.0\%  &0.5\%  \\
& & \text{BIC}  & 100.0\%  &0.4\%  &100.0\%   &0.6\%   &100.0\%   &0.1\%   &100.0\%   &0.2\%   \\
& & \text{GIC}  &100.0\%  &0.2\%   &100.0\%   &0.4\%   &100.0\%   &0.1\%   &100.0\%   &0.2\%  \\
& \text{Unif} & 5-CV  &92.1\%  &0.8\%  &94.3\%  & 4.4\%  &89.9\%  &0.4\%  &89.8\%  &0.4\%  \\ 
& & BIC  &82.7\%  &0.2\%  &89.7\%  &0.4\%   &88.9\%   &0.1\%  &88.8\%   &0.2\% \\
& & GIC  &82.3\% & 0.1\% &86.7\%   &0.3\%   &88.5\%   &0.1\%   &88.2\%  &0.2\%  \\
 \hline\hline
 \end{tabular}
 \label{tab:top}
 \end{table}

Furthermore, we find that PROD by random generations RGZ and RGB are performing very promisingly. PROD with random generations provide better true positive rates than the other two methods in the Toeplitz design setting, and reduce the false positive rates better than the other two methods in several cases. This shows that PROD with LICM method is advantageous under the factor model, while the random generation method is favored under some general settings. This matches our theoretical analysis. We finally point out that the RGZ method provides slightly better empirical performance than the RGB method. 

\begin{figure}
 \begin{center}
 \subfigure{
  \includegraphics[width=0.3\textwidth,height=0.2\textheight]{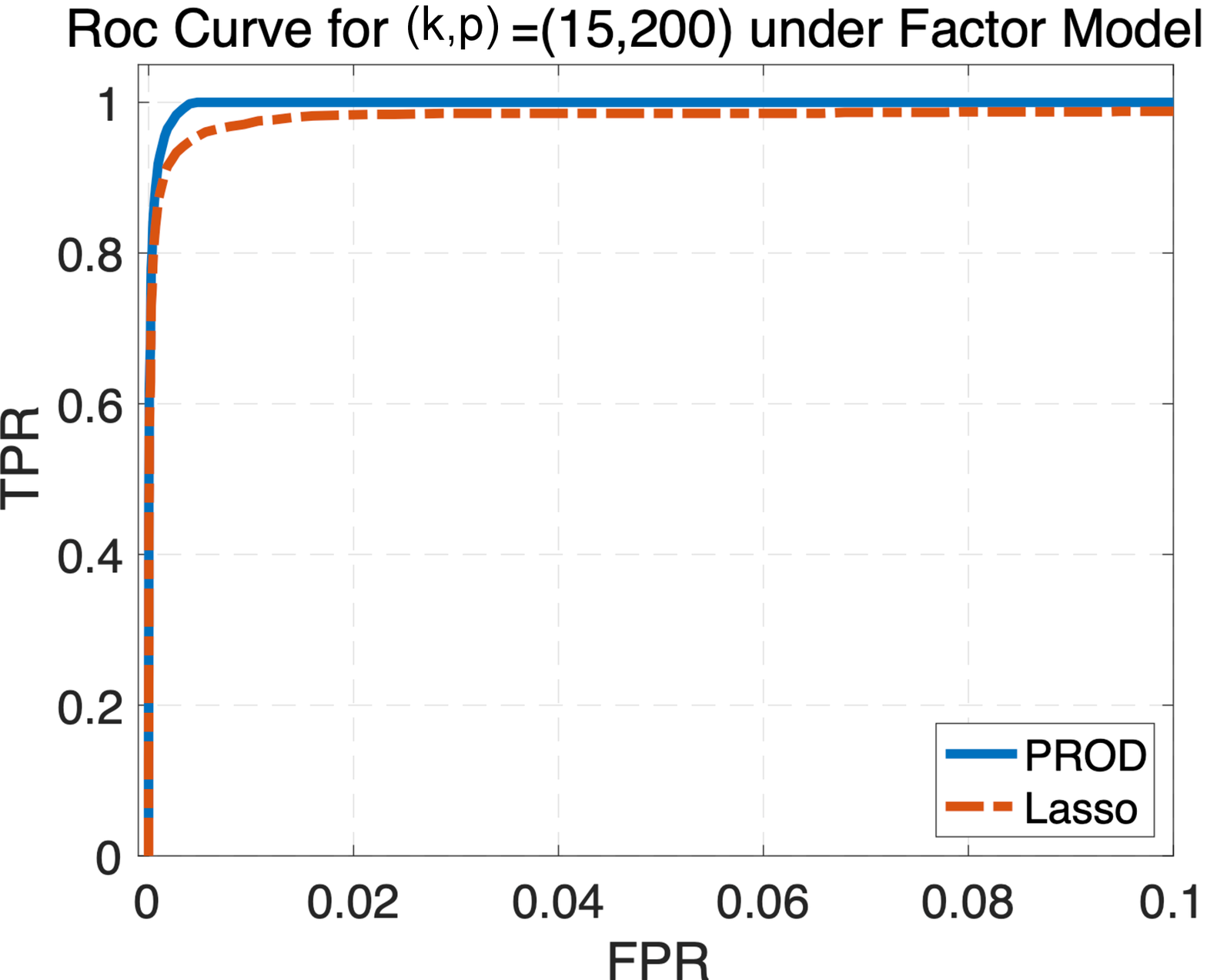}
  }
   \subfigure{
  \includegraphics[width=0.3\textwidth,height=0.2\textheight]{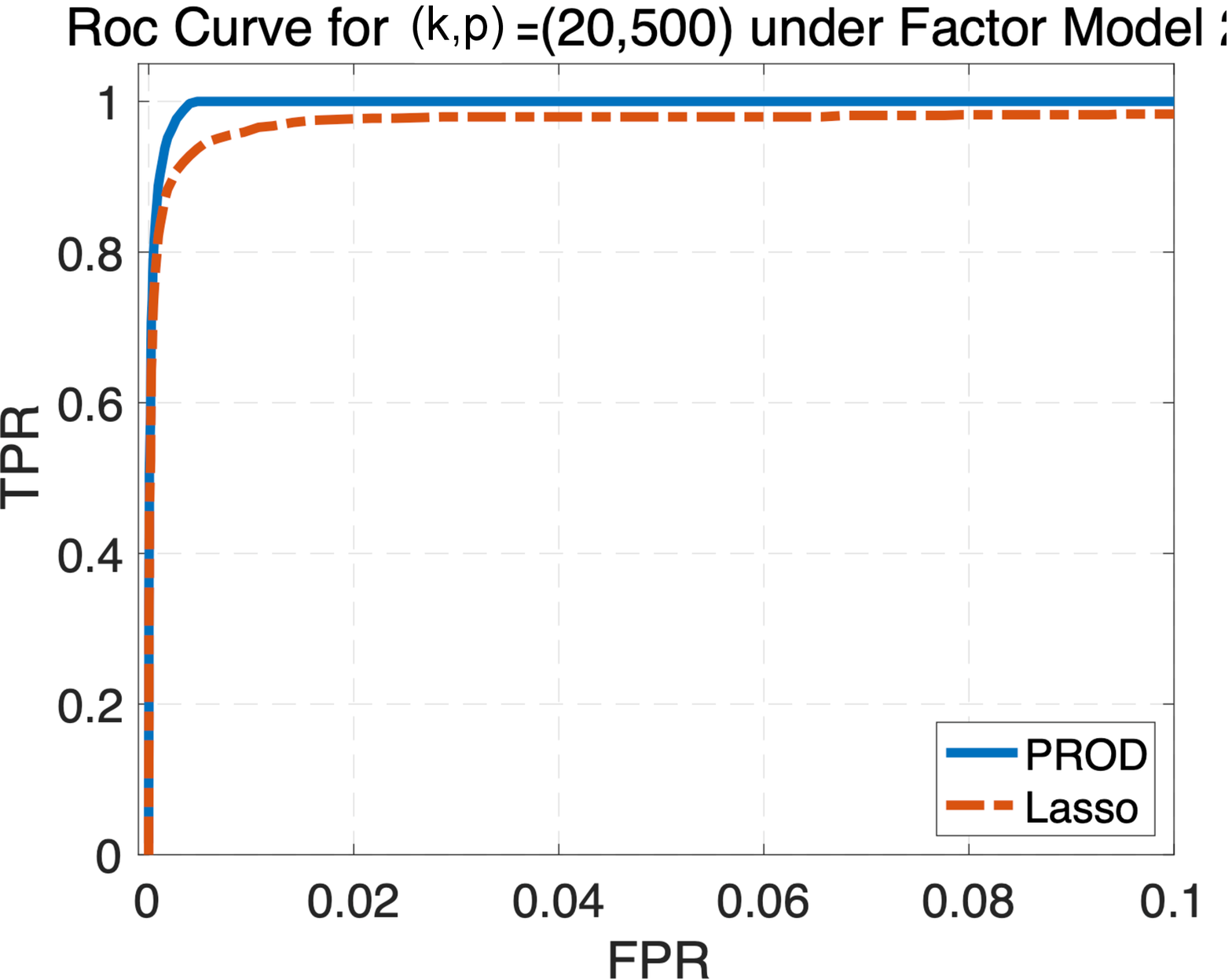}
  }
   \subfigure{
  \includegraphics[width=0.3\textwidth,height=0.2\textheight]{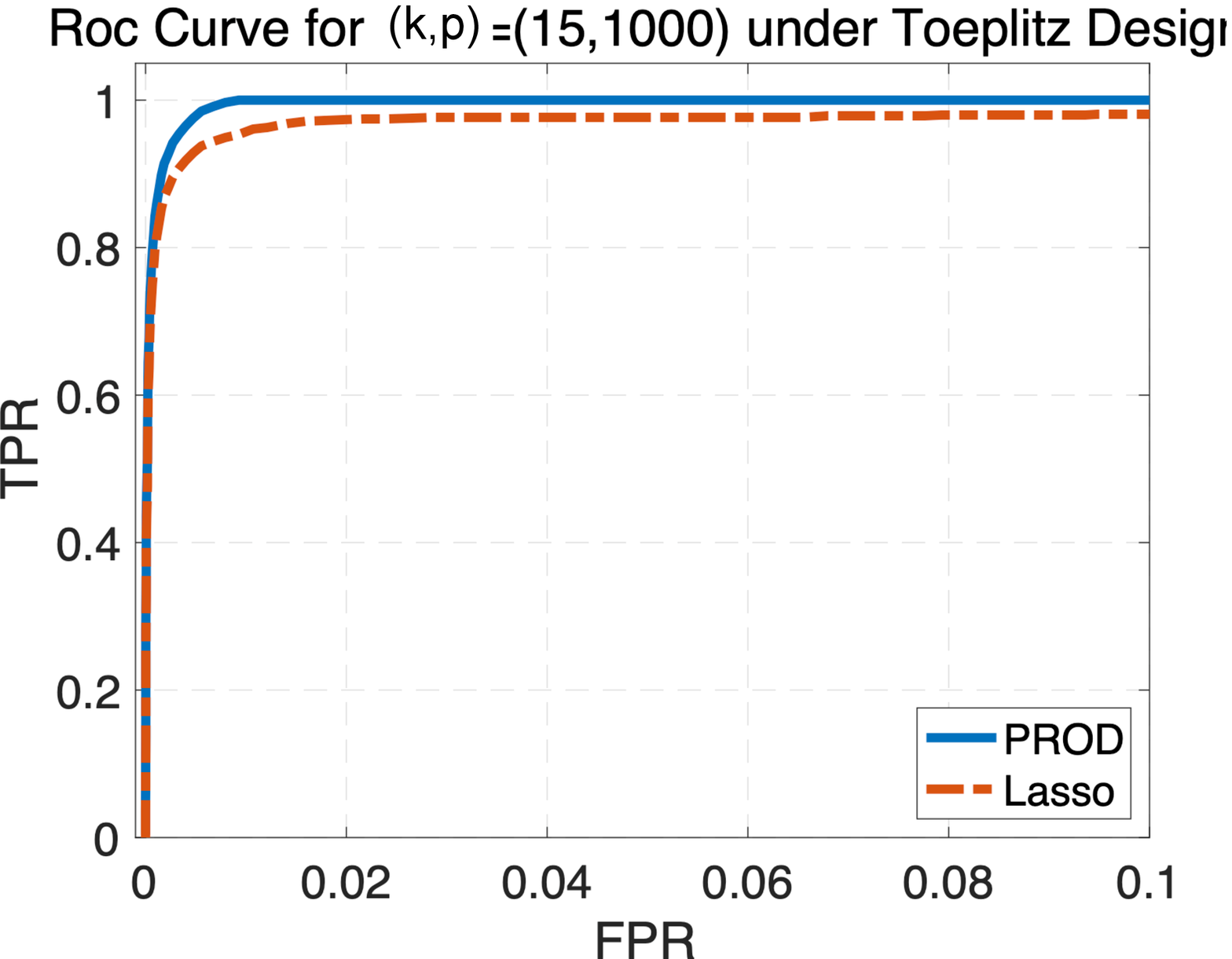}
  }
 \end{center}
 \vspace{-12pt}
 \caption{\footnotesize ROC Curves under three generating schemes for different settings. From left to right, the plots correspond to Factor Model 1, Factor Model 2 and Toeplitz Model, respectively. For the Toeplitz model, we choose $\rho = 0.7$.}
 \label{fig:roc}
 \end{figure}

We then provide receiver operating characteristic (ROC) curves in Figure \ref{fig:roc} to further compare the overall profiles of the methods by LASSO and LICM. where we choose one simulation setup from each of the three generating schemes. In the plots of ROC curves, the $x$-axis and $y$-axis represent the false and true positive rates (FPR and TPR), respectively. Within a certain range of true positive rate, for the same true positive rate, Lasso produces a larger false positive rate compared with PROD. For the same false positive rate, PROD produces a larger true positive rate compared with Lasso. Overall, it is clear that the proposed PROD method outperforms Lasso method. 

\subsection{Comparison with Other Penalized Methods}
We  compare the numerical performance of Lasso \citep{Tibshirani_1996_JRSSB}, SCAD in \citep{FanLi_2001_JASA}, Adaptive Lasso \citep{Zou_2006_JASA}, and MCP in \citep{Zhang_2010} with their combinations of PROD. The detailed methods will be presented in Section \ref{s44}.  To simplify the presentation, we only consider LICM for PROD in Section 3.2. 

{\bf Model 4}: Let $\bB$ be a $p\times 3$ matrix and each element is randomly generated from $Uniform(-0.5,0.5)$. Let $\bSigma$ be the correlation matrix of $\bB\bB'+\bI_p$. Let $\bx_i'$ be the $i$th row of $\bX$. Then $\bx_i$ is randomly generated from $N(0,\bSigma)$. Let $y_i=\bx_i'\bbeta+\epsilon_i$ with $\epsilon_i\sim N(0,1)$ for $i=1,\cdots, n$. For $\bbeta$, the sparsity is 10 and the nonzero $\beta$'s are 1. 

{\bf Model 5}: It is the same as Model 3 but we set nonzero $\beta$'s as 1. 

In Table 4, all the methods  correctly identify  true positives. However, the performances in terms of false positives are quite different. Generally speaking, methods with combination of PROD can beat the corresponding original methods. The non convex penalty methods such as SCAD and MCP perform better than Lasso and AdaLasso. Furthermore, PROD-SCAD and PROD-MCP can even reduce the number of the false positives significantly compared with SCAD and MCP respectively. 

In Table 5, SCAD and MCP can not correctly identify the true positives, and they  mis-select some true variables. On the other hand, Lasso and Adaptive Lasso  correctly select the true positives. PROD with different methods produce fewer false positives compared with the corresponding original methods. 

\begin{table}[ht!]
\centering
\caption{\footnotesize Quantitative comparisons of the penalized methods and their combination with PROD. We report the averaged true positive and false positive rates after repeating each simulation setup 100 times with $n=400$ under Model 4. }
\footnotesize
\begin{tabular}{r*{11}{r}}
\hline\hline
 &   & \multicolumn{9}{c}{$(k,p)$} \\
 \cline{3-11}
 &  &\multicolumn{3}{c}{(10,1000)}  &\multicolumn{3}{c}{(10,2000)}  &\multicolumn{3}{c}{(20,2000)}\\
\cline{3-11} 
  &  & 5-CV  &BIC   &GIC   &5-CV  &BIC  &GIC  &5-CV  &BIC   &GIC\\
\cline{3-11}
Lasso & TPR &100.0\%  &100.0\%   &100.0\%   &100.0\%   &100.0\%   &100.0\%   &100.0\%   &100.0\%  &100.0\% \\
   &FPR  &4.61\%     &0.89\%     &0.58\%   &2.67\%     &0.47\%     &0.26\%      &5.24\%    &1.19\%    &0.80\%   \\
PROD-Lasso &TPR  &100.0\%    &100.0\%    &100.0\%   &100.0\%    &100.0\%    &100.0\%    &100.0\%    &100.0\%    &100.0\% \\
   &FPR  &0.64\%     &0.47\%   &0.43\%    &0.34\%     &0.23\%     &0.21\%    &0.53\%    &0.44\%    &0.40\%   \\
SCAD  &TPR  &100.0\%    &100.0\%    &100.0\%    &100.0\%    &100.0\%    &100.0\%    & 100.0\%   &100.0\%    &100.0\% \\
   &FPR  &0.357\%    &0.002\%    &0.000\%   &0.18\%    &0.001\%     &0.000\%    &0.20\%    &0.00\%    &0.00\%  \\
PROD-SCAD  &TPR  & 100.0\%   &100.0\%    &100.0\%    &100.0\%    &100.0\%    &100.0\%    &100.0\%    &100.0\%    &100.0\% \\
   &FPR  &0.006\%    & 0.004\%   & 0.004\%   &0.004\%     &0.003\%     &0.003\%     &0.001\%    &0.001\%    &0.001\%  \\
MCP & TPR &100.0\%  &100.0\%   &100.0\%   &100.0\%   &100.0\%   &100.0\%   &100.0\%   &100.0\%  &100.0\% \\
   &FPR  &0.129\%    &0.002\%    &0.000\%    &0.095\%     &0.002\%      &0.000\%    &0.046\%    &0.001\%    &0.000\%  \\
PROD-MCP & TPR &100.0\%  &100.0\%   &100.0\%   &100.0\%   &100.0\%   &100.0\%   &100.0\%   &100.0\%  & 100.0\%\\
   &FPR  &0.003\%   &0.001\%    &0.001\%    &0.001\%     & 0.001\%    &0.001\%    &0\%   &0\%   &0\%  \\
AdaLasso & TPR &100.0\%  &100.0\%   &100.0\%   &100.0\%   &100.0\%   &100.0\%   &100.0\%   &100.0\%  &100.0\% \\
   &FPR  &3.884\%    & 0.045\%   &0.004\%    & 2.603\%    &0.034\%     &0.004\%    &4.306\%   &0.038\%    &0.003\%  \\
PROD-AdaLasso  &TPR  &100.0\%    &100.0\%    &100.0\%    &100.0\%    &100.0\%    &100.0\%    &100.0\%    &100.0\%    &100.0\% \\
   &FPR  &0.318\%    & 0.085\%  & 0.051\%   &0.150\%    &0.046\%     &0.022\%    &0.005\%    &0.000\%   &0.000\%  \\
\hline\hline\\
\end{tabular}
\end{table}
\begin{table}[ht!]
\centering
\caption{\footnotesize Quantitative comparisons of the penalized methods and their combination with PROD. We report the averaged true positive and false positive rates after repeating each simulation setup 100 times with $n=400$ under Model 5. }
\footnotesize
\begin{tabular}{r*{11}{r}}
\hline\hline
 &   & \multicolumn{9}{c}{$(k,p,\rho)$} \\
 \cline{3-11}
 &  &\multicolumn{3}{c}{(10,1000,0.8)}  &\multicolumn{3}{c}{(10,2000,0.9)}  &\multicolumn{3}{c}{(20,2000,0.9)}\\
\cline{3-11} 
  &  & 5-CV  &BIC   &GIC   &5-CV  &BIC  &GIC  &5-CV  &BIC   &GIC\\
\cline{3-11}
Lasso & TPR &100.0\%  &100.0\%   &100.0\%   &100.0\%   &100.0\%   &100.0\%   &100.0\%   &100.0\%  &100.0\% \\
   &FPR  &1.02\%     &0.05\%     &0.02\%     &0.428\%     &0.020\%     &0.008\%    &0.56\%    &0.03\%    &0.02\%  \\
PROD-Lasso  &TPR  &100.0\%    &100.0\%    &100.0\%    &100.0\%    &100.0\%    &100.0\%    &100.0\%   &100.0\%    &100.0\% \\
   &FPR  &0.06\%     &0.02\%     &0.01\%     &0.023\%    &0.009\%     &0.009\%    & 0.01\%   &0.01\%    &0.01\%   \\
SCAD  &TPR  &65.3\%     & 64.7\%    &64.4\%     &47.9\%     &46.7\%     &45.4\%     &52.55\%    &48.30\%   &42.75\% \\
   &FPR  &0.007\%     &0.006\%     &0.006\%     &0.002\%     &0.002\%   &0.001\%    &0.002\%    &0.003\%    &0.003\%   \\
PROD-SCAD  &TPR  &65.9\%    & 65.6\%   &65.6\%    &47.8\%    &47.0\%    &46.3\%    &46.25\%    &45.55\%    &41.35\% \\
   &FPR  &0.004\%    &0.003\%    &0.003\%    &0.002\%   &0.002\%     &0.003\%    &0.003\%    &0.003\%    &0.003\%   \\
MCP & TPR &74.9\%   &70.2\%    &67.6\%   & 49.2\% & 47.7\%   &47.6\%   &34.25\%   &33.65\%   &33.45\% \\
   &FPR  & 0\%   &0\%    &0\%    &0\%    &0\%    &0\%    &0.001\%    &0.001\%    &0.001\%   \\
PROD-MCP & TPR &73.4\%   &70.6\%    &68.1\%    & 49.9\%   &49.3\%    &49.3\%    &34.35\%    &33.80\%   &33.65\% \\
   &FPR  &0.001\%    &0.001\%     &0.000\%    &0\%    &0\%    &0\%    & 0.0005\%   &0.0005\%    &0.0005\%  \\
AdaLasso & TPR &100.0\%  & 100.0\%  &100.0\%   &100.0\%   &100.0\%   &100.0\%   & 100.0\%  & 100.0\% & 100.0\%\\
   &FPR  &0.681\%     & 0.104\%   & 0.008\%   &0.255\%    &0.042\%    &0.004\%    &0.364\%    &0.063\%    &0.003\% \\
PROD-AdaLasso  &TPR  &100.0\%    &100.0\%    &100.0\%    &100.0\%    &100.0\%    &100.0\%    &99.95\%    & 99.95\%    & 99.80\% \\
   &FPR  &0.15\%   &0.07\%    &0.05\%    & 0.09\%   &0.06\%    &0.06\%    &0.001\%   & 0.000\%   &0.000\%   \\
\hline\hline\\
\end{tabular}
\end{table}

\subsection{Comparison with Other Spectral Transformation Methods}

In the following Table \ref{tab:transform}, we will compare our PROD-LICM method with Trim transform \citep{cevid2020spectral}, Puffer transform \citep{Jia_2015}, Lava \citep{chernozhukov2017lava} under Models 1-4 in the current paper with various settings of $n$ and $p$. We consider the model size as $k=10$. The nonzero $\beta$'s are set as 1 and we use the five-fold cross validation to select the tuning parameter $\lambda$. 

\begin{table}[ht!]
\centering
\caption{\footnotesize Quantitative comparisons of PROD, Puffer, Trim and Lava. We report the averaged true positive and false positive rates after repeating each simulation setup 100 times under Models 1-4. For Model 3 (Toeplitz model), we sep $\rho=0.9$.  }
\footnotesize
\begin{tabular}{r*{10}{r}}
\hline\hline
&   \  & \multicolumn{2}{c}{PROD}  & \multicolumn{2}{c}{Puffer} & \multicolumn{2}{c}{Trim} & \multicolumn{2}{c}{Lava}\\
 \cline{3-10}
Model &$(n,p)$  & TPR &  FPR & TPR &  FPR & TPR &  FPR & TPR &  FPR \\
\hline
Model 1 &(250,1000) &100.0\%   &0.23\%   &100.0\%   & 0.88\%  &100.0\%    &0.66\%   & 100.0\% &0.53\% \\
               & (400,1000)&100.0\%   &0.11\%   &100.0\%    &0.90\%    &100.0\%   &0.65\%   &100.0\%   &0.60\%\\
               &(400,2000)&100.0\%   &0.10\%   &100.0\%   &0.40\%   &100.0\%   &0.33\%   &100.0\%    &0.24\%  \\
Model 2 &(250,1000) &100.0\%   &0.06\%  & 100.0\%   &0.84\%   &100.0\%   &0.84\%   &100.0\%  &0.58\%  \\
               &(400,1000) &100.0\%   &0.02\%   &100.0\%    &0.61\%   &100.0\%    &0.37\%   &100.0\%   &0.28\%  \\
               &(400,2000)&100.0\%   &0.02\%   &100.0\%    &0.33\%    &100.0\%    &0.34\%    &100.0\%   &0.24\%  \\
Model 3&(250,1000) &99.9\%   &0.01\%   &99.7\%   &0.07\%    &100.0\%    &0.01\%    &100.0\%   &0.01\%  \\
              &(400,1000)&100.0\%   &0.01\%     &99.9\%    &0.10\%    &100.0\%    &0.01\%    &100.0\%    &0.01\%   \\
              &(400,2000) &100.0\%   &0.01\%   &100.0\%    &0.03\%    &100.0\%    &0.01\%   &100.0\%   &0.01\%  \\
Model 4 &(250,1000)&99.5\%   &0.11\%    &100.0\%    &0.90\%    &100.0\%   & 0.65\%  &100.0\%   &0.60\%   \\
               &(400,1000)&100.0\%   &0.07\%    &100.0\%    &0.50\%    &100.0\%    &0.36\%   &100.0\%   &0.26\%   \\
               &(400,2000)&100.0\%   &0.03\%    &100.0\%    &0.40\%    &100.0\%    &0.32\%    &100.0\%   &0.29\%  \\                                            
\hline\hline
 \end{tabular}
 \label{tab:transform}
 \end{table}
Table \ref{tab:transform} shows that PROD-LICM achieves superior performance in terms of producing a smaller false positive rate under various settings. For the true positive rate, when $n$ is relatively small, PROD and Puffer may mis-select some true variables occasionally. 

\subsection{Heteroscedastic setting}

We further compare the performance of PROD and Lasso under the heteroscedastic settings. In particular, we consider the Models 1-3, but we let the error $\epsilon_i$ depend on the covariate $\bx_i$ such that $\epsilon_i\sim N(0,\sigma_i^2)$, where $\sigma_i=\|\bx_i/p\|_2$. We present the results in Table \ref{tab:heteroscedastic}, and we find that the methods behave similarly as in the previous cases. 
\begin{table}[ht!]
\centering
\caption{\footnotesize Quantitative comparisons of the LICM, Lasso, RGZ and RGB under heteroscedastic settings. We report the averaged true positive and false positive rates after repeating each simulation setup 100 times with $n=250$ under Factor Model 1 (FM1), Factor Model 2 (FM2), and Toeplitz Model (TM). For Toeplitz model, we sep $\rho=0.5$.  }
\footnotesize
\begin{tabular}{r*{12}{r}}
\hline\hline
& & &  \  & \multicolumn{2}{c}{LICM}  & \multicolumn{2}{c}{Lasso} & \multicolumn{2}{c}{RGZ} & \multicolumn{2}{c}{RGB}\\
 \cline{4-12}
Model &$(k,p)$ & \text{Coef} &  \text{TM}  & TPR &  FPR & TPR &  FPR & TPR &  FPR & TPR &  FPR \\
\hline
FM1 &(15,200) & \text{Dirac} & 5-CV  &100.0\%   & 5.6\%  & 100.0\%  & 14.8\%  & 100.0\%& 7.5\%  & 100.0\% & 6.2\%\\
& & & \text{BIC} &100.0\%  & 5.48\%  &100.0\%  & 8.71\%  & 100.0\%  & 3.29\%  &100.0\%  &3.94\%  \\
& & & \text{GIC}  & 100.0\% &3.0\%  &100.0\%  &4.7\%  &100.0\%  &1.8\% &100.0\% &2.1\%  \\ 
& & \text{Unif} & 5-CV  & 82.6\%   & 3.2\%  & 85.8\% & 10.9\%  & 92.1\%  & 7.7\%  & 90.2\%  & 5.2\%  \\ 
& & & BIC  &85.2\%   & 1.1\%  & 89.3\%  & 3.5\%   & 90.3\%  & 2.1\% & 88.9\%  & 2.8\%\\
& & & GIC  &78.2\%  &1.0\%  & 83.3\%  & 2.9\%  & 87.3\% & 1.8\% & 86.5\%  & 2.1\% \\ 
& (15,500) & \text{Dirac} & 5-CV  &100.0\%   & 1.2\%   &100.0\%  &6.7\%   &100.0\%  &3.5\% & 100.0\%& 2.3\%\\
& & & \text{BIC}  &100.0\%   &0.6\%   &100.0\%   &1.0\%   &100.0\%   &0.8\%   &100.0\%   &1.0\%  \\
& & & \text{GIC}  &100.0\%   & 0.4\%  & 100.0\%  &0.8\%   &100.0\%   &0.6\%   &100.0\%  &0.8\% \\ 
& & \text{Unif} & 5-CV  & 88.1\%   &1.0\%   &86.3\%   &6.1\%   &92.3\%  & 3.4\%  &90.1\%  & 2.3\% \\
& & & BIC  &82.2\%   &0.5\%  &89.3\%  & 1.0\%  & 90.0\%  &0.8\%   & 88.4\% &1.1\% \\
& & & GIC  &81.8\%  &0.2\%   &87.9\%   & 0.9\%  & 89.1\%  &0.7\%   &87.4\%  & 1.1\%\\
FM2 &(15,500) & \text{Dirac} & 5-CV  &100.0\%   &1.2\%  &100.0\%  &6.9\%   &100.0\%  &4.6\%  &100.0\%  &3.2\% \\
& & & \text{BIC} &100.0\%  &0.4\%  &100.0\%  &1.6\%   &100.0\%  &0.6\%  &100.0\%  &0.7\%  \\
& & & \text{GIC}  &100.0\%  & 0.4\% & 100.0\% &1.6\%  & 100.0\% &0.5\% & 100.0\%&0.6\%  \\ 
& & \text{Unif} & 5-CV  &93.5\%   & 0.3\% &92.5\%  &5.4\%   &95.6\%  & 4.0\%  &94.5\% &3.3\%  \\ 
& & & BIC  &90.5\%   &0.2\%  & 88.5\% &1.5\% &91.7\%  &0.9\%  & 89.2\% &0.9\% \\
& & & GIC  & 88.8\% &0.2\%  & 89.4\%  &1.7\%   &90.0\%  &0.6\%  &88.3\%  &0.9\% \\ 
& (10,1000) & \text{Dirac} & 5-CV  & 100.0\%  &1.2\%   &100.0\%  & 4.3\%  & 100.0\% &2.0\% &100.0\% &2.0\% \\
& & & \text{BIC}  & 100.0\%  &0.3\%   &100.0\%   &0.4\%   &100.0\%   &0.4\%   &100.0\%   &0.3\%  \\
& & & \text{GIC}  & 100.0\%  &0.3\%   & 100.0\%  &0.3\%   &100.0\%   &0.2\%   &100.0\%  &0.2\% \\ 
& & \text{Unif} & 5-CV  &94.1\%   &1.2\%   & 97.2\%  &3.3\%   &95.9\%  &1.8\%   &93.2\%  &1.6\% \\
& & & BIC  &88.7\%   &0.1\%  &90.4\%  & 0.4\%  &88.3\%   &0.2\%   &87.9\%  &0.3\% \\
& & & GIC  &84.0\%  &0.1\%   &86.9\%   &0.2\%   &87.9\%   &0.1\%   &87.2\%  & 0.1\%\\
TM &(15,1000) & \text{Dirac} & 5-CV  & 100.0\%  &1.1\%  &100.0\%  & 3.8\%  &100.0\%  &0.6\%  & 100.0\% &0.6\% \\
& & & \text{BIC} &100.0\%  &0.5\%  &100.0\%  &0.8\%   &100.0\%  &0.1\%  & 100.0\% & 0.2\% \\
& & & \text{GIC}  &100.0\%  &0.3\%  &100.0\%  &0.9\%  &100.0\%  &0.1\% & 100.0\%& 0.2\% \\ 
& & \text{Unif} & 5-CV  & 89.8\%  & 0.9\%  & 95.5\%  &4.6\%   &86.8\%  &0.5\%  &86.2\%  & 0.6\%  \\ 
& & & BIC  &82.1\%   &0.3\%  &86.5\%  &0.4\%    &85.6\%  &0.1\%  &84.8\%  &0.2\% \\
& & & GIC  &81.9\%  &0.2\%  &86.7\%   &0.3\%   &85.0\%  &0.1\%  &84.3\%  &0.3\% \\ 
\hline\hline
 \end{tabular}
 \label{tab:heteroscedastic}
 \end{table}

\subsection{Real data analysis}
In this section, we apply different methods to analyze a genomic dataset \citep{mccall2011gene,fang2016mining}. The data set  is generated from Affymetrix Human 133A (GPL96) arrays. The data was downloaded from GEO, preprocessed and normalized using frozen RMA  \citep{mccall2010frozen}. The data set has more than 10,000 samples spanning different biological contests, and each sample contains 20,248 probes, corresponding to 12,704 genes. 

We focus our study on the samples of  breast cancer, where we have $n = 89$ samples. Our goal is to find which genes are associated with an  important gene BRCA1, which is of great importance for breast cancer \citep{ford1998genetic}. In particular, we take the gene expression levels of BRCA1 as response variable $Y$, and we aim to find which genes are most associated with it. Due to the low sample size $n=89$, and there are 12,703 other remaining genes, we do some prescreening as the first step. Specifically, using the 89 breast cancer samples, we regress each gene's expression levels on BRCA1's expression level. We keep $p=1,000$ genes with strongest marginal significance levels.

We then apply the proposed LICM, RGZ and RGB methods compared with Lasso for the data to find which genes are most associated with BRCA1. For ease of presentation, we only report the findings using BIC as tuning parameter selection criterion in Table \ref{tab:gene}. We first find that all four methods identify that BRCA2 is closely associated with BRCA1, which is well known in literature \citep{king2003breast}. Then, we find that the proposed three methods give quite similar results, and most of the findings by these three methods are known to be associated with BRCA1, such as TP53 \citep{lakhani2002pathology}, RAD51C \citep{wong2011brip1} and RRM1 \citep{su2011ercc1}. While the significant genes found by Lasso are different, and we do not find existing literature to support these associations. In addition, we validate the models  using only  genes selected by different methods and compute the 5-fold cross validation mean squared prediction error (CMSE). In particular, we divide the sample into five parts, and every time we use four parts to fit the model using only selected genes and compute the mean squared prediction error using the rest part of the samples, and we compute the averaged five errors for the five parts. We provide CMSE in Table \ref{tab:gene}. It is seen that Lasso gives higher CMSE than the other three methods, and RGZ method produces smallest CMSE. We also apply SCAD, MCP and Adaptive Lasso for selecting the genes. It is seen that the CMSE of SCAD, MCP and Adaptive Lasso are in between the proposed methods and Lasso. The discovered genes by SCAD and MCP are the same. Among them, BRCA2, RAD51C and TP53 coincide with the proposed methods, and GINS1 is also discovered by RGZ. FANCA is known to be associated with BRCA1 \citep{folias2002brca1}. For the other two discoveries, SPN1 and STRC, we do not find supporting literature to show their interactions with BRCA1. The genes selected by Adaptive Lasso have a large overlap with those selected by Lasso. The only difference is STRC, which is also selected by SCAD and MCP. However, we do not find evidence to support STRC. We take a further investigation of the data by looking at marginal correlations between  RPA3 or ELAC2, which are only selected by Lasso, and some other genes selected by our proposed methods, where we report the marginal correlations between expression levels of RPA3 or ELAC2 and some other genes in Table \ref{tab:gene2}. It is seen that RPA3 and ELAC2 both have strong marginal correlations with some genes only selected by our proposed methods. Furthermore, we regress the expression levels of genes selected by Lasso on the expression level of GINS1, which is not selected by Lasso, but by RGZ method. We find that the resulting R-square is 0.8013. Thus, we may conjecture that the reason Lasso selects some genes such as ELAC2 and RPA3 might be due to these genes' high  correlations with some other genes, which are associated with BRCA1. Hence,  this analysis clearly demonstrate the benefit of using our PROD method especially to avoid false inclusions.  
\begin{table}[ht!]
\centering
\caption{\footnotesize Genes associated with BRCA1 selected by LICM, RGZ, RGB, Lasso, SCAD, MCP and Adaptive Lasso, where the tuning parameter is selected using BIC criterion. }
\footnotesize
\begin{tabular}{c*{8}{c}}
\hline\hline
 & LICM & Lasso & RGZ & RGB & SCAD & MCP & AdaLasso\\
\hline
Genes & BRCA2 &   BRCA2 & BRCA2 & BRCA2 & BRCA2 & BRCA2 & BRCA2\\
 & RAD51C &  TP53 & RAD51C & RAD51C &RAD51C &RAD51C &TP53\\
 &  TP53 &  EWSR1  &  TP53 & TP53 &TP53 & TP53 & EWSR1\\
 & PSME3 &   ELAC2 & PSME3 & PSME3 &GINS1 &GINS1&ELAC2\\
 &  RRM1 &  NECTIN3 & RRM1 & RRM1 &FANCA &FANCA &RPA3 \\
 & SPAG5 & RPA3 & CCNA1&   TK1 &SPN1 &SPN1 &STRC\\
 & $\times$ & $\times$ & GINS1 & PFAS &STRC &STRC & $\times$\\
\hline
CMSE  & 0.1205 & 0.1504 & 0.1115 & 0.1229 &0.1380 & 0.1398 & 0.1417 \\
\hline\hline
\end{tabular}
\label{tab:gene}
\end{table}

\begin{table}[ht!]
\centering
\caption{\footnotesize Marginal correlations between expression levels of Lasso's selected genes RPA3 or ELAC2  and other genes selected by our proposed methods.}
\footnotesize
\begin{tabular}{c*{8}{c}}
\hline\hline
& RRM1 &  SPAG5 & GINS1 & TK1 & PFAS  \\
RPA3 & 0.5314 & 0.5293 & 0.4688 & 0.5250 & 0.4139 \\
\hline
  & PSME3 & RRM1 &  SPAG5 & TK1  & PFAS \\
ELAC2  & 0.5315 & 0.4463 &  0.6319 & 0.5599 &  0.5712 \\
\hline\hline
\end{tabular}
\label{tab:gene2}
\end{table}


\section{Discussions} \label{s5}

We conclude our paper with some further discussions. 

\subsection{Rotational Irrepresentable Condition}

Our PROD can be conducted with any given $\bZ$.  If we assume that the columns of $\bZ$ are orthogonal, Condition \ref{m} can be considered as a rotational version of the Irrepresentable Condition (Condition \ref{ir}). More specifically, let $\bZ=(\sqrt{\eta_1}\bgamma_1,\cdots,\sqrt{\eta_q}\bgamma_q)$, where $q$ is an integer, $\eta_1,\cdots,\eta_q$ are positive constants, $\bgamma_1,\cdots,\bgamma_q$ are $q$ $n$-dimensional unit vectors that are orthogonal of each other. 

\begin{condition}[Rotational Irrepresentable Condition]\label{rir}
There exists an integer $q$, $0\leq q\leq n-1$ and $n-$dimensional orthogonal unit vectors $\bgamma_{q+1},\cdots,\bgamma_{n}$ such that 
\begin{equation}\label{eq11}
\max_{j\in S^c}\|(\bA_S'\bA_S)^{-1}\bA_S'\ba_j\|_1\leq1-\gamma
\end{equation}
for some $0<\gamma\leq 1$ where $\bA_S=(\bgamma_{q+1},\cdots,\bgamma_n)'\bX_S$ and $\ba_j=(\bgamma_{q+1},\cdots,\bgamma_n)'\bx_j$. 
\end{condition}

Condition \ref{rir} is a general one containing a few important cases as given below.
\begin{enumerate}
\item Let $q=0$ and $(\bgamma_{q+1},\cdots,\bgamma_n)=\bI_{n\times n}$, then (\ref{eq11}) is equivalent to  the Irrepresentable Condition (\ref{eq1}). Correspondingly, $\bZ$ is a null matrix. 
\item Let $\bgamma_{i}$  $(i=1,\dots, q)$ be the eigenvector associated with the $i$th largest eigenvalue of the sample covariance matrix of $\bX$,  it corresponds to PROD with least important components.
\item Let each element of $\bgamma_i$ be an independent realization from a given distribution with zero mean, then the vectors are nearly orthogonal. This corresponds to PROD with random generations. By Gram-Schmidt process, the constructed unit vectors can be orthogonal.  
\end{enumerate}
Note that $\bP_Z=\bZ(\bZ'\bZ)^{-1}\bZ'=\sum_{i=1}^q\bgamma_i\bgamma_i'$. There exist orthogonal vectors $\bgamma_{q+1},\cdots,\bgamma_n$ such that $\sum_{i=1}^n\bgamma_i\bgamma_i'=\bI$. Therefore, $(\bI-\bP_Z)\bX_S=(\sum_{i=q+1}^n\bgamma_i\bgamma_i')\bX_S$. Thus, if $\bZ$ has orthogonal columns, then Condition \ref{rir} is equivalent to Condition \ref{m}.

\subsection{Comparison with Principal Component Regression and Cluster Representative Lasso}\label{s253}

It should be pointed out that although our PROD can be constructed based on principal component analysis, it is fundamentally different from the principal component regression in \cite{Kendall_1957}. When $n>p$, the principal component regression (PCR) was designed to overcome the multicollinearity problem. The PCR estimator also enjoys the advantage of reducing variance compared with the conventional ordinary least squares estimator. More specifically, let $\bX=\bU\bD\bV'$ be the singular value decomposition as in Section 2.3 where $\bV=(\bv_1,\cdots,\bv_r)$. Let $\bV_q=(\bv_1,\cdots,\bv_q)$, then the PCR regresses on $\bX\bV_q$ instead of the data matrix $\bX$, focusing on the first $q$ principal components. Despite the difference in the dimensionality, our PROD addresses a different problem. We aim to recover the support under the correlated covariates, especially correctly identifying the zero components in the true $\bbeta_0$ as zeroes. Correspondingly, our PROD implement on $(\bI-\bP_Z)\bX$, focusing on the last $r-q$ principal components. Furthermore, the detailed formula of $(\bI-\bP_Z)\bX$ is also different from $\bX\bV_q$ in the PCR. 

When $p>n$, \cite{Buhlmannetal_2013} is also noteworthy here. When many variables within a group of correlated variables belong to the active set, conventional Lasso tends to only select one of these active variables in the group. \cite{Buhlmannetal_2013} considered a cluster representative Lasso estimator (CRL) to address this issue. More specifically, it first partition the covariates into clusters. Then the CRL regresses on $\overline{\bX}$, where the $r$th column of $\overline{\bX}$ is calculated as the average of covariates in the $r$th cluster. The CRL estimator produces better estimation error and prediction error compared with the conventional Lasso, especially when the correlations among different clusters are weak. Note that the $\overline{\bX}$ used in the CRL can be viewed as $\bA\bX$ for a transformation matrix $\bA$. Our PROD uses $(\bI-\bP_Z)\bX$, which is also a transformation of the model matrix $\bX$ but with a different transformation matrix. Furthermore, although both CRL and our PROD handles the challenging situation with correlated covariates, we have a different aim. In our setting, the active variables and the inactive variables can be strongly correlated, and we want to recover the support.

\subsection{Information Loss}\label{s254}

One concern of applying our PROD is on the information loss if we only use one portion $(\bI-\bP_Z)\bX$ of $\bX$ as in (\ref{eq:decom}) while the other portion $\bP_Z\bX$ is ignored. In the PROD with least important components, such a step can be justified since the components in $\bP_Z\bX$ are highly correlated which will cause trouble of false positives. However, in the PROD with random generations, the roles of $\bP_Z\bX$ and $(\bI-\bP_Z)\bX$ are similar for the regularized regression methods. An immediate remedy to the concern of information loss in the PROD with random generations could be using both $\bP_Z\bX$ and $(\bI-\bP_Z)\bX$. We actually treat $\bP_Z\bX$ and $(\bI-\bP_Z)\bX$ equally, and to explore an opportunity for combining them.   In our work, we  propose to estimate $\bbeta$ by regressing $\bY$ on $\bP_Z\bX$ and $(\bI-\bP_Z)\bX$ with penalized least squares, respectively, and then combine their support.  Specifically, we compute
\begin{eqnarray*}
\widehat\btheta_1 &=& \argmin_{ \btheta\in\mathbb{R}^p}\{n^{-1}\|\bY-\bP_Z\bX\btheta\|_2^2 + \lambda_1\|\btheta\|_1\},\\ 
\widehat\btheta_2 &=& \argmin_{\btheta\in\mathbb{R}^p}\{n^{-1}\|\bY-(\bI-\bP_Z)\bX\btheta\|_2^2 + \lambda_2\|\btheta\|_1\}.
\end{eqnarray*}
Then, we select the support by taking the union of $\widehat\btheta_1$ and $\widehat\btheta_2$, i.e., $\text{supp}(\widehat\btheta_1)\cup \text{supp}(\widehat\btheta_2)$. Note that $\widehat{\btheta}_1$ is also a Lasso-type estimator of the true value $\bbeta_0$ in model (\ref{lm}) due to the orthogonal decomposition of $\bX$. 

\subsection{Other Penalization Methods}\label{s44}
We will illustrate three successful methods in this section, and interested readers should be able to extend to other favorable procedures. 

\cite{FanLi_2001_JASA} argued that a good sparse estimator should possess unbiasedness property, which Lasso estimator is lack of. To address this issue, they proposed the following SCAD procedure which enjoys the oracle property, meaning that the SCAD estimator performs as well as if the zero components in $\bbeta$ were known. The SCAD estimator is defined as

\begin{equation*}
\widehat{\bbeta}_{SCAD}=\argmin_{\bbeta\in \mathbb{R}^p}n^{-1}\|\bY-\bX\bbeta\|_2^2+\sum_{j=1}^p \rho_\lambda (|\beta_j|), 
\end{equation*}
where 
\begin{equation*}
\rho'_{\lambda}(\theta)=\lambda\{I(\theta<\lambda)+\frac{(a\lambda-\theta)_{+}}{(a-1)\lambda}I(\theta>\lambda)\}
\end{equation*}
for some $a>2$ and $\theta>0$. In practice, we set $a=3.7$. 

Our PROD method for SCAD can be defined as follows:
\begin{equation*}
\widehat{\bbeta}_{PROD-SCAD}=\argmin_{\bbeta\in\mathbb{R}^p}n^{-1}\|\bY-(\bI-\bP_Z)\bX\bbeta\|_2^2+\sum_{j=1}^p \rho_\lambda (|\beta_j|). 
\end{equation*}

\cite{Zou_2006_JASA} found out that although the Lasso estimator is biased, it can be easy to make some modifications for Lasso by some weighted penalties such that the estimator is unbiased. \cite{Zou_2006_JASA} proposed an estimator named adaptive Lasso, which also possesses the oracle property. More specifically, 

\begin{equation*}
\widehat{\bbeta}_{AdaLasso}=\argmin_{\bbeta\in\mathbb{R}^p}n^{-1}\|\bY-\bX\bbeta\|_2^2+\lambda\sum_{j=1}^p \frac{|\beta_j|}{|\widehat{\beta}_{init,j}|},
\end{equation*}
where $\widehat{\bbeta}_{init}$ is some initial estimator. We will consider the Lasso estimator for $\widehat{\bbeta}_{init}$ here. 

Applying PROD, we obtain the PROD-Adaptive Lasso estimator as follows:

\begin{equation*}
\widehat{\bbeta}_{PROD-AdaLasso}=\argmin_{\bbeta\in\mathbb{R}^p}n^{-1}\|\bY-(\bI-\bP_Z)\bX\bbeta\|_2^2+\lambda\sum_{j=1}^p \frac{|\beta_j|}{|\widehat{\beta}_{init,j}|}. 
\end{equation*}

\cite{Zhang_2010} proposed a MCP estimator, which considers a minimax concave penalty. The MCP estimator also enjoys the oracle property as well as many other good statistical properties. Mathematically, MCP estimator is defined as

\begin{equation*}
\widehat{\bbeta}_{MCP}=\argmin_{\bbeta\in\mathbb{R}^p}n^{-1}\|\bY-\bX\bbeta\|_2^2+\lambda\sum_{j=1}^p\int_0^{|\beta_j|}(1-\frac{x}{\gamma\lambda})_{+}dx. 
\end{equation*}

Combining with PROD, we have 

\begin{equation*}
\widehat{\bbeta}_{PROD-MCP}=\argmin_{\bbeta\in\mathbb{R}^p}n^{-1}\|\bY-(\bI-\bP_Z)\bX\bbeta\|_2^2+\lambda\sum_{j=1}^p\int_0^{|\beta_j|}(1-\frac{x}{\gamma\lambda})_{+}dx.
\end{equation*}









\bibliography{Hbib,bib}

\newpage

\begin{appendix}

\section{Proofs for Theorem 1-5}
For notational convenience, we will use $\bM$ to denote $(\bI-\bP_Z)\bX$ in the following theoretical proof. 

The following Assumptions \ref{a1}-\ref{a5} are adopted from \cite{Fanetal_2013} for factor model analysis, and will only be applied for the proof of Theorem \ref{th2}, Theorem \ref{th5} and Theorem \ref{thm5}.
 
\begin{assumption}\label{a1}
Sparsity condition on $\bSigma_K=(\sigma_{ij})_{p\times p}$: 
\begin{equation*}
m_p=\max_{i\leq p}\sum_{j\leq p}|\sigma_{ij}|^m\quad\text{for some}\quad m\in[0,1], \quad\text{and}\quad m_p=o(p). 
\end{equation*}
\end{assumption}

\begin{assumption}\label{a2}
$\|p^{-1}\bB'\bB-\bOmg\|=o(1)$ for some $k\times k$ symmetric positive definite matrix $\bOmg$ such that $\lambda_{\min}(\bOmg)$ and $\lambda_{\max}(\bOmg)$ are bounded away from both zero and infinity.
\end{assumption}

\begin{assumption}\label{a3}
(i)  $Ek_{il}=Ek_{il}f_{jl}=0$ for all $i\leq p, j\leq k$ and $l\leq n$. \\
(ii) There exist positive constants $c_1$ and $c_2$ such that $\lambda_{\min}(\bSigma_K)>c_1$, $\|\bSigma_K\|_1<c_2$, and 
$\min_{i,j} var(k_{il}k_{jl})>c_1$. \\
(iii) There exist positive constants $r_1$,  $r_2$, $b_1$, and $b_2$ such that for any $s>0$, $i\leq p$ and $j\leq k$,
\begin{equation*}
P(|k_{il}|>s)\leq\exp(-(s/b_1)^{r_1}), \quad\quad\quad P(|f_{jl}|>s)\leq\exp(-(s/b_2)^{r_2}).
\end{equation*}
\end{assumption}

\begin{assumption}\label{a5}
Regularity conditions: There exists $M>0$  such that  for all $i\leq p$, $t\leq n$ and $s\leq n$,\\
(i) $\|\bb_{j}\|_{\max}<M$,\\
(ii)  $\bE[p^{-1/2}({\bk}_s'{\bk}_t-\bE{\bk}_s'{\bk}_t)]^4<M$,\\
(iii) $\bE\|p^{-1/2}\sum_{i=1}^p\bb_ik_{it}\|^4<M$.
\end{assumption}

Here Assumptions \ref{a1}-\ref{a5} are from the study in \cite{Fanetal_2013} for factor models. Assumption \ref{a1} is a sparsity condition, imposing constraints such that the covariance matrix between the idiosyncratic components is sparse following the convention of existing studies as in, for example, \cite{Bickel2008}. As in \cite{Fanetal_2013}, Assumption \ref{a2}-\ref{a5} are needed to consistently estimate the common factors as well as the factor loadings. For more discussions, please see \cite{Fanetal_2013}. 
   
\textbf{Proof of Theorem \ref{th2}:} Without loss of generality, assume that the mean of each variable $\{X_{ji}\}_{i=1}^n$ has been removed.  First of all, let us consider an alternative optimization problem: 
\begin{eqnarray*}
(\widehat{\bB},\widehat{\bF})&\equiv&\argmin_{\bB, \bF}\|\bX'-\bB\bF'\|_F^2\\
\text{subject to}&& n^{-1}\bF'\bF=\bI_s, \quad \bB'\bB\quad\text{is diagonal}
\end{eqnarray*}
where $\bX=(\bX_1,\cdots,\bX_p)$, $\bX_j=(X_{j1},\cdots,X_{jn})'$ for $1\leq j\leq p$, $\bF'=(\bff_1,\cdots,\bff_n)$. The minimizer is given as follows: the columns of $\widehat{\bF}/\sqrt{n}$ are the eigenvectors corresponding to the $s$ largest eigenvalues of the $n\times n$ matrix $\bX\bX'$ and $\widehat{\bB}=n^{-1}\bX'\widehat{\bF}$. See \cite{StockWatson_2002_JASA}. 

Let $\widehat{k}_{ji}=x_{ji}-\widehat{\bb}_j'\widehat{\bff}_i$, $\widetilde{\sigma}_{ij}=n^{-1}\sum_{l=1}^n\widehat{k}_{il}\widehat{k}_{jl}$ and $\widetilde{\Lambda}\equiv(\widetilde{\sigma}_{ij})$, then we have 
\begin{eqnarray*}
\widetilde{\Lambda}&=&\frac{1}{n}(\bX'-\widehat{\bB}\widehat{\bF}')(\bX-\widehat{\bF}\widehat{\bB}')\\
        &=&\frac{1}{n}\bX'\bX-\frac{1}{n}\widehat{\bB}\widehat{\bF}'\bX-\frac{1}{n}\bX'\widehat{\bF}\widehat{\bB}'+\frac{1}{n}\widehat{\bB}\widehat{\bF}'\widehat{\bF}\widehat{\bB}'
\end{eqnarray*}
Plug in $\widehat{\bB}=n^{-1}\bX'\widehat{\bF}$ and $n^{-1}\widehat{\bF}'\widehat{\bF}=\bI_s$, we further have
\begin{eqnarray*}
\widetilde{\Lambda}&=&\frac{1}{n}\bX'\bX-\frac{1}{n}\bX'\frac{1}{n}\widehat{\bF}\widehat{\bF}'\bX
\end{eqnarray*}
Since $\bX=(\bu_1,\cdots,\bu_r)\diag(d_1,\cdots,d_r)(\bv_1,\cdots,\bv_r)'$, correspondingly, $n^{-1/2}\widehat{\bF}=(\bu_1,\cdots,\bu_s)$. Therefore, 
\begin{eqnarray*}
\widetilde{\Lambda}&=&n^{-1}\bX'(\sum_{i=s+1}^r\bu_i\bu_i')\bX\\
                    &=&n^{-1}\bV\bD(\bu_1,\cdots,\bu_r)'(\bu_{s+1},\cdots,\bu_r)(\bu_{s+1},\cdots,\bu_r)'(\bu_1,\cdots,\bu_r)\\
                    &=&n^{-1}\bV\bD\left(\begin{array}{cc}0 & 0 \\0 & \bI_{r-s}\end{array}\right)\left(\begin{array}{cc}0 & 0 \\0 & \bI_{r-s}\end{array}\right)\bD\bV'\\
                    &=&n^{-1}\sum_{i=s+1}^rd_i^2\bv_i\bv_i'. 
\end{eqnarray*}

By Lemma C.11 in \cite{Fanetal_2011}, we have 
\begin{equation*}
\max_{i\leq p}n^{-1}\sum_{i=1}^n|k_{ji}-\widehat{k}_{ji}|^2=O_p(\omega_p^2)
\end{equation*}
and 
\begin{equation*}
\max_{j,i}|k_{ji}-\widehat{k}_{ji}|=o_p(1),
\end{equation*}
where $\omega_p=\frac{1}{\sqrt{p}}+\sqrt{\frac{\log p}{n}}$. Therefore, by Lemma A.3 in \cite{Fanetal_2011}, for any $\epsilon>0$, there is positive constant $M$ such that the event $\{\max_{i\leq p, j\leq p}|\widetilde{\sigma}_{ij}-\sigma_{K,ij}|<M\omega_p\}$ occurs with probability at least $1-\epsilon$. 

Let $\bC=(c_{ij})_{p\times p}$ be the correlation matrix of $\bSigma_K$. Let $\widehat{\bC}=(\widehat{c}_{ij})_{p\times p}$ be the correlation matrix of $\widetilde{\Lambda}$. We want to show that under some regularity conditions, $\max_{i\leq p,j\leq p}|\widehat{c}_{ij}-c_{ij}|=o_p(1)$. Therefore, if the population correlations of $\bK$ are bounded by $\frac{c}{2q-1}$ for some $0\leq c<1$, then for sufficiently large $p$ and $n$, the elements in the correlation matrix of $\widetilde{\Lambda}$ are bounded by $\frac{c^*}{2q-1}$ for some $0\leq c^*<1$ with high probability.  

Next we will connect the above results with the correlation matrix, which is crucial for the Irrepresentable Condition. Note that
\begin{eqnarray*}
|\widehat{c}_{ij}-c_{ij}|&=&|\frac{\widetilde{\sigma}_{ij}}{\sqrt{\widetilde{\sigma}_{K,ii}\widetilde{\sigma}_{jj}}}-\frac{\sigma_{K,ij}}{\sqrt{\sigma_{K,ii}\sigma_{K,jj}}}|\\
 &=&|\frac{\widetilde{\sigma}_{ij}\sqrt{\sigma_{K,ii}\sigma_{K,jj}}-\sigma_{ij}\sqrt{\widetilde{\sigma}_{ii}\widetilde{\sigma}_{jj}}}{\sqrt{\widetilde{\sigma}_{ii}\widetilde{\sigma}_{jj}\sigma_{K,ii}\sigma_{K,jj}}}|\\
 &\leq&|\frac{\widetilde{\sigma}_{ij}-\sigma_{K,ij}}{\sqrt{\widetilde{\sigma}_{ii}\widetilde{\sigma}_{jj}}}|+|\frac{\sigma_{K,ij}\sqrt{\sigma_{K,ii}\sigma_{K,jj}}-\sigma_{K,ij}\sqrt{\widetilde{\sigma}_{ii}\widetilde{\sigma}_{jj}}}{\sqrt{\widetilde{\sigma}_{ii}\widetilde{\sigma}_{jj}\sigma_{K,ii}\sigma_{K,jj}}}|\\
 &\equiv&S_1+S_2. 
\end{eqnarray*}
When $\sigma_{K,ii}$ are all bounded away from zero, $\widetilde{\sigma}_{ii}$ are all bounded away from zero with high probability, Therefore, the event $\{\max_{i,j}S_1\leq M^*\omega_p\}$ for some positive constant $M^*$ occurs with high probability. For $S_2$, with high probability, 
\begin{eqnarray*}
\max_{i,j}S_2&=&\max_{i,j}|\frac{\sigma_{K,ij}(\sigma_{K,ii}\sigma_{K,jj}-\widetilde{\sigma}_{ii}\widetilde{\sigma}_{jj})}{\sqrt{\widetilde{\sigma}_{ii}\widetilde{\sigma}_{jj}\sigma_{K,ii}\sigma_{K,jj}}(\sqrt{\sigma_{K,ii}\sigma_{K,jj}}+\sqrt{\widetilde{\sigma}_{ii}\widetilde{\sigma}_{jj}})}|\\
  &=&\max_{i,j}|\frac{\sigma_{K,ij}}{\sqrt{\widetilde{\sigma}_{ii}\widetilde{\sigma}_{jj}\sigma_{K,ii}\sigma_{K,jj}}(\sqrt{\sigma_{K,ii}\sigma_{K,jj}}+\sqrt{\widetilde{\sigma}_{ii}\widetilde{\sigma}_{jj}})}|\\
  &&\times|(\widetilde{\sigma}_{ii}-\sigma_{K,ii})(\widetilde{\sigma}_{jj}-\sigma_{K,jj})+\sigma_{K,ii}(\widetilde{\sigma}_{jj}-\sigma_{K,jj})+\sigma_{jj}(\widetilde{\sigma}_{ii}-\sigma_{K,ii})|\\
  &\leq&M^{**}\omega_p
\end{eqnarray*}
for some positive constant $M^{**}$. Therefore, 
\begin{equation*}
\max_{i\leq p,j\leq p}|\widehat{c}_{ij}-c_{ij}|=O_p(\omega_p). 
\end{equation*}

On the other hand, 
\begin{eqnarray*}
\frac{1}{n}\bM'\bM&=&\frac{1}{n}(\sum_{i=q+1}^rd_i\bv_i\bu_i')(\sum_{i=q+1}^rd_i\bu_i\bv_i')\\
                         &=&\frac{1}{n}\sum_{i=q+1}^rd_i^2\bv_i\bv_i'\\
                         &=&\widetilde{\Lambda}
\end{eqnarray*}
under the assumption that $q=s$. By Corollary 2 in \cite{ZhaoYu_2006_JMLR}, Condition \ref{m} is satisfied with high probability. This completes the proof.  

\textbf{Proof of Theorem \ref{th5}:} For notational convenience, let $\bB_S$ denote the sub matrix of $\bB$ with rows corresponding to the set $S$. Let $\bF=(\bff_1,\cdots,\bff_n)$ where $\bff_1,\cdots,\bff_n$ are independent sample data of the $s-$dimensional factors $\bff$. Let $\bK=(\bk_1,\cdots,\bk_n)$ where $\bk_1,\cdots,\bk_n$ are independent sample data of $p-$dimensional error $\bk$. Let $\bK_S$ denote the sub matrix of $\bK$ with rows corresponding to the set $S$. Let $\bB_j$ and $\bK_j$ be the $j$th row of $\bB$ and $\bK$ respectively. Note that 
\begin{eqnarray*}
\bX_S'&=&\bB_S\bF+\bK_S\\
\bx_j'&=&\bB_j\bF+\bK_j. 
\end{eqnarray*}
Therefore, we have 
\begin{eqnarray*}
\bX_S'\bX_S&=&(\bB_S\bF+\bK_S)(\bF'\bB_S'+\bK_S')\\
                       &=&\bB_S\bF\bF'\bB_S'+\bK_S\bF'\bB_S'+\bB_S\bF\bK_S'+\bK_S\bK_S'. 
\end{eqnarray*}
Since $n^{-1}\bF\bF'\stackrel{{\mathcal{P}}}\rightarrow\bSigma_f$, $n^{-1}\bK_S\bF'\stackrel{{\mathcal{P}}}\rightarrow\bzero$, and $n^{-1}\bK_S\bK_S'\stackrel{{\mathcal{P}}}\rightarrow\bSigma_{K,SS}$. Hence,
\begin{equation*}
n^{-1}\bX_S'\bX_S\stackrel{{\mathcal{P}}}\rightarrow\bB_S\bSigma_f\bB_S'+\bSigma_{K,SS}. 
\end{equation*}
Similarly, 
\begin{eqnarray*}
\bX_S'\bx_j&=&(\bB_S\bF+\bK_S)(\bB_j\bF+\bK_j)'\\
             &=&\bB_S\bF\bF'\bB_j+\bK_S\bF'\bB_j+\bB_S\bF\bK_j'+\bK_S\bK_j, 
\end{eqnarray*}
and 
\begin{equation*}
n^{-1}\bX_S'\bx_j\stackrel{{\mathcal{P}}}\rightarrow\bB_S\bSigma_f\bB_j'+\bSigma_{K,Sj}. 
\end{equation*}
Therefore, when $\bSigma_K$ is diagonal, $\bSigma_{K,Sj}=0$, 
\begin{equation*}
\max_{j\in S^c}\|(\bX_S'\bX_S)^{-1}\bX_S'\bx_j\|_1\stackrel{{\mathcal{P}}}\rightarrow\max_{j\in S^c}\|(\bB_S\bSigma_f\bB_S'+\bSigma_{K, SS})^{-1}\bB_S\bSigma_f\bB_j'\|_1. 
\end{equation*}
The proof of the first conclusion is complete. 

Next we want to show that 
\begin{equation*}
(\bM_S'\bM_S)^{-1}\bM_S'\bm_j\stackrel{{\mathcal{P}}}\rightarrow\bSigma_{K,SS}^{-1}\bSigma_{K,Sj}
\end{equation*}
In the proof of Theorem \ref{th2}, $\bM$ can be written as 
\begin{equation*}
\bM=\bX-\widehat{\bF}\widehat{\bB}'
\end{equation*}
where $\widehat{\bB}=n^{-1}\bX'\widehat{\bF}$, and $n^{-1/2}\widehat{\bF}=(\bu_1,\cdots,\bu_s)$. Based on the argument in the proof of Theorem \ref{th2}, for any $\epsilon>0$, there is positive constant $M$ such that the event $\{\max_{i\leq p, j\leq p}|\widetilde{\sigma}_{ij}-\sigma_{K,ij}|<M\omega_p\}$ occurs with probability at least $1-\epsilon$. This means that $\widetilde{\Lambda}_{ij}\stackrel{{\mathcal{P}}}\rightarrow\bSigma_{K,ij}$ for $1\leq i,j\leq p$. Equivalently, 
\begin{equation*}
n^{-1}\bM'\bM=\widetilde{\Lambda}\stackrel{{\mathcal{P}}}\rightarrow\bSigma_K. 
\end{equation*}
Correspondingly, we have 
\begin{eqnarray*}
&&n^{-1}\bM_S'\bM_S=\widetilde{\Lambda}_{SS}\stackrel{{\mathcal{P}}}\rightarrow\bSigma_{K,SS}\\
&&n^{-1}\bM_S'\bm_j=\widetilde{\Lambda}_{Sj}\stackrel{{\mathcal{P}}}\rightarrow\bSigma_{K,Sj}. 
\end{eqnarray*}
By Slustky's lemma for random matrices, 
\begin{equation*}
(\bM_S'\bM_S)^{-1}(\bM_S'\bm_j)\stackrel{{\mathcal{P}}}\rightarrow\bSigma_{K,SS}^{-1}\bSigma_{K,Sj}=0. 
\end{equation*}
The proof of the second conclusion is now complete. 

Let $A_1=\max_{j\in S^c}\|(\bB_S\bSigma_f\bB_S'+\bSigma_{K,SS})^{-1}\bB_S\bSigma_f\bB_j'\|_1$ and $A_2=0$, we will prove that $A_1>A_2$. Note that $A_1\geq A_2$, so we only need to prove that $A_1\neq A_2$. If $A_1=0$, then for all $j\in S^c$, $(\bB_S\bSigma_f\bB_S'+\bI_S)^{-1}\bB_S\bSigma_f\bB_j'=0$, which implies that $\bB_S\bSigma_f\bB_j'=0$ for $j\in S^c$. However, this is a contradiction to the condition that $\bB_S\bSigma_f\bB_{S^c}'\neq 0$. Therefore, $A_1\neq 0$. The proof of the third conclusion is now complete. 

\textbf{Proof of Theorem \ref{thm5}:} Let 
\begin{equation*}
\Delta\equiv \Big|\max_{j\in S^c}\|(\bM_S'\bM_S)^{-1}\bM_S'\bm_j\|_1-\max_{j\in S^c}\|\bSigma_{K,SS}^{-1}\bSigma_{K,Sj}\|_1\Big|. 
\end{equation*}
By triangle inequality for $l_1$ norm, we have 
\begin{eqnarray*}
\Delta&=&\Big|\|(\bM_S'\bM_S)^{-1}\bM_S'\bM_{S^{c}}\|_1-\|\bSigma_{K,SS}^{-1}\bSigma_{K,SS^{c}}\|_1\Big|\\
&\leq&\|(\frac{1}{n}\bM_S'\bM_S)^{-1}\frac{1}{n}\bM_S'\bM_{S^c}-\bSigma_{K,SS}^{-1}\bSigma_{K,SS^{c}}\|_1\\
&=&\|(\frac{1}{n}\bM_S'\bM_S)^{-1}\frac{1}{n}\bM_S'\bM_{S^c}-\bSigma_{K,SS}^{-1}\frac{1}{n}\bM_S'\bM_{S^c}+\bSigma_{K,SS}^{-1}\frac{1}{n}\bM_S'\bM_{S^c}-\bSigma_{K,SS}^{-1}\bSigma_{K,SS^{c}}\|_1\\
&\leq&\|(\frac{1}{n}\bM_S'\bM_S)^{-1}\frac{1}{n}\bM_S'\bM_{S^c}-\bSigma_{K,SS}^{-1}\frac{1}{n}\bM_S'\bM_{S^c}\|_1+\|\bSigma_{K,SS}^{-1}\frac{1}{n}\bM_S'\bM_{S^c}-\bSigma_{K,SS}^{-1}\bSigma_{K,SS^{c}}\|_1. 
\end{eqnarray*}
Then by the Cauchy Schwartz inequality for $l_1$ norm,
\begin{equation*}
\Delta\leq\|(\frac{1}{n}\bM_S'\bM_S)^{-1}-\bSigma_{K,SS}^{-1}\|_1\|\frac{1}{n}\bM_S'\bM_{S^c}\|_1+\|\bSigma_{K,SS}^{-1}\|_1\|\frac{1}{n}\bM_S'\bM_{S^c}-\bSigma_{K,SS^{c}}\|_1.\\
\end{equation*}
Recall in the proof of Theorem \ref{th2}, $\max_{i\leq p, j\leq p}|\widetilde{\sigma}_{ij}-\sigma_{ij}|=O_p(\frac{1}{\sqrt{p}}+\sqrt{\frac{\log p}{n}})$. Therefore, 
\begin{eqnarray*}
\|\frac{1}{n}\bM_S'\bM_S-\bSigma_{K,SS}\|_1&=&\max_{j\in S}\sum_{i=1}^k|\widetilde{\sigma}_{ij}-\sigma_{ij}|\\
&\leq&\max_{i\leq p,j\leq p}|\widetilde{\sigma}_{ij}-\sigma_{ij}|\times k\\
&=&O_p(\frac{k}{\sqrt{p}}+k\sqrt{\frac{\log p}{n}})
\end{eqnarray*}
Since for a symmetric matrix $\bA$, $\|\bA\|_2\leq\|\bA\|_1$, we have
\begin{equation*}
\|\frac{1}{n}\bM_S'\bM_S-\bSigma_{K,SS}\|_2=O_p(\frac{k}{\sqrt{p}}+k\sqrt{\frac{\log p}{n}}). 
\end{equation*}
When $\lambda_{\min}(\bSigma_{K,SS})\geq c_{\min}>0$, by Lemma A.1 in \cite{Fanetal_2011}, 
\begin{equation*}
\|(\frac{1}{n}\bM_S'\bM_S)^{-1}-\bSigma_{K,SS}^{-1}\|_2=\Omega_p(\|\frac{1}{n}\bM_S'\bM_S-\bSigma_{K,SS}\|_2)
\end{equation*}
where $A=\Omega_p(B)$ means that $A=O_p(B)$ and $B=O_p(A)$. Here we adopt the notation $\Omega_p$ from \cite{Bickel2008}. 

Correspondingly, we have 
\begin{equation*}
\|(\frac{1}{n}\bM_S'\bM_S)^{-1}-\bSigma_{K,SS}^{-1}\|_1\leq\sqrt{k}\|(\frac{1}{n}\bM_S'\bM_S)^{-1}-\bSigma_{K,SS}^{-1}\|_2=O_p(\frac{k^{3/2}}{\sqrt{p}}+k^{3/2}\sqrt{\frac{\log p}{n}}). 
\end{equation*}

On the other hand, 
\begin{eqnarray*}
\|\frac{1}{n}\bM_S'\bM_{S^{c}}-\bSigma_{K,SS^{c}}\|_1&=&\max_{j\in S^c}\sum_{i=1}^k|\widetilde{\sigma}_{ij}-\sigma_{ij}|\\
                    &\leq&\max_{i\leq p, j\leq p}|\widetilde{\sigma}_{ij}-\sigma_{ij}|\times k\\
                    &=&O_p(\frac{k}{\sqrt{p}}+k\sqrt{\frac{\log p}{n}})
\end{eqnarray*}
Furthermore, since $\bSigma_{K,SS^{c}}=\bzero$, we have 
\begin{equation*}
\|\frac{1}{n}\bM_S'\bM_{S^{c}}\|_1=O_p(\frac{k}{\sqrt{p}}+k\sqrt{\frac{\log p}{n}}). 
\end{equation*}
Finally, we need to evaluate $\|\bSigma_{K,SS}^{-1}\|_1$. Note that 
\begin{equation*}
\|\bSigma_{K,SS}^{-1}\|_1\leq\sqrt{k}\|\bSigma_{K,SS}^{-1}\|_2\leq c_{\min}^{-1}\sqrt{k}. 
\end{equation*}
Hence, 
\begin{eqnarray*}
&&\Big|\|(\bM_S'\bM_S)^{-1}\bM_S'\bM_{S^c}\|_1-\|\bSigma_{K,SS}^{-1}\bSigma_{K,SS^{c}}\|_1\Big|\\
&=&O_p(\frac{k^{3/2}}{\sqrt{p}}+k^{3/2}\sqrt{\frac{\log p}{n}})O_p(\frac{k}{\sqrt{p}}+k\sqrt{\frac{\log p}{n}})+\sqrt{k}O_p(\frac{k}{\sqrt{p}}+k\sqrt{\frac{\log p}{n}})\\
&=&O_p(\frac{k^{3/2}}{\sqrt{p}}+k^{3/2}\sqrt{\frac{\log p}{n}})\\
&=&o_p(1)
\end{eqnarray*}
if $k=o(p^{1/3})$ and $k^3\log p=o(n)$. The proof of the first conclusion is now complete. 

For the second conclusion, note that $\max_{j\in S^c}\|(\bX_S'\bX_S)^{-1}\bX_S'\bx_j\|_1\geq 0$. So we only need to prove that 
\begin{equation*}
\max_{j\in S^c}\|(\bX_S'\bX_S)^{-1}\bX_S'\bx_j\|_1\neq 0. 
\end{equation*}
If $\max_{j\in S^c}\|(\bX_S'\bX_S)^{-1}\bX_S'\bx_j\|_1= 0$, then for all $j\in S^c$, $(\bX_S'\bX_S)^{-1}\bX_S'\bx_j=\bzero$. Since $(\bX_S'\bX_S)$ is invertible, this implies that $\bX_S'\bx_j=\bzero$ for all $j\in S^c$, which is a contradiction to the condition that $\bX_S'\bX_{S^c}\neq\bzero$. The proof of the second conclusion is now complete. 

\begin{lemma}\label{l1}
Consider $\bZ$ with orthogonal columns. If $(\bM_S'\bM_S)$ in Condition \ref{m} is invertible, then we have
\begin{eqnarray} \label{eq17}
&&(\bM_S'\bM_S)^{-1}\bM_S'\bm_j\nonumber\\
&=&(\bX_S'\bX_S)^{-1}\bX_S'\bx_j-(\bX_S'\bX_S)^{-1}\bX_S'(\sum_{i=1}^q\bgamma_i\bgamma_i')\bx_j\nonumber\\
&&-(\bX_S'\bX_S)^{-1}\bX_S'(\sum_{i=1}^q\bgamma_i\bgamma_i')\bX_S(\bX_S'\bX_S-\bX_S'(\sum_{i=1}^q\bgamma_i\bgamma_i')\bX_S)^{-1}\bX_S'(\sum_{i=1}^q\bgamma_i\bgamma_i')\bx_j\nonumber\\
&&+(\bX_S'\bX_S)^{-1}\bX_S'(\sum_{i=1}^q\bgamma_i\bgamma_i')\bX_S(\bX_S'\bX_S-\bX_S'(\sum_{i=1}^q\bgamma_i\bgamma_i')\bX_S)^{-1}\bX_S'(\sum_{i=1}^q\bgamma_i\bgamma_i')\nonumber\\
&&\quad\quad\times\bX_S(\bX_S'\bX_S)^{-1}\bX_S'\bx_j.\nonumber 
\end{eqnarray}
\end{lemma}
\noindent\textbf{Proof of Lemma \ref{l1}: } Note that 
\begin{equation}\label{eq21}
(\bM_S'\bM_S)^{-1}\bM_S'\bm_j=(\bX_S'(\bI_{n\times n}-\sum_{i=1}^q\bgamma_i\bgamma_i')\bX_S)^{-1}\bX_S'(\bI_{n\times n}-\sum_{i=1}^q\bgamma_i\bgamma_i')\bx_j. 
\end{equation}
By Sherman-Woodbury formula, for generic matrices, 
\begin{equation*}
(\bA+\bU\bC\bV)^{-1}=\bA^{-1}-\bA^{-1}\bU(\bC^{-1}+\bV\bA^{-1}\bU)^{-1}\bV\bA^{-1}. 
\end{equation*}
Let $\bA=\bX_S'\bX_S$, $\bU=-\bX_S'(\bgamma_1,\cdots,\bgamma_q)$, $\bV=\bX_S'(\bgamma_1,\cdots,\bgamma_q)$, $\bC=\bI_{q\times q}$, then the right handed side of (\ref{eq21}) can be further expressed as 
\begin{eqnarray}\label{eq14}
&&\Big\{(\bX_S'\bX_S)^{-1}+(\bX_S'\bX_S)^{-1}\bX_S'(\bgamma_1,\cdots,\bgamma_q)\nonumber\\
&&\times(\bI_{q\times q}-(\bgamma_1,\cdots,\bgamma_q)'\bX_S(\bX_S'\bX_S)^{-1}\bX_S'(\bgamma_1,\cdots,\bgamma_q))^{-1}\nonumber\\
&&\times(\bgamma_1,\cdots,\bgamma_q)'\bX_S(\bX_S'\bX_S)^{-1}\Big\}\times [\bX_S'\bx_j-\bX_S'(\bgamma_1,\cdots,\bgamma_q)(\bgamma_1,\cdots,\bgamma_q)'\bx_j]
\end{eqnarray}
Let $\eta_1,\cdots,\eta_q$ be the nondecreasing eigenvalues of the matrix 
\begin{equation*}
(\bgamma_1,\cdots,\bgamma_q)'\bX_S(\bX_S'\bX_S)^{-1}\bX_S'(\bgamma_1,\cdots,\bgamma_q),
\end{equation*}
and $\bxi_1,\cdots,\bxi_q$ be the corresponding eigenvectors. We have
\begin{eqnarray*}
&&(\bI_{q\times q}-(\bgamma_1,\cdots,\bgamma_q)'\bX_S(\bX_S'\bX_S)^{-1}\bX_S'(\bgamma_1,\cdots,\bgamma_q))^{-1}\\
&=&(\bI_{q\times q}-\sum_{i=1}^q\eta_i\bxi_i\bxi_i')^{-1}\\
  &=&(\sum_{i=1}^q\bxi_i\bxi_i'-\sum_{i=1}^q\eta_i\bxi_i\bxi_i')^{-1}\\
  &=&(\sum_{i=1}^q(1-\eta_i)\bxi_i\bxi_i')^{-1}\\
  &=&\sum_{i=1}^q\frac{1}{1-\eta_i}\bxi_i\bxi_i'. 
\end{eqnarray*}
Therefore, (\ref{eq14}) can be further expressed as 
\begin{eqnarray}\label{eq15}
&&(\bX_S'\bX_S)^{-1}\bX_S'\bx_j+(\bX_S'\bX_S)^{-1}\bX_S'(\bgamma_1,\cdots,\bgamma_q)(\sum_{i=1}^q\frac{1}{1-\eta_i}\bxi_i\bxi_i')(\bgamma_1,\cdots,\bgamma_q)'\nonumber\\
&&\quad\quad\times\bX_S(\bX_S'\bX_S)^{-1}\bX_S'\bx_j\nonumber\\
&&-(\bX_S'\bX_S)^{-1}\bX_S'(\bgamma_1,\cdots,\bgamma_q)(\sum_{i=1}^q\frac{1}{1-\eta_i}\bxi_i\bxi_i')(\bgamma_1,\cdots,\bgamma_q)'\bX_S(\bX_S'\bX_S)^{-1}\bX_S'(\bgamma_1,\cdots,\bgamma_q)\nonumber\\
&&\quad\times(\bgamma_1,\cdots,\bgamma_q)'\bx_j-(\bX_S'\bX_S)^{-1}\bX_S'(\bgamma_1,\cdots,\bgamma_q)(\bgamma_1,\cdots,\bgamma_q)'\bx_j. 
\end{eqnarray}
Note that in the third term of (\ref{eq15}), we further apply 
\begin{equation*}
(\bgamma_1,\cdots,\bgamma_q)'\bX_S(\bX_S'\bX_S)^{-1}\bX_S'(\bgamma_1,\cdots,\bgamma_q)=\sum_{i=1}^q\eta_i\bxi_i\bxi_i',
\end{equation*}
then the third term can be simplified as 
\begin{eqnarray*}
&&-(\bX_S'\bX_S)^{-1}\bX_S'(\bgamma_1,\cdots,\bgamma_q)(\sum_{i=1}^q\frac{1}{1-\eta_i}\bxi_i\bxi_i')(\sum_{i=1}^q\eta_i\bxi_i\bxi_i')(\bgamma_1,\cdots,\bgamma_q)'\bx_j\\
&=&-(\bX_S'\bX_S)^{-1}\bX_S'(\bgamma_1,\cdots,\bgamma_q)(\sum_{i=1}^q\frac{\eta_i}{1-\eta_i}\bxi_i\bxi_i')(\bgamma_1,\cdots,\bgamma_q)'\bx_j. 
\end{eqnarray*}
Further combine with the fourth term in (\ref{eq15}), we have 
\begin{equation}\label{eq16}
-(\bX_S'\bX_S)^{-1}\bX_S'(\bgamma_1,\cdots,\bgamma_q)(\sum_{i=1}^q\frac{1}{1-\eta_i}\bxi_i\bxi_i')(\bgamma_1,\cdots,\bgamma_q)'\bx_j. 
\end{equation}
Combine (\ref{eq16}) and the second term in (\ref{eq15}). Note that 
\begin{eqnarray*}
\sum_{i=1}^q\frac{1}{1-\eta_i}\bxi_i\bxi_i'&=&[\bI_{p\times p}-(\bgamma_1,\cdots,\bgamma_q)'\bX_S(\bX_S'\bX_S)^{-1}\bX_S'(\bgamma_1,\cdots,\bgamma_q)]^{-1}
\end{eqnarray*}
Therefore, we have obtained that 
\begin{eqnarray*}
&&(\bM_S'\bM_S)^{-1}\bM_S'\bm_j\\
&=&(\bX_S'\bX_S)^{-1}\bX_S'\bx_j-\\
&&(\bX_S'\bX_S)^{-1}\bX_S'(\bgamma_1,\cdots,\bgamma_q)(\bI_{q\times q}-(\bgamma_1,\cdots,\bgamma_q)'\bX_S(\bX_S'\bX_S)^{-1}\bX_S'(\bgamma_1,\cdots,\bgamma_q))^{-1}\\
&&\quad\times(\bgamma_1,\cdots,\bgamma_q)'(\bI_{n\times n}-\bX_S(\bX_S'\bX_S)^{-1}\bX_S')\bx_j. 
\end{eqnarray*}
Apply Sherman-Woodbury formula to $(\bI_{q\times q}-(\bgamma_1,\cdots,\bgamma_q)'\bX_S(\bX_S'\bX_S)^{-1}\bX_S'(\bgamma_1,\cdots,\bgamma_q))^{-1}$, let $A=\bI_{q\times q}$, $U=-(\bgamma_1,\cdots,\bgamma_q)'\bX_S$, $C=(\bX_S'\bX_S)^{-1}$, and $V=\bX_S'(\bgamma_1,\cdots,\bgamma_q)$, we can obtain the result in Lemma \ref{l1}.  

Finally, the condition that $(\bM_S'\bM_S)$ is invertible is equivalent to the condition that all eigenvalues $\eta_i$s are less than 1. The result in Lemma \ref{l1} is valid. The proof is now complete. 

\noindent\textbf{Proof of Theorem \ref{thm3}: }

To simplify the notation, without loss of generality, consider $X_1$ in $\bX_S$, and $\cov(X_1,X_j)=\bSigma_{1,j}$ for any $j\in S^c$. We want to show 
\begin{equation}\label{eq18}
\frac{1}{n}\sum_{l=1}^q\{(\sum_{i=1}^nX_{1i}\gamma_{li})(\sum_{i=1}^nX_{ji}\gamma_{li})\}\rightarrow\xi\bSigma_{1,j} \text{  in probability}  
\end{equation}
as $n\rightarrow\infty$. Note that 
\begin{eqnarray*}
&&E\{(\sum_{i=1}^nX_{1i}\gamma_{li})(\sum_{i=1}^nX_{ji}\gamma_{li})\}\\
&=&E(\sum_{i=1}^n\sum_{h=1}^nX_{1i}X_{jh}\gamma_{li}\gamma_{lh})\\
&=&\sum_{i=1}^nE(X_{1i}X_{ji})E\gamma_{li}^2\\
&=&\bSigma_{1j}\sum_{i=1}^nE\gamma_{li}^2\\
&=&\bSigma_{1j}. 
\end{eqnarray*}
In the third step, because of independent random sample copies, $X_{1i}$ is independent of $X_{jh}$ when $i\neq h$, then the second term in the third step is 0. In the last step, we have applied $\sum_{i=1}^n\gamma_{li}^2=1$. 

Next, we want to show that $Var((\sum_{i=1}^nX_{1i}\gamma_{li})(\sum_{i=1}^nX_{ji}\gamma_{li}))$ is finite. It is sufficient to show that the second moment is finite. More specifically,
\begin{eqnarray*}
&&E\{(\sum_{i=1}^nX_{1i}\gamma_{li})^2(\sum_{i=1}^nX_{ji}\gamma_{li})^2\}\\
&=&E\{(\sum_{i=1}^nX_{1i}^2\gamma_{li}^2)(\sum_{h=1}^nX_{ji}^2\gamma_{lh}^2)\}+E\{(\sum\sum_{i\neq m}X_{1i}\gamma_{li}X_{1m}\gamma_{lm})(\sum_{h=1}^nX_{jh}^2\gamma_{lh}^2)\}\\
&&+E\{(\sum_{i=1}^nX_{1i}^2\gamma_{li}^2)(\sum\sum_{h\neq r}X_{jh}\gamma_{lh}X_{jr}\gamma_{lr})\}\\
&&+E\{(\sum\sum_{i\neq m}X_{1i}\gamma_{li}X_{1m}\gamma_{lm})(\sum\sum_{h\neq r}X_{jh}\gamma_{lh}X_{jr}\gamma_{lr})\}\\
&\equiv&E1+E2+E3+E4. 
\end{eqnarray*}
For $E1$, we have 
\begin{eqnarray*}
&&E\{(\sum_{i=1}^nX_{1i}^2\gamma_{li}^2)(\sum_{h=1}^nX_{ji}^2\gamma_{lh}^2)\}\\
&=&E(\sum_{i=1}^nX_{1i}^2X_{ji}^2\gamma_{li}^4)+E(\sum\sum_{i\neq h}X_{1i}^2\gamma_{li}^2X_{jh}^2\gamma_{lh}^2)\\
       &\leq&E(X_{1}^2X_{j}^2)+EX_{1}^2EX_{j}^2\\
       &<&\infty. 
\end{eqnarray*}
In the second step, because of independent sample copies, $X_{1i}$ is independent of $X_{jh}$ when $i\neq h$. In the third step, we have applied $\gamma_{li}^4\leq\gamma_{li}^2$, $\sum_{i=1}^n\gamma_{li}^2=1$ and $\sum\sum_{i\neq h}E(\gamma_{li}^2\gamma_{lh}^2)\leq E(\sum_{i=1}^n\gamma_{li}^2)(\sum_{h=1}^n\gamma_{lh}^2)=1$. In the last step, we have applied Cauchy Schwartz inequality $E(X_1^2X_j^2)\leq(EX_1^4)^{1/2}(EX_j^4)^{1/2}$. 

For $E2$, we have 
\begin{eqnarray*}
&&E\{(\sum\sum_{i\neq m}X_{1i}\gamma_{li}X_{1m}\gamma_{lm})(\sum_{h=1}^nX_{jh}^2\gamma_{lh}^2)\}\\
&=&E(\sum\sum\sum_{i\neq m\neq h}X_{1i}\gamma_{li}X_{1m}\gamma_{lm}X_{jh}^2\gamma_{lh}^2)+E(\sum\sum_{i\neq m}X_{1i}\gamma_{li}X_{1m}\gamma_{lm}X_{jm}^2\gamma_{lm}^2)\\
&=&\sum\sum\sum_{i\neq m\neq h}EX_{1i}EX_{1m}EX_{jh}^2E(\gamma_{li}\gamma_{lm}\gamma_{lh}^2)+\sum\sum_{i\neq m}EX_{1i}E(X_{1m}X_{jm}^2)E(\gamma_{li}\gamma_{lm}^3)\\
&=&0. 
\end{eqnarray*}
Similarly, we can show $E3=0$. For $E4$, we have
\begin{eqnarray*}
&&E\{(\sum\sum_{i\neq m}X_{1i}\gamma_{1i}X_{1m}\gamma_{lm})(\sum\sum_{h\neq r}X_{jh}\gamma_{lh}X_{jr}\gamma_{lr})\}\\
&=&\sum\sum\sum\sum_{i\neq m\neq h\neq r}EX_{1i}EX_{1m}EX_{jh}EX_{jr}E(\gamma_{li}\gamma_{lm}\gamma_{lh}\gamma_{lr})\\
&&+\sum\sum\sum_{i\neq m\neq r}EX_{1i}E(X_{1m}X_{jm})EX_{jr}E(\gamma_{li}\gamma_{lm}^2\gamma_{lr})\\
&=&0. 
\end{eqnarray*}
Based on the above analysis, $Var((\sum_{i=1}^nX_{1i}\gamma_{li})(\sum_{i=1}^nX_{ji}\gamma_{li}))$ is finite. 

Finally, for any $1\leq l\neq m\leq q$, we want to analyze the covariance terms. 
\begin{eqnarray*}
&&cov((\sum_{i=1}^nX_{1i}\gamma_{li})(\sum_{i=1}^nX_{ji}\gamma_{li}),(\sum_{i=1}^nX_{1h}\gamma_{mh})(\sum_{h=1}^nX_{jh}\gamma_{mh}))\\
&=&cov(\sum_{i=1}^nX_{1i}X_{ji}\gamma_{li}^2,\sum_{h=1}^nX_{1h}X_{jh}\gamma_{mh}^2)+cov(\sum\sum_{i\neq r}X_{1i}X_{jr}\gamma_{li}\gamma_{lr},\sum_{h=1}^nX_{1h}X_{jh}\gamma_{mh}^2)\\
&&+cov(\sum_{i=1}^nX_{1i}X_{ji}\gamma_{li}^2,\sum\sum_{h\neq s}X_{1h}X_{js}\gamma_{mh}\gamma_{ms})\\
&&+cov(\sum\sum_{i\neq r}X_{1i}X_{jr}\gamma_{li}\gamma_{lr},\sum\sum_{h\neq s}X_{1h}X_{js}\gamma_{mh}\gamma_{ms})\\
&\equiv&Cov1+Cov2+Cov3+Cov4. 
\end{eqnarray*}
For $Cov1$, we have
\begin{eqnarray*}
&&cov(\sum_{i=1}^nX_{1i}X_{ji}\gamma_{li}^2,\sum_{h=1}^nX_{1h}X_{jh}\gamma_{mh}^2)\\
&=&\sum_{i=1}^n\{E(X_{1i}^2X_{ji}^2)E(\gamma_{li}^2\gamma_{mi}^2)-(EX_{1i}X_{ji})^2E\gamma_{li}^2E\gamma_{mi}^2\}\\
&=&\sum_{i=1}^n\{(E(X_{1}^2X_{j}^2)-EX_{1}^2EX_{j}^2)E(\gamma_{li}^2\gamma_{mi}^2)+EX_{1}^2EX_{j}^2cov(\gamma_{li}^2,\gamma_{mi}^2)\}. 
\end{eqnarray*}
When $\sum_{i=1}^nE\gamma_{li}^2\gamma_{mi}^2=O(n^{-\delta_1})$ and $\sum_{i=1}^ncov(\gamma_{li}^2,\gamma_{mi}^2)=O(n^{-\delta_2})$, the last line converges to zero as $n\rightarrow\infty$. 

For $Cov2$, we have
\begin{eqnarray*}
Cov2&=&\sum\sum_{i\neq r}\sum_{h=1}^n\big\{E(X_{1i}X_{jr}\bgamma_{li}\bgamma_{lr}X_{1h}X_{jh}\bgamma_{mh}^2)-E(X_{1i}X_{jr}\bgamma_{li}\bgamma_{lr})E(X_{1h}X_{jh}\bgamma_{mh}^2)\big\}\\
    &=&\sum\sum_{i\neq r}\sum_{h=1}^n\big\{E(X_{1i}X_{jr}X_{1h}X_{jh})E(\bgamma_{li}\bgamma_{lr}\bgamma_{mh}^2)\\
    &&\quad\quad\quad\quad\quad\quad-E(X_{1i})E(X_{jr})E(X_{1h}X_{jh})E(\bgamma_{li}\bgamma_{lr})E(\bgamma_{mh}^2)\big\}\\
    &=&\sum\sum_{i\neq r}\sum_{h=1}^nE(X_{1i}X_{jr}X_{1h}X_{jh})E(\bgamma_{li}\bgamma_{lr}\bgamma_{mh}^2)
\end{eqnarray*}
In the second step, $i\neq r$ implies the independence of $X_{1i}$ and $X_{jr}$. The third step is because $E(X_{1i})=0$. Furthermore, if $h=i$, then $E(X_{1i}X_{jr}X_{1h}X_{jh})=E(X_{1i}X_{1h}X_{jh})E(X_{jr})=0$. If $h=r$, then $E(X_{1i}X_{jr}X_{1h}X_{jh})=E(X_{1i})E(X_{jr}X_{1h}X_{jh})=0$. If $h\neq i\neq r$, then $E(X_{1i}X_{jr}X_{1h}X_{jh})=E(X_{1i})E(X_{jr})E(X_{1h}X_{jh})=0$. Therefore, $Cov2=0$. 

Similarly, we can show that $Cov3=0$. For $Cov4$, we have 
\begin{eqnarray*}
Cov4&=&\sum\sum_{i\neq r}\sum\sum_{h\neq s}cov(X_{1i}X_{jr}\bgamma_{li}\bgamma_{lr}, X_{1h}X_{js}\bgamma_{mh}\bgamma_{ms})\\
      &=&\sum\sum_{i\neq r}\sum\sum_{h\neq s}\big\{E(X_{1i}X_{jr}\bgamma_{li}\bgamma_{lr}X_{1h}X_{js}\bgamma_{mh}\bgamma_{ms})\\
      &&\quad\quad\quad\quad\quad-E(X_{1i})E(X_{jr})E(\bgamma_{li}\bgamma_{lr})E(X_{1h})E(X_{js})E(\bgamma_{mh}\bgamma_{ms})\big\}\\
      &=&\sum\sum_{i\neq r}\sum\sum_{h\neq s}E(X_{1i}X_{jr}X_{1h}X_{js})E(\bgamma_{li}\bgamma_{lr}\bgamma_{mh}\bgamma_{ms}). 
\end{eqnarray*}
Let $S_1=\{i\neq r, h\neq s, h=i, r=s\}$, $S_2=\{i\neq r, h\neq s, h=r, s=i\}$, $S_3=\{i\neq r, h\neq s, h\neq i, h\neq r\}$, $S_4=\{i\neq r, h\neq s, h=i, r\neq s\}$, $S_5=\{i\neq r, h\neq s, h=r, s\neq i\}$, then 
\begin{eqnarray*}
Cov4&=&\sum\sum_{S_1}E(X_{1i}X_{jr}X_{1h}X_{js})E(\bgamma_{li}\bgamma_{lr}\bgamma_{mh}\bgamma_{ms})\\
&&+\sum\sum_{S_2}E(X_{1i}X_{jr}X_{1h}X_{js})E(\bgamma_{li}\bgamma_{lr}\bgamma_{mh}\bgamma_{ms})\\
&&+\sum\sum\sum\sum_{S_3}E(X_{1i}X_{jr}X_{1h}X_{js})E(\bgamma_{li}\bgamma_{lr}\bgamma_{mh}\bgamma_{ms})\\
&&+\sum\sum\sum_{S_4}E(X_{1i}X_{jr}X_{1h}X_{js})E(\bgamma_{li}\bgamma_{lr}\bgamma_{mh}\bgamma_{ms})\\
&&+\sum\sum\sum_{S_5}E(X_{1i}X_{jr}X_{1h}X_{js})E(\bgamma_{li}\bgamma_{lr}\bgamma_{mh}\bgamma_{ms}). 
\end{eqnarray*} 
Note that on $S_3$, $E(X_{1i}X_{jr}X_{1h}X_{js})=E(X_{1h})E(X_{1i}X_{jr}X_{js})=0$, hence the third term above is 0. On $S_4$, $E(X_{1i}X_{jr}X_{1h}X_{js})=E(X_{jr})E(X_{1i}X_{1h}X_{js})=0$, hence the fourth term above is 0. On $S_5$, $E(X_{1i}X_{jr}X_{1h}X_{js})=E(X_{1i})E(X_{jr}X_{1h}X_{js})=0$, hence the fifth term above is 0. The discussion for $S_1$ and $S_2$ is more delicate. 
\begin{eqnarray*}
&&\sum\sum_{S_1}E(X_{1i}X_{jr}X_{1h}X_{js})E(\bgamma_{li}\bgamma_{lr}\bgamma_{mh}\bgamma_{ms})\\
&=&\sum\sum_{i\neq r}E(X_{1i}^2)E(X_{jr}^2)E(\bgamma_{li}\bgamma_{lr}\bgamma_{mi}\bgamma_{mr})\\
&=&\sigma^4\{E(\sum_{i=1}^n\bgamma_{li}\bgamma_{mi})^2-\sum_{i=1}^nE(\bgamma_{li}^2\bgamma_{mi}^2)\}\\
&=&-\sigma^4\sum_{i=1}^nE(\bgamma_{li}^2\bgamma_{mi}^2)\\
&=&O(n^{-\delta_1}). 
\end{eqnarray*}
In the third step, we use the fact that $\bgamma_l$ and $\bgamma_m$ are orthogonal and hence $\sum_{i=1}\bgamma_{li}\bgamma_{mi}=0$. Similarly, we can show that 
\begin{eqnarray*}
&&\sum\sum_{S_1}E(X_{1i}X_{jr}X_{1h}X_{js})E(\bgamma_{li}\bgamma_{lr}\bgamma_{mi}\bgamma_{mr})\\
&=&-\sum\sum_{i\neq r}\rho_{1j}^2E(\bgamma_{li}^2\bgamma_{mi}^2)\\
&=&O(n^{-\delta_1}). 
\end{eqnarray*}

Therefore, (\ref{eq18}) is correct. 

In Lemma \ref{l1}, let 
\begin{eqnarray*}
A&\equiv&(\frac{1}{n}\bX_S'\bX_S)^{-1}\frac{1}{n}\bX_S'\bx_j\\
B&\equiv&(\frac{1}{n}\bX_S'\bX_S)^{-1}\frac{1}{n}\bX_S'(\sum_{i=1}^q\bgamma_i\bgamma_i')\bx_j\\
C&\equiv&(\frac{1}{n}\bX_S'\bX_S)^{-1}\frac{1}{n}\bX_S'(\sum_{i=1}^q\bgamma_i\bgamma_i')\bX_S(\frac{1}{n}\bX_S'\bX_S-\frac{1}{n}\bX_S'(\sum_{i=1}^q\bgamma_i\bgamma_i')\bX_S)^{-1}\frac{1}{n}\bX_S'(\sum_{i=1}^q\bgamma_i\bgamma_i')\bx_j\\
D&\equiv&(\frac{1}{n}\bX_S'\bX_S)^{-1}\frac{1}{n}\bX_S'(\sum_{i=1}^q\bgamma_i\bgamma_i')\bX_S(\frac{1}{n}\bX_S'\bX_S-\frac{1}{n}\bX_S'(\sum_{i=1}^q\bgamma_i\bgamma_i')\bX_S)^{-1}\\
&&\times\frac{1}{n}\bX_S'(\sum_{i=1}^q\bgamma_i\bgamma_i')\bX_S(\frac{1}{n}\bX_S'\bX_S)^{-1}\frac{1}{n}\bX_S'\bx_j. 
\end{eqnarray*}
Then we have 
\begin{equation*}
(\bM_S'\bM_S)^{-1}\bM_S'\bm_j=A-B-C+D. 
\end{equation*}
By Slutsky's Lemma for random matrix, 
\begin{eqnarray*}
A&\stackrel{{\mathcal{P}}}\rightarrow&\bSigma_{SS}^{-1}\bSigma_{Sj},\\
B&\stackrel{{\mathcal{P}}}\rightarrow&\bSigma_{SS}^{-1}\xi\bSigma_{Sj},\\
C&\stackrel{{\mathcal{P}}}\rightarrow&\bSigma_{SS}^{-1}\xi\bSigma_{SS}(\bSigma_{SS}-\xi\bSigma_{SS})^{-1}\xi\bSigma_{Sj}=\frac{\xi^2}{1-\xi}\bSigma_{SS}^{-1}\bSigma_{Sj},\\
D&\stackrel{{\mathcal{P}}}\rightarrow&\bSigma_{SS}^{-1}\xi\bSigma_{SS}(\bSigma_{SS}-\xi\bSigma_{SS})^{-1}\xi\bSigma_{SS}\bSigma_{SS}^{-1}\bSigma_{Sj}=\frac{\xi^2}{1-\xi}\bSigma_{SS}^{-1}\bSigma_{Sj}. 
\end{eqnarray*}
Therefore, 
\begin{equation*}
(\bM_S'\bM_S)^{-1}\bM_S'\bm_j\stackrel{{\mathcal{P}}}\rightarrow(1-\xi)\bSigma_{SS}^{-1}\bSigma_{Sj}. 
\end{equation*}

Based on the above argument, 
\begin{equation}\label{eq23}
\max_{j\in S^c}\|(\bM_S'\bM_S)^{-1}\bM_S'\bm_j\|_1\stackrel{{\mathcal{P}}}\rightarrow(1-\xi)\max_{j\in S^c}\|\bSigma_{SS}^{-1}\bSigma_{Sj}\|_1. 
\end{equation}
Combining with the result that 
\begin{equation*}
\max_{j\in S^c}\|(\bX_S'\bX_S)^{-1}\bX_S'\bx_j\|_1\stackrel{{\mathcal{P}}}\rightarrow\max_{j\in S^c}\|\bSigma_{SS}^{-1}\bSigma_{Sj}\|_1
\end{equation*}
by the Slutsky's Theorem, we have the conclusion in Theorem \ref{thm3}. The proof is complete. 


\begin{lemma}\label{l4}
Suppose $\{\bgamma_i\}_{i=1}^q$ are orthonormal and $\{\bgamma_i\}_{i=1}^q$ are independent of $\bX$. \\If $(X_1,\cdots,X_p)'\sim N_p(\bzero,\bSigma)$, $q/n=\xi$, and let $\bX_i=(X_{i1},\cdots, X_{in})'$ be the $n$ independent sample data of $X_i$ for $i=1,\cdots, p$, then 
\begin{equation*}
\max_{1\leq i\leq p,1\leq j\leq p}|\frac{1}{n}\bX_i'(\bgamma_1,\cdots,\bgamma_q)(\bgamma_1,\cdots,\bgamma_q)'\bX_j-\xi\Sigma_{ij}|=O_p(\sqrt{\frac{\log p}{n}}). 
\end{equation*}
\end{lemma}

\textbf{Proof of Lemma \ref{l4}:} To simplify the notation, we use $\Sigma_{ij}=\rho_{ij}$. By Gaussian distribution, we can write 
\begin{equation}\label{eq25}
X_j=\rho_{ij} X_i+\sqrt{1-\rho_{ij}^2}W
\end{equation}
where $W\sim N(0,1)$ and $W$ is independent of $X_i$. 

Since linear combination of independent Gaussian random variables still follows Gaussian distribution, we have $\bX_i'\bgamma_l\Big|\bgamma_1,\cdots,\bgamma_q\sim N(0,1)$ for $l=1,\cdots,q$. Furthermore, 
\begin{equation*}
E[\bX_i'\bgamma_i\bgamma_j'\bX_i\Big|\bgamma_1,\cdots,\bgamma_q]=E[\bgamma_i'\bX_i\bX_i'\bgamma_j\Big|\bgamma_1,\cdots,\bgamma_q]=\bgamma_i'E[\bX_i\bX_i']\bgamma_j=\bgamma_i'\bI_n\bgamma_j=0
\end{equation*}
Therefore, $cov(\bX_i'\bgamma_i,\bX_i'\bgamma_j\Big|\bgamma_1,\cdots,\bgamma_q)=0$ for $1\leq i\neq j\leq q$. Because of Gaussian distribution, we have 
\begin{equation*}
(\bX_i'\bgamma_1,\cdots,\bX_i'\bgamma_q)'\Big|\bgamma_1,\cdots,\bgamma_q\sim N(\bzero,\bI_q). 
\end{equation*}
Similarly, if we write $\bW=(W_1,\cdots,W_n)'$ as the $n$ independent sample data of $W$, then $(\bW'\bgamma_1,\cdots,\bW'\bgamma_q)'\Big|\bgamma_1,\cdots,\bgamma_q\sim N(\bzero,\bI_q)$. 

Note that for any positive $t$, we want to evaluate
\begin{eqnarray*}
&&P(|\frac{1}{n}\bX_i'(\bgamma_1,\cdots,\bgamma_q)(\bgamma_1,\cdots,\bgamma_q)'\bX_j-\xi\rho_{ij}|\geq t)\\
&\leq&P(\frac{1}{n}\bX_i'(\bgamma_1,\cdots,\bgamma_q)(\bgamma_1,\cdots,\bgamma_q)'\bX_j-\xi\rho_{ij}\geq t)\\
&&+P(\frac{1}{n}\bX_i'(\bgamma_1,\cdots,\bgamma_q)(\bgamma_1,\cdots,\bgamma_q)'\bX_j-\xi\rho_{ij}\leq -t)\\
&\equiv&\Delta_1+\Delta_2. 
\end{eqnarray*}
We will derive the tail probability for the first term above. The derivation for the second term is similar. 

Recall (\ref{eq25}), $\Delta_1$ can be written as
\begin{eqnarray*}
\Delta_1&=&P\Big(\frac{\rho_{ij}}{n}\bX_i'(\bgamma_1,\cdots,\bgamma_q)(\bgamma_1,\cdots,\bgamma_q)'\bX_i\\
&&\quad\quad\quad\quad+\frac{\sqrt{1-\rho_{ij}^2}}{n}\bX_i'(\bgamma_1,\cdots,\bgamma_q)(\bgamma_1,\cdots,\bgamma_q)'\bW-\xi\rho_{ij}\geq t\Big). 
\end{eqnarray*}
Note that for two generic variables $A$ and $B$, 
\begin{eqnarray*}
P(A+B\geq t)&=&P(A+B\geq t, B\geq \frac{t}{2})+P(A+B\geq t, B<\frac{t}{2})\\
               &\leq&P(B\geq \frac{t}{2})+P(A\geq \frac{t}{2},B<\frac{t}{2})\\
               &\leq&P(B\geq \frac{t}{2})+P(A\geq \frac{t}{2}). 
\end{eqnarray*}

Therefore, we have 
\begin{eqnarray*}
\Delta_1&\leq& P\Big(\frac{\rho_{ij}}{n}\bX_i'(\bgamma_1,\cdots,\bgamma_q)(\bgamma_1,\cdots,\bgamma_q)'\bX_i-\xi\rho_{ij}\geq \frac{t}{2}\Big)\\
&&\quad+P\Big(\frac{\sqrt{1-\rho_{ij}^2}}{n}\bX_i'(\bgamma_1,\cdots,\bgamma_q)(\bgamma_1,\cdots,\bgamma_q)'\bW\geq \frac{t}{2}\Big)\\
    &\equiv&\Delta_{11}+\Delta_{12}. 
\end{eqnarray*}
We will analyze $\Delta_{11}$ and $\Delta_{12}$ separately. For $\Delta_{11}$, let $Z_l=\bX_i'\bgamma_l$, we have 
\begin{eqnarray*}
\Delta_{11}&=&E\{P(\frac{\rho_{ij}}{n}\bX_i'(\bgamma_1,\cdots,\bgamma_q)(\bgamma_1,\cdots,\bgamma_q)'\bX_i-\xi\rho_{ij}\geq \frac{t}{2}|\bgamma_1,\cdots,\bgamma_q)\}\\
   &=&E\{P(\frac{\rho_{ij}}{n}\sum_{l=1}^qZ_l^2-\xi\rho_{ij}\geq \frac{t}{2}|\bgamma_1,\cdots,\bgamma_q)\}\\
   &=&E\{P(\frac{q}{n}\frac{1}{q}\sum_{l=1}^qZ_l^2-\xi\geq \frac{t}{2\rho_{ij}}|\bgamma_1,\cdots,\bgamma_q)\}
\end{eqnarray*}
By the condition that $q/n=\xi$, we have
\begin{equation*}
\Delta_{11}\leq E\{P(\frac{1}{q}\sum_{l=1}^qZ_l^2-1\geq \frac{t}{2\rho_{ij}\xi}|\bgamma_1,\cdots,\bgamma_q)\}.
\end{equation*}
Note that conditional on $\bgamma_1,\cdots,\bgamma_q$, $Z_l^2\stackrel{i.i.d.}\sim \chi_1^2$, which is sub exponential variable. By the concentration inequality, 
\begin{equation*}
P(\frac{1}{q}\sum_{l=1}^qZ_l^2-1\geq \frac{t}{2\rho_{ij}\xi}|\bgamma_1,\cdots,\bgamma_q)\leq \exp(-\frac{qt^2}{32\rho_{ij}^2\xi^2})
\end{equation*}
for all $t\in(0,1)$, and correspondingly, we have
\begin{equation*}
\Delta_{11}\leq  \exp(-\frac{qt^2}{32\rho_{ij}^2\xi^2}). 
\end{equation*}

For $\Delta_{12}$, if we let $V_l=\bgamma_l'W$, then we have
\begin{equation*}
\Delta_{12}=E\{P(\frac{\sqrt{1-\rho_{ij}^2}}{n}\sum_{l=1}^qZ_lV_l\geq \frac{t}{2}|\bgamma_1,\cdots,\bgamma_q)\}.
\end{equation*}
Note that $Z_lV_l=\frac{1}{4}[(Z_l+V_l)^2-(Z_l-V_l)^2]$. Therefore, 
\begin{eqnarray*}
\Delta_{12}&=&E\{P(\frac{\sqrt{1-\rho_{ij}^2}}{n}\sum_{l=1}^q\frac{(Z_l+V_l)^2}{2}-\sqrt{1-\rho_{ij}^2}\xi\\
&&\quad\quad\quad+\sqrt{1-\rho_{ij}^2}\xi-\frac{\sqrt{1-\rho_{ij}^2}}{n}\sum_{l=1}^q\frac{(Z_l-V_l)^2}{2}\geq t|\bgamma_1,\cdots,\bgamma_q)\}\\
  &\leq&E\{P(\sqrt{1-\rho_{ij}^2}(\frac{1}{n}\sum_{l=1}^q\frac{(Z_l+V_l)^2}{2}-\xi)\geq \frac{t}{2}|\bgamma_1,\cdots,\bgamma_q)\}\\
  &&\quad\quad+E\{P(\sqrt{1-\rho_{ij}^2}(\xi-\frac{1}{n}\sum_{l=1}^q\frac{(Z_l-V_l)^2}{2})\geq \frac{t}{2}|\bgamma_1,\cdots,\bgamma_q)\}\\
  &\equiv&\Delta_{121}+\Delta_{122}. 
\end{eqnarray*}
For $\Delta_{121}$, we have 
\begin{eqnarray*}
\Delta_{121}&=&E\{P(\frac{q}{n}\frac{1}{q}\sum_{l=1}^q\frac{(Z_l+V_l)^2}{2}-\xi\geq \frac{t}{2\sqrt{1-\rho_{ij}^2}}|\bgamma_1,\cdots,\bgamma_q)\}\\
    &\leq&E\{P(\frac{1}{q}\sum_{l=1}^q\frac{(Z_l+V_l)^2}{2}-1\geq \frac{t}{2\xi\sqrt{1-\rho_{ij}^2}}|\bgamma_1,\cdots,\bgamma_q)\}.
\end{eqnarray*}
Note that conditional on $\bgamma_1,\cdots,\bgamma_q$, $Z_l$ and $V_l$ are independent, and $\frac{(Z_l+V_l)^2}{2}\stackrel{i.i.d.}\sim \chi_1^2$. By concentration inequality, 
\begin{equation*}
\Delta_{121}\leq\exp(-\frac{qt^2}{32\xi^2(1-\rho_{ij}^2)}). 
\end{equation*}
Similarly, for $\Delta_{122}$, we have 
\begin{eqnarray*}
\Delta_{122}&\leq&E\{P(\frac{1}{q}\sum_{l=1}^q\frac{(Z_l-V_l)^2}{2}-1\leq-\frac{t}{2\xi\sqrt{1-\rho_{ij}^2}}|\bgamma_1,\cdots,\bgamma_q)\}\\
           &\leq&\exp(-\frac{qt^2}{32\xi^2(1-\rho_{ij}^2)}). 
\end{eqnarray*}
Combining the results above, by the union bound, we have for any positive $t$,  
\begin{equation}\label{eq26}
P(\max_{1\leq i\leq p,1\leq j\leq p}|\frac{1}{n}\bX_i'(\bgamma_1,\cdots,\bgamma_q)(\bgamma_1,\cdots,\bgamma_q)'\bX_j-\xi\Sigma_{ij}|\geq t)\leq C_1p^2\exp(-C_2\frac{qt^2}{\xi^2})
\end{equation}
for some positive constants $C_1$ and $C_2$. By the condition that $q/n=\xi$, the right handed side can be written as $C_1p^2\exp(-C_2\frac{nt^2}{\xi})$. There exists a positive constant $C_3$ such that $t>C_3\sqrt{\frac{\log p}{n}}$, the right handed side of (\ref{eq26}) is o(1), which implies the conclusion in the lemma. The proof is now complete. 

\textbf{Proof of Theorem \ref{thm6}:} Note that 
\begin{eqnarray*}
&&|\|(\bX_S'\bX_S)^{-1}\bX_S'\bX_{S^{c}}\|_1-\|\bSigma_{SS}^{-1}\bSigma_{SS^{c}}\|_1|\\
&\leq&\|(\bX_S'\bX_S)^{-1}\bX_S'\bX_{S^{c}}-\bSigma_{SS}^{-1}\bSigma_{SS^{c}}\|_1\\
&=&\|[(\frac{1}{n}\bX_S'\bX_S)^{-1}-\bSigma_{SS}^{-1}]\frac{1}{n}\bX_S'\bX_{S^{c}}+\bSigma_{SS}^{-1}(\frac{1}{n}\bX_S'\bX_{S^{c}}-\bSigma_{SS^{c}})\|_1\\
&\leq&\|(\frac{1}{n}\bX_S'\bX_S)^{-1}-\bSigma_{SS}^{-1}\|_1\|\frac{1}{n}\bX_S'\bX_{S^{c}}\|_1+\|\bSigma_{SS}^{-1}\|_1\|\frac{1}{n}\bX_S'\bX_{S^{c}}-\bSigma_{SS^{c}}\|_1
\end{eqnarray*} 
By Lemma 9(b) in \cite{Wainwright_2009}, we have 
\begin{equation*}
\|(\frac{1}{n}\bX_S'\bX_S)^{-1}-\bSigma_{SS}^{-1}\|_2=O_p(\sqrt{\frac{k}{n}}). 
\end{equation*}
Therefore, 
\begin{equation*}
\|(\frac{1}{n}\bX_S'\bX_S)^{-1}-\bSigma_{SS}^{-1}\|_1\leq \sqrt{k}\|(\frac{1}{n}\bX_S'\bX_S)^{-1}-\bSigma_{SS}^{-1}\|_2=O_p(\frac{k}{\sqrt{n}}). 
\end{equation*}
To evaluate $\|\frac{1}{n}\bX_S'\bX_{S^{c}}-\bSigma_{SS^{c}}\|_1$, note that by \cite{Bickel2008}, 
\begin{equation*}
\max_{1\leq i\leq p,1\leq j\leq p}|\widehat{\sigma}_{ij}-\sigma_{ij}|=O_p(\sqrt{\frac{\log p}{n}}). 
\end{equation*}
Correspondingly, 
\begin{eqnarray*}
\|\frac{1}{n}\bX_S'\bX_{S^{c}}-\bSigma_{SS^{c}}\|_1&=&\max_{j\in S^c}\sum_{i=1}^k|\widehat{\sigma}_{ij}-\sigma_{ij}|\\
              &\leq&\max_{1\leq i\leq p,1\leq j\leq p}|\widehat{\sigma}_{ij}-\sigma_{ij}|\times k\\
              &=&O_p(k\sqrt{\frac{\log p}{n}}). 
\end{eqnarray*}
Furthermore, we have 
\begin{equation*}
\|\frac{1}{n}\bX_S'\bX_{S^{c}}\|_1=\|\bSigma_{SS^{c}}\|_1+O_p(k\sqrt{\frac{\log p}{n}}). 
\end{equation*}
We also have 
\begin{equation*}
\|\bSigma_{SS}^{-1}\|_1\leq\sqrt{k}\|\bSigma_{SS}^{-1}\|_2\leq c_{\min}^{-1}\sqrt{k}. 
\end{equation*}
Finally, if $k^{3/2}\sqrt{\frac{\log p}{n}}=o(1)$ and $\frac{k}{\sqrt{n}}=o(1)$,  we obtain 
\begin{equation*}
|\|(\bX_S'\bX_S)^{-1}\bX_S'\bX_{S^{c}}\|_1-\|\bSigma_{SS}^{-1}\bSigma_{SS^{c}}\|_1|=O_p(\frac{k}{\sqrt{n}}\|\bSigma_{SS^{c}}\|_1+k^{3/2}\sqrt{\frac{\log p}{n}}). 
\end{equation*}
The proof of the first conclusion is now complete. 

For the second conclusion, note that by Lemma \ref{l1} and triangle inequality, 
\begin{eqnarray*}
&&|\|(\bM_S'\bM_S)^{-1}\bM_S'\bM_{S^c}\|_1-\|(1-\xi)\bSigma_{SS}^{-1}\bSigma_{SS^c}\|_1|\\
&\leq&\|(\bX_S'\bX_S)^{-1}\bX_S'\bX_{S^c}-\bSigma_{SS}^{-1}\bSigma_{SS^c}\|_1+\|(\bX_S'\bX_S)^{-1}\bX_S'(\sum_{i=1}^q\bgamma_i\bgamma_i')\bX_{S^c}-\xi\bSigma_{SS}^{-1}\bSigma_{SS^c}\|_1\\
&&+\|(\bX_S'\bX_S)^{-1}\bX_S'(\sum_{i=1}^q\bgamma_i\bgamma_i')\bX_S(\bX_S'\bX_S-\bX_S'(\sum_{i=1}^q\bgamma_i\bgamma_i')\bX_S)^{-1}\bX_S'(\sum_{i=1}^q\bgamma_i\bgamma_i')\bX_{S^c}\\
&&\quad\quad-\bSigma_{SS}^{-1}\xi\bSigma_{SS}(\bSigma_{SS}-\xi\bSigma_{SS})^{-1}\xi\bSigma_{SS^c}\|_1\\
&&+\|(\bX_S'\bX_S)^{-1}\bX_S'(\sum_{i=1}^q\bgamma_i\bgamma_i')\bX_S(\bX_S'\bX_S-\bX_S'(\sum_{i=1}^q\bgamma_i\bgamma_i')\bX_S)^{-1}\\
&&\quad\quad\quad\times\bX_S'(\sum_{i=1}^q\bgamma_i\bgamma_i')\bX_S(\bX_S'\bX_S)^{-1}\bX_S'\bX_{S^c}\\
&&\quad\quad-\bSigma_{SS}^{-1}\xi\bSigma_{SS}(\bSigma_{SS}-\xi\bSigma_{SS})^{-1}\xi\bSigma_{SS}\bSigma_{SS}^{-1}\bSigma_{SS^c}\|_1\\
&\equiv&\Lambda_1+\Lambda_2+\Lambda_3+\Lambda_4. 
\end{eqnarray*}
We have analyzed $\Lambda_1$ for the first conclusion. We now evaluate $\Lambda_2$. Note that
\begin{eqnarray*}
\Lambda_2&=&\|(\frac{1}{n}\bX_S'\bX_S)^{-1}-\bSigma_{SS}^{-1})\frac{1}{n}\bX_S'(\sum_{i=1}^q\bgamma_i\bgamma_i')\bX_{S^c}+\bSigma_{SS}^{-1}(\frac{1}{n}\bX_S'(\sum_{i=1}^q\bgamma_i\bgamma_i')\bX_{S^c}-\xi\bSigma_{SS^c})\|_1\\
&\leq&\|(\frac{1}{n}\bX_S'\bX_S)^{-1}-\bSigma_{SS}^{-1}\|_1\|\frac{1}{n}\bX_S'(\sum_{i=1}^q\bgamma_i\bgamma_i')\bX_{S^c}\|_1+\|\bSigma_{SS}^{-1}\|_1\|\frac{1}{n}\bX_S'(\sum_{i=1}^q\bgamma_i\bgamma_i')\bX_{S^c}-\xi\bSigma_{SS^c}\|_1
\end{eqnarray*}
By Lemma \ref{l4}, 
\begin{eqnarray*}
&&\|\frac{1}{n}\bX_S'(\sum_{i=1}^q\bgamma_i\bgamma_i')\bX_{S^c}-\xi\bSigma_{SS^c}\|_1\\
&=&\max_{j\in S^c}\sum_{i\in S}|\frac{1}{n}\bX_i'(\sum_{l=1}^q\bgamma_l\bgamma_l')\bX_{j}-\xi\bSigma_{ij}|\\
&\leq&\max_{1\leq i\leq p,1\leq j\leq p}|\frac{1}{n}\bX_i'(\sum_{l=1}^q\bgamma_l\bgamma_l')\bX_{j}-\xi\bSigma_{ij}|\times k\\
&=&O_p(k\sqrt{\frac{\log p}{n}}). 
\end{eqnarray*}
Hence, if $k^{3/2}\sqrt{\frac{\log p}{n}}=o(1)$ and $\frac{k}{\sqrt{n}}=o(1)$, 
\begin{eqnarray*}
\Lambda_2&=&O_p(\frac{k}{\sqrt{n}}\xi\|\bSigma_{SS^c}\|_1+\frac{k^2}{n}\sqrt{\log p}+k^{3/2}\sqrt{\frac{\log p}{n}})\\
       &=&O_p(\frac{k}{\sqrt{n}}\xi\|\bSigma_{SS^c}\|_1+k^{3/2}\sqrt{\frac{\log p}{n}}). 
\end{eqnarray*}
Next, for $\Lambda_3$, let 
\begin{eqnarray*}
A&=&(\frac{1}{n}\bX_S'\bX_S)^{-1}\\
B&=&\frac{1}{n}\bX_S'(\sum_{i=1}^q\bgamma_i\bgamma_i')\bX_S\\
C&=&(\frac{1}{n}\bX_S'\bX_S-\frac{1}{n}\bX_S'(\sum_{i=1}^q\bgamma_i\bgamma_i')\bX_S)^{-1}\\
D&=&\frac{1}{n}\bX_S'(\sum_{i=1}^q\bgamma_i\bgamma_i')\bX_{S^c}. 
\end{eqnarray*}
then by triangle inequality, we have
\begin{equation*}
\Lambda_3\leq\|AB-\bSigma_{SS}^{-1}\xi\bSigma_{SS}\|_1\|CD\|_1+\xi\|CD-(\bSigma_{SS}-\xi\bSigma_{SS})^{-1}\xi\bSigma_{SS^c}\|_1
\end{equation*}
For the second term from the last line, 
\begin{eqnarray*}
&&\|CD-(\bSigma_{SS}-\xi\bSigma_{SS})^{-1}\xi\bSigma_{SS^c}\|_1\\
&\leq&\|C-(\bSigma_{SS}-\xi\bSigma_{SS})^{-1}\|_1\|D\|_1+\|(\bSigma_{SS}-\xi\bSigma_{SS})^{-1}\|_1\|D-\xi\bSigma_{SS^c}\|_1. 
\end{eqnarray*}
By Lemma A.1 in \cite{Fanetal_2011}, when $\lambda_{min}(\bSigma_{SS})>c_{min}>0$, 
\begin{eqnarray*}
\|C-(\bSigma_{SS}-\xi\bSigma_{SS})^{-1}\|_1&\leq& k^{1/2}\|C-(\bSigma_{SS}-\xi\bSigma_{SS})^{-1}\|_2\\
            &=&k^{1/2}\Omega_p(\|\frac{1}{n}\bX_S'\bX_S-\frac{1}{n}\bX_S'(\sum_{i=1}^q\bgamma_i\bgamma_i')\bX_S-\bSigma_{SS}+\xi\bSigma_{SS}\|_2)\\
            &=&O_p(k^{3/2}\sqrt{\frac{\log p}{n}}), 
\end{eqnarray*}
and 
\begin{equation*}
\|D-\xi\bSigma_{SS}\|_1=O_p(k\sqrt{\frac{\log p}{n}}). 
\end{equation*}
Therefore, if $k^{3/2}\sqrt{\frac{\log p}{n}}=o(1)$ and $\frac{k}{\sqrt{n}}=o(1)$, 
\begin{eqnarray*}
&&\|CD-(\bSigma_{SS}-\xi\bSigma_{SS})^{-1}\xi\bSigma_{SS^c}\|_1\\
&=&O_p(k^{3/2}\sqrt{\frac{\log p}{n}}\|\bSigma_{SS^c}\|_1+k^{5/2}\frac{\log p}{n}+k^{3/2}\sqrt{\frac{\log p}{n}})\\
&=&O_p(k^{3/2}\sqrt{\frac{\log p}{n}}\|\bSigma_{SS^c}\|_1+k^{3/2}\sqrt{\frac{\log p}{n}}). 
\end{eqnarray*}
Besides, from $\Lambda_2$, we have
\begin{equation*}
\|AB-\bSigma_{SS}^{-1}\xi\bSigma_{SS}\|_1=O_p(\frac{k}{\sqrt{n}}\|\bSigma_{SS}\|_1+k^{3/2}\sqrt{\frac{\log p}{n}}).
\end{equation*}
Hence, if $k^{3/2}\sqrt{\frac{\log p}{n}}=o(1)$, we have
\begin{eqnarray*}
\Lambda_3&=&O_p(\frac{k^{3/2}}{\sqrt{n}}\|\bSigma_{SS}\|_1\|\bSigma_{SS^c}\|_1+\frac{k^{5/2}}{n}\sqrt{\log p}\|\bSigma_{SS}\|_1+k^2\sqrt{\frac{\log p}{n}}\|\bSigma_{SS^c}\|_1+k^{3/2}\sqrt{\frac{\log p}{n}}). 
\end{eqnarray*}

Finally, we will analyze $\Lambda_4$. Adopting the notations $A$, $B$ and $C$ in $\Lambda_3$, we have 
\begin{eqnarray*}
\Lambda_4&=&\|ABCB(\frac{1}{n}\bX_S'\bX_S)^{-1}\frac{1}{n}\bX_S'\bX_{S^c}-\frac{\xi^2}{1-\xi}\bSigma_{SS}^{-1}\bSigma_{SS^c}\|_1\\
           &\leq&\|ABCB-\frac{\xi^2}{1-\xi}\|_1\|(\frac{1}{n}\bX_S'\bX_S)^{-1}\frac{1}{n}\bX_S'\bX_{S^c}\|_1\\
           &&\quad+\frac{\xi^2}{1-\xi}\|(\frac{1}{n}\bX_S'\bX_S)^{-1}\frac{1}{n}\bX_S'\bX_{S^c}-\bSigma_{SS}^{-1}\bSigma_{SS^c}\|_1. 
\end{eqnarray*}
Note that based on the analysis in the first conclusion, we have
\begin{equation*}
\|(\frac{1}{n}\bX_S'\bX_S)^{-1}\frac{1}{n}\bX_S'\bX_{S^c}-\bSigma_{SS}^{-1}\bSigma_{SS^c}\|_1=O_p(\frac{k}{\sqrt{n}}\|\bSigma_{SS^{c}}\|_1+k^{3/2}\sqrt{\frac{\log p}{n}}), 
\end{equation*}
and 
\begin{equation*}
\|(\frac{1}{n}\bX_S'\bX_S)^{-1}\frac{1}{n}\bX_S'\bX_{S^c}\|_1=O_p(k^{1/2}\|\bSigma_{SS^c}\|_1+k^{3/2}\sqrt{\frac{\log p}{n}}). 
\end{equation*}
Similar to the discussion in $\Lambda_3$, 
\begin{equation*}
\|ABCB-\frac{\xi^2}{1-\xi}\|_1=O_p(\frac{k^{3/2}}{\sqrt{n}}\|\bSigma_{SS}\|_1^2+k^2\sqrt{\frac{\log p}{n}}\|\bSigma_{SS}\|_1+k^{3/2}\sqrt{\frac{\log p}{n}}),  
\end{equation*}
except that we replace $\|\bSigma_{SS^c}\|_1$ by $\|\bSigma_{SS}\|_1$. Combining the results above, we have
\begin{eqnarray*}
\Lambda_4&=&O_p(\frac{k^2}{\sqrt{n}}\|\bSigma_{SS}\|_1^2\|\bSigma_{SS^c}\|_1+k^{5/2}\sqrt{\frac{\log p}{n}}\|\bSigma_{SS}\|_1\|\bSigma_{SS^c}\|_1\\
&&+k^2\sqrt{\frac{\log p}{n}}\|\bSigma_{SS^c}\|_1+\frac{k^3}{n}\sqrt{\log p}\|\bSigma_{SS}\|_1^2+k^{7/2}\frac{\log p}{n}\|\bSigma_{SS}\|_1+k^{3/2}\sqrt{\frac{\log p}{n}}). 
\end{eqnarray*}

Combining the results for $\Lambda_1$, $\Lambda_2$, $\Lambda_3$ and $\Lambda_4$, After some algebra, we have 
\begin{eqnarray*}
&&\big|\|(\bM_S'\bM_S)^{-1}\bM_S'\bM_{S^{c}}\|_1-(1-\xi)\|\bSigma_{SS}^{-1}\bSigma_{SS^{c}}\|_1\big|\\
&=&O_p\Big((\frac{k}{\sqrt{n}}+k^2\sqrt{\frac{\log p}{n}})\|\bSigma_{SS^c}\|_1+k^{3/2}\sqrt{\frac{\log p}{n}}+(\frac{k^{5/2}}{n}\sqrt{\log p}+\frac{k^{7/2}}{n}\sqrt{\log p})\|\bSigma_{SS}\|_1\\
&&\quad+(\frac{k^{3/2}}{\sqrt{n}}+k^{5/2}\sqrt{\frac{\log p}{n}})\|\bSigma_{SS}\|_1\|\bSigma_{SS^c}\|_1+\frac{k^2}{\sqrt{n}}\|\bSigma_{SS}\|_1^2\|\bSigma_{SS^c}\|_1+\frac{k^3}{n}\sqrt{\log p}\|\bSigma_{SS}\|_1^2\Big). 
\end{eqnarray*}
When $k^{5/2}\sqrt{\frac{\log p}{n}}=o(1)$ and $\frac{k^2}{\sqrt{n}}=o(1)$, the right handed side of the last expression can be expressed as $o_p(\|\bSigma_{SS^c}\|_1+\|\bSigma_{SS}\|_1(\|\bSigma_{SS}\|_1+1)(\|\bSigma_{SS^c}\|_1+1))$. The proof is now complete.

\section{Proofs for Examples 1 and 2 in Section 2.5}

{\bf Proof of Example 1: } First, for the PROD with least important component method, we consider the conditions in the current Theorem \ref{th5}. Note that $n^{-1}\bX_S'(\bI-\bP_Z)\bX_S\stackrel{{\mathcal{P}}}\rightarrow\bSigma_{K,SS}=\bI_S$ since we assume $\bSigma_K=\bI$. Thus, $\|(n^{-1}\bX_S'(\bI-\bP_Z)\bX_S)^{-1}\|_{\infty}\stackrel{{\mathcal{P}}}\rightarrow 1$, and $C_{\min}^*=1$. In addition, Theorem \ref{th5} has shown that for $(\bI-\bP_Z)\bX$, the irrepresentable condition converges to 0, thus the corresponding $\gamma_{PROD}$ is 1. Overall, we can evaluate the required minimum signal strength for $(\bI-\bP_Z)\bX$ by   
\begin{equation}\label{sigM}
q_2=1\Big[1+\frac{4\sigma}{1}\Big].
\end{equation}
For $\bX$, Theorem \ref{th5} has shown that $(n^{-1}\bX_S'\bX_S)^{-1}\stackrel{{\mathcal{P}}}\rightarrow(\bB_S\bSigma_f\bB_S'+\bI_S)^{-1}$ and $(\bX_S'\bX_S)^{-1}\bX_S'\bX_{S^c}$ $\stackrel{{\mathcal{P}}}\rightarrow(\bB_S\bSigma_f\bB_S'+\bI_S)^{-1}(\bB_S\bSigma_f\bB_{S^c})$. To simplify the discussion, we further assume that both $\bU$ and $\widetilde{\bU}$ are identity matrices, and $\bV=\widetilde{\bV}$. Thus, for $\bX$, we can obtain $\|(\bB_S\bB_S'+\bI_S)^{-1}\|_{\infty}=1$, $\Lambda_{\min}(\bB_S\bB_S'+\bI_S)=1$, and $\|(\bB_S\bB_S'+\bI_S)^{-1}\bB_S\bB_{S^c}'\|_1=\|(\bLambda^2+\bI)^{-1}\bLambda\widetilde{\bLambda}\|_1>0$. Thus, the $\gamma_X$ in the irrepresentable condition for $\bX$ is smaller than 1. We can evaluate the required minimum signal strength for $\bX$ by 
\begin{eqnarray}\label{sigX}
q_1&=&\frac{1}{\gamma_X}\Big[\|(\bB_S\bB_S'+\bI_S)^{-1}\|_{\infty}+\frac{4\sigma}{\sqrt{\Lambda_{\min}(\bB_S\bB_S'+\bI_S)}}\Big]\nonumber\\
&=&\frac{1}{\gamma_X}\Big[1+\frac{4\sigma}{1}\Big]. 
\end{eqnarray}
Comparison of (\ref{sigM}) with (\ref{sigX}) shows that $q_2<q_1$. 

\noindent{\bf Proof of Example 2:} Next, for the PROD with random generation methods, let $\Delta\equiv(\bX_S(\bI-\bP_Z)\bX_S)^{-1}$. By Sherman-Woodbury formula, for generic matrices, 
\begin{equation*}
(\bA+\bU\bC\bV)^{-1}=\bA^{-1}-\bA^{-1}\bU(\bC^{-1}+\bV\bA^{-1}\bU)^{-1}\bV\bA^{-1}. 
\end{equation*}
Let $\bA=\bX_S'\bX_S$, $\bU=-\bX_S'(\bgamma_1,\cdots,\bgamma_q)$, $\bV=(\bgamma_1,\cdots,\bgamma_1)'\bX_S$ and $\bC=\bI$, then 
\begin{eqnarray*}
\Delta&=&(\bX_S'\bX_S)^{-1}+(\bX_S'\bX_S)^{-1}\bX_S'(\bgamma_1,\cdots,\bgamma_q)\\
   &&\quad\quad\times(\bI-(\bgamma_1,\cdots,\bgamma_q)'\bX_S(\bX_S'\bX_S)^{-1}\bX_S'(\bgamma_1,\cdots,\bgamma_q))^{-1}(\bgamma_1,\cdots,\bgamma_q)'\bX_S(\bX_S'\bX_S)^{-1}. 
\end{eqnarray*}
We apply Sherman-Woodbury formula for $(\bI-(\bgamma_1,\cdots,\bgamma_q)'\bX_S(\bX_S'\bX_S)^{-1}\bX_S'(\bgamma_1,\cdots,\bgamma_q))^{-1}$, and let $\bA=\bI$, $\bU=-(\bgamma_1,\cdots,\bgamma_q)'\bX_S$, $\bC=(\bX_S'\bX_S)^{-1}$, $\bV=\bX_S'(\bgamma_1,\cdots,\bgamma_q)$. Then we can write 
\begin{eqnarray*}
n\Delta&=&(\frac{1}{n}\bX_S'\bX_S)^{-1}+(\frac{1}{n}\bX_S'\bX_S)^{-1}\frac{1}{n}\bX_S'(\bgamma_1,\cdots,\bgamma_q)(\bgamma_1,\cdots,\bgamma_q)'\bX_S(\frac{1}{n}\bX_S'\bX_S)^{-1}\\
&&+(\frac{1}{n}\bX_S'\bX_S)^{-1}\frac{1}{n}\bX_S'(\bgamma_1,\cdots,\bgamma_q)(\bgamma_1,\cdots,\bgamma_q)'\bX_S\\
&&\times (\frac{1}{n}\bX_S'\bX_S-\frac{1}{n}\bX_S'(\bgamma_1,\cdots,\bgamma_q)(\bgamma_1,\cdots,\bgamma_q)'\bX_S)^{-1}\\
&&\times \frac{1}{n}\bX_S'(\bgamma_1,\cdots,\bgamma_q)(\bgamma_1,\cdots,\bgamma_q)'\bX_S(\frac{1}{n}\bX_S'\bX_S)^{-1}. 
\end{eqnarray*}
Under the conditions of Theorem \ref{thm3}, we have 
\begin{equation*} 
n\Delta\stackrel{{\mathcal{P}}}\rightarrow\bSigma_{SS}^{-1}+\bSigma_{SS}^{-1}\xi\bSigma_{SS}\bSigma_{SS}^{-1}+\bSigma_{SS}^{-1}\xi\bSigma_{SS}(\bSigma_{SS}-\xi\bSigma_{SS})^{-1}\xi\bSigma_{SS}\bSigma_{SS}^{-1}=\frac{1}{1-\xi}\bSigma_{SS}^{-1}. 
\end{equation*}
For the irrepresentable condition, by Theorem \ref{thm3}, we have
\begin{equation*}
\|(\bX_S'(\bI-\bP_Z)\bX_S)^{-1}\bX_S'(\bI-\bP_Z)\bX_{S^c}\|_1-(1-\xi)\|(\bX_S'\bX_S)^{-1}\bX_S'\bX_{S^c}\|_1\stackrel{{\mathcal{P}}}\rightarrow 0. 
\end{equation*}
Therefore, in the asymptotic sense, the requirement for $\gamma_{PROD}$ is 
\begin{equation*}
\frac{1-\gamma_{PROD}}{1-\xi}=1-\gamma_X, 
\end{equation*}
which is equivalent to $\gamma_{PROD}=\xi+\gamma_X-\xi\gamma_X$. 

In the proof of Theorem \ref{thm3}, we have also shown that $n^{-1}\bX_S'(\bI-\bP_Z)\bX_S-(1-\xi)n^{-1}\bX_S'\bX_S\stackrel{{\mathcal{P}}}\rightarrow 0$. Thus, $C_{\min}^*=(1-\xi)C_{\min}$. Now considering the minimum signal strength requirement, for PROD method, we evaluate 
\begin{equation}\label{rg}
q_2=\frac{1}{\gamma_X(1-\xi+\frac{\xi}{\gamma_X})}\Big[\frac{1}{1-\xi}\|\bSigma_{SS}^{-1}\|_{\infty}+\frac{4\sigma}{\sqrt{(1-\xi)C_{\min}}}\Big],
\end{equation}
while for the original $\bX$, we look at 
\begin{equation}\label{sigRG}
q_1=\frac{1}{\gamma_X}\Big[\|\bSigma_{SS}^{-1}\|_{\infty}+\frac{4\sigma}{\sqrt{C_{\min}}}\Big]. 
\end{equation}
For the first term in (\ref{rg}), to achieve a reduction compared with (\ref{sigRG}) for the original $\bX$, we need
\begin{equation*}
\frac{1}{(1-\xi+\frac{\xi}{\gamma_X})(1-\xi)}\leq 1
\end{equation*}
which is equivalent to $\gamma_X\leq \frac{1-\xi}{2-\xi}$. 

For the second term in (\ref{rg}), to achieve a reduction compared with $\bX$, we need
\begin{equation*}
\frac{1}{1-\xi+\frac{\xi}{\gamma_X}}\frac{1}{\sqrt{1-\xi}}\leq 1
\end{equation*}
which is equivalent to $\gamma_X\leq\frac{\xi\sqrt{1-\xi}}{1-(1-\xi)^{3/2}}$. 

Thus, when $\gamma_X\leq \min(\frac{1-\xi}{2-\xi},\frac{\xi\sqrt{1-\xi}}{1-(1-\xi)^{3/2}})$, PROD with random generation can achieve a reduction in the required minimum signal strength along with a reduction in the irrepresentable condition compared with the original $\bX$. 
\end{appendix}

\end{document}